\documentclass[journal]{IEEEtran}
\usepackage[T1]{fontenc}
\usepackage{caption}
\usepackage{times}
\usepackage{epsfig}
\usepackage{graphicx}
\usepackage{amsmath}
\usepackage{amssymb}
\usepackage{verbatim}
\usepackage[numbers,sort]{natbib}
\usepackage{diagbox}
\usepackage{subfigure}
\usepackage{upgreek}
\usepackage{multirow}
\usepackage{color}
\usepackage{bm}

\usepackage{verbatim}
\usepackage{amsthm}

\usepackage{booktabs}       
\usepackage{latexsym}
\usepackage{makecell}
\usepackage{boldline}
\setcellgapes{3pt}
\usepackage{amsfonts}       
\usepackage{microtype}      

\usepackage{arydshln}
\usepackage{tabularx}
\newcommand{\etal}{\textit{et\ al}.}

\ifCLASSINFOpdf

\else

\fi

\begin{document}

\title{CameraNet: A Two-Stage Framework for Effective Camera ISP Learning}

\author{Zhetong~Liang,
	Jianrui Cai,
	Zisheng Cao,
	and~Lei~Zhang,~\IEEEmembership{Fellow,~IEEE}
	\thanks{Zhetong Liang, Jianrui Cai and Lei Zhang are with the Department
		of Computing, The Hong Kong Polytechnic University, Hong Kong. E-mail: \{csztliang, csjcai, cslzhang\}@comp.polyu.edu.hk.}
	\thanks{Zisheng Cao is with DJI Co.,Ltd. E-mail: zisheng.cao@dji.com,}
}

\markboth{}%
{Shell \MakeLowercase{\textit{et al.}}: Bare Demo of IEEEtran.cls for IEEE Journals}

\maketitle

\begin{abstract}
Traditional image signal processing (ISP) pipeline consists of a set of individual image processing components onboard a camera to reconstruct a high-quality sRGB image from the sensor raw data. Due to the hand-crafted nature of the ISP components, traditional ISP pipeline has limited reconstruction quality under challenging scenes. Recently, the convolutional neural networks (CNNs) have demonstrated their competitiveness in solving many individual image processing problems, such as image denoising, demosaicking, white balance and contrast enhancement. However, it remains a question whether a CNN model can address the multiple tasks inside an ISP pipeline simultaneously. We make a good attempt along this line and propose a novel framework, which we call CameraNet, for effective and general ISP pipeline learning. The CameraNet is composed of two CNN modules to account for two sets of relatively uncorrelated subtasks in an ISP pipeline: restoration and enhancement. To train the two-stage CameraNet model, we specify two groundtruths that can be easily created in the common workflow of photography. CameraNet is trained to progressively address the restoration and the enhancement subtasks with its two modules. Experiments show that the proposed CameraNet achieves consistently compelling reconstruction quality on three benchmark datasets and outperforms traditional ISP pipelines. 

\end{abstract}


\IEEEpeerreviewmaketitle

\begin{figure*}[ht]
	\begin{center}
		\includegraphics[width=7in]{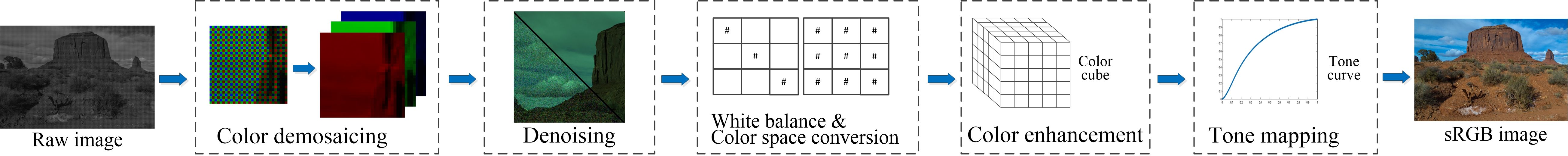}\hfill
	\end{center}
	\vspace{-0.1in}
	\caption{Major imaging components/stages in a traditional camera image processing pipeline}
	\label{fig1}
	\vspace{-0.15in}
\end{figure*}

\section{Introduction}

\IEEEPARstart{T}{he} raw image data captured by camera sensors are typically red, green and blue channel-mosaiced irradiance signals containing noise, incorrect colors and tones \cite{Ramanath2005,karaimer2016}. To reconstruct a displayable high-quality sRGB image, a camera image signal processing (ISP) pipeline is generally required, which consists of a set of cascaded components, including color demosaicking, denoising, white balance, color space conversion, tone mapping and color enhancement, etc. The performance of an ISP pipeline plays the key role for the quality of sRGB images output from a camera.

Traditionally, the ISP pipeline is designed as a set of hand-crafted modules that are performed in sequence \cite{Ramanath2005}. For instance, the denoising component may be designed as local filtering and the color enhancement may be modeled as a 3D lookup table search \cite{karaimer2016}. In such traditional designs, the ISP components are usually designed separately without considering the interaction between them, which may cause cumulative errors in the final output \cite{PulliTOG14}. This drawback impedes the reconstruction quality for challenging scenarios, resulting in images with high noise level, low dynamic range and less vivid color. Moreover, the development of traditional ISP pipeline can be time-consuming because each component needs to be tuned for satisfactory reconstruction quality. Besides the basic ISP pipeline, there are also some complex design based on multi-image capture in the literature \cite{hasinoff2016,liu2014,Mildenhall2018CVPR}. They leverage the information of multiple captures of a scene to improve the dynamic range. However, these methods are subject to image alignment techniques, which could lead to ghost artifacts caused by object motion. 

Recently, deep learning based methods have shown leading performance on low-level vision tasks, some of which are closely related to ISP problems, including denoising \cite{Zhang2018tip,zhang2017beyond}, white balance \cite{Hu2017fc4,Simone2015CC}, color demosaicking \cite{Tan2017ColorID,Gharbi2016}, color enhancement \cite{Sheng2018deepphoto,gharbi2017deep,CaiGZ18}, etc. In these methods, task-specific datasets with image-pairs are leveraged to train the deep convolutional neural network (CNN) in an end-to-end manner in contrast to the hand-crafted design of different components in traditional ISP methods. Inspired by the success of CNN in these single tasks, one natural idea is to exploit deep-learning techniques to improve the ISP pipeline design.  Specifically, given a raw image as input, we want to train a CNN model to produce a high-quality sRGB image under challenging imaging environments, hence improving the photo-finishing mode of a camera. 

One straightforward way for deep-learning-based ISP pipeline design is to train a single CNN model as a unified ISP pipeline in an end-to-end manner. Actually, some recent works follow this idea to develop ISP pipelines for low-light denoising \cite{Chen2018CVPR} and exposure correction \cite{Schwartz2019tip}. However, this one-stage design strategy may result in limited learning capability of the CNN model because the learning of different ISP subtasks are mixed together. This arrangement results in weak network representation of the ISP subtasks that have diverse algorithm characteristics. Another approach is to employ a parametric model for each stage of an ISP pipeline and train them separately. However, it is difficult, even impossible, to obtain the groundtruth images for each stage of the ISP pipeline, and this will make it cumbersome to train an ISP model.

In this paper, we propose an effective and general framework for deep-learning-based ISP pipeline design, which includes a two-stage CNN called CameraNet and an associated training scheme. Specifically, based on the functionality of different components inside an ISP pipeline, we group them into two weakly correlated clusters, namely, the restoration and enhancement clusters. The two clusters, which are weakly correlated with each other, are respectively embodied by two different subnetworks. Accordingly, a restoration and an enhancement groundtruths are specified and used to train the CameraNet in three steps. The first two steps perform the major trainings of the two subnetworks in a separate manner, followed by a mild joint fine-tuning step. With this arrangement, the two-stage CameraNet allows collaborative processing of correlated ISP subtasks while avoiding mixed treatment of weakly correlated subtasks, which leads to high quality sRGB image reconstruction in various ISP learning tasks. In our experiments, CameraNet outperforms the state-of-the-art methods and obtains consistently compelling results on three benchmark datasets, including HDR+ \cite{hasinoff2016}, SID \cite{Chen2018CVPR} and FiveK datasets \cite{Vladimir2011fivek}. 

The rest of the paper is organized as follows. Section II reviews the related work. Section III presents the framework of CameraNet, including the CNN architecture and training scheme. Section IV presents the experimental results. Section V is the conclusion.

\section{Related Work}

Our work is related to camera ISP pipeline design and deep learning for low level vision, which are reviewed briefly as follows.

\subsection{Image Signal Processing Pipeline}

There exist various types of image processing components inside the ISP pipeline of a camera. The major ones include demosaicking, noise reduction, white balancing, color space conversion, tone mapping and color enhancement, as shown in Fig.\ \ref{fig1}. The demosaicking operation interpolates the single-channel raw image with repetitive mosaic pattern (e.g., Bayer pattern) into a full color image \cite{Zhang2011ColorDB}, followed by a denoising step to enhance the signal-to-noise ratio \cite{Dabov2007BM3d}. White balancing corrects the color that is shifted by illumination according to human perception \cite{Cheng2015BYW}. Color space conversion usually involves two steps of matrix multiplication. It firstly transforms the raw image in camera color space to an intermediate color space (e.g., CIE XYZ) for processing and then transforms the image to sRGB space for display. Tone mapping compresses the dynamic range of the raw image and enhances the image details \cite{YuanECCV2012}. Color enhancement operation manipulates the color style of an image, usually in the form of 3D lookup table (LUT) search. A detailed survey of the ISP components can be found in \cite{Ramanath2005,karaimer2016}.

In the design of a traditional ISP pipeline, each algorithm component is usually developed and optimized independently without knowing the effect to its successors. This may cause error accumulation along the algorithm flow in the pipeline \cite{PulliTOG14}.  Moreover, each step inside an ISP pipeline is characterized by simple algorithms which are not able to tackle the challenging imaging requirement by ubiquitous cellphone photography. Recently, there are a few works that apply the learning-based approach to the ISP pipeline design. One pioneering work of this type is Jiang \etal’s affine mapping framework \cite{Jiang2017}. In this work, the raw image patches are clustered based on simple features and then a per-class affine mapping is learned to map the raw patches to the sRGB patches. This learning-based approach has limited regression performance due to the use of simple parametric model. Chen \etal\ proposed a multiscale CNN for nighttime denoising \cite{Chen2018CVPR}. They constructed a denoising dataset and the CNN model is trained to convert a noisy raw image to a clean sRGB image. Schwartz \etal\ proposed a CNN architecture called DeepISP that learns to correct the exposure \cite{Schwartz2019tip}. One common limitation of Chen’s and Schwartz’s models is that they only considered one single aspect of ISP pipeline. Their task-specific architectures are not general enough to model the various components inside an ISP pipeline. 

There exist a few datasets with different imaging scenarios that can be used for ISP pipeline learning \cite{hasinoff2016,Chen2018CVPR,Vladimir2011fivek}. These datasets contain raw images and the corresponding groundtruth sRGB images that are manually processed and retouched in a controlled setting. The HDR+ dataset is featured with burst denoising and sophisticated style retouching \cite{hasinoff2016}; the SID dataset is featured with nighttime denoising \cite{Chen2018CVPR}; and the FiveK dataset contains groundtruth images that are retouched by five photographers to have different color styles \cite{Vladimir2011fivek}.

\subsection{Deep Learning For Low-level Vision}

The deep learning techniques have been widely used in low-level vision research, including image restoration \cite{zhang2017beyond,Hu2017fc4,Dong2014eccv,Zhang2018tip} and enhancement \cite{gharbi2017deep,Sheng2018deepphoto,Chen2017ICCV,CaiGZ18}. Many task-specific CNNs have been proposed to address the single imaging tasks. Zhang \etal\ proposed a deep CNN featured with batch normalization to address the denoising task \cite{zhang2017beyond}. Dong \etal\ proposed a three layer CNN for super-resolution task \cite{Dong2014eccv}. Gharbi \etal\ addressed the photographic style enhancement task by learning to estimate the per-pixel affine mapping in the data structure of bilateral grid \cite{gharbi2017deep}. 

There also exist some joint solutions for image restoration tasks \cite{Gharbi2016,Zhou2018dmsr,Qian2019Trinity,VandewalleKAS07}. Gharbi \etal\ proposed a feedforward CNN for effective joint denoising and demosaicking \cite{Gharbi2016}. Zhou \etal\ developed a residual network for joint demosaicking and super-resolution \cite{Zhou2018dmsr}. Recently, Qian \etal\ proposed a joint solution for denoising, super-resolution and demosaicking starting from raw images \cite{Qian2019Trinity}. Despite these successes in joint restoration tasks, little work has been conducted on the ISP pipeline learning, which is a complex mixture of image restoration and enhancement tasks. In this paper, we address this problem and unify the whole ISP pipeline with a two-stage CNN model.

\begin{figure}[t]
	\begin{center}
		\subfigure{
			\centering
			\includegraphics[width=3in]{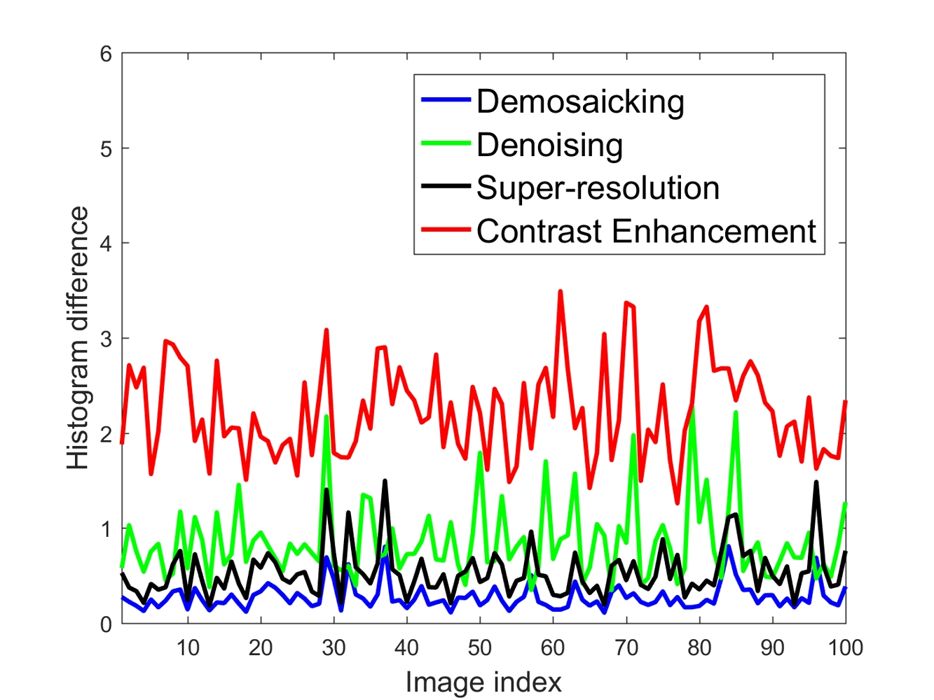}\hfill
		}	
	\end{center}
	\vspace{-0.1in}
	\caption{The image histogram change caused by various image operators. Individual operators, including demosaicking, denoising (with noise $\sigma=20$), 4$\times$ super-resolution and contrast enhancement \cite{WangTIP2013}, are evaluated on the images in BSD100 dataset \cite{MartinICCV2001bsd}. The $\ell_{1}$ norm of histogram differences of each image for each operator is plotted.}
	\label{fig2}
	\vspace{-0.15in}
\end{figure}

\section{Framework}

In this section we describe our learning framework that consists of a two-stage CNN system and the associated training scheme for ISP pipeline design, which aims to achieve high learning performance in various imaging environments. 

\vspace{-0.05in}

\subsection{Problem Formulation}

Suppose there are $N$ essential subtasks in an ISP pipeline, which includes but is not restricted to demosaicking, white balance, denoising, tone mapping and color enhancement and color conversion. The traditional ISP pipeline employs $N$ cascaded hand-crafted algorithm components to address these subtasks, respectively. Let $I_{raw}^{cfa}$ and $I^{srgb}$ be the raw image with specific color filter array (CFA) pattern and the output sRGB image, respectively. The traditional ISP can be denoted as $I^{srgb}=f_N(f_{N-1}(\dots(f_1(I_{raw}^{cfa})\dots))$, where $f_i, 1\leq i\leq N$, denotes the $i$th algorithm component. The main drawback of such traditional ISP design is that each algorithm component is usually hand-crafted independently without considering much its interaction with other components, which limits the quality of the reconstructed sRGB image. In addition, it is time-consuming to develop the $N$ components because they require careful design and parameter tuning for a specific camera sensor. 

In contrast to the traditional ISP design, we adopt the data-driven method and model an ISP as a deep CNN system to address the $N$ subtasks:

\vspace{-0.1in}
\begin{equation}
\label{fn1}
\begin{split}
I^{srgb}
=M_{isp}(I_{raw}^{cfa},\omega;\theta),
\end{split}
\end{equation} 

\noindent where $M_{isp}(.;\theta)$ refers to the CNN model with parameters $\theta$ to be optimized. $\omega$ denotes the optional input camera metadata that can be used to help the training and inference. We leverage a dataset $S$ to train $M_{isp}$ in a supervised manner. The dataset contains for each scene a group of images, including the input raw image $I_{raw}^{cfa}$ and $K$ groundtruth images $G_i,1\leq i\leq K$. The case $K=1$ indicates that there is only one final groundtruth output. The case $K>1$ indicates that multiple groundtruths are leveraged to train the network in different aspects. $G_K$ indicates the groundtruth for the final sRGB reconstruction. 

To design a CNN model with high learning capability, the network structure should ensure good representation of the task at hand. For example, in the task of image classification, the network structures are usually designed to characterize the feature hierarchy of the semantic object \cite{ZeilerF14}. In the design of the CNN model $M_{isp}(.;\theta)$, it is desirable that the CNN model can explicitly address different ISP tasks, while keeping the network as simple as possible. To this end, we propose an effective yet compact two-stage CNN architecture, which is described in the following sections. 

\begin{figure*}[t]
	\begin{center}
		\includegraphics[width=7.2in]{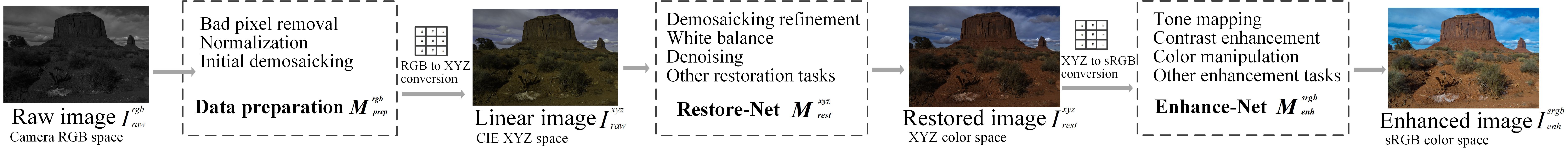}\hfill
	\end{center}
	\vspace{-0.1in}
	\caption{The proposed CameraNet system for ISP pipeline.}
	\label{fig3}
	\vspace{-0.05in}
\end{figure*}

\begin{figure*}[t]
	\begin{center}
		\includegraphics[width=6.8in]{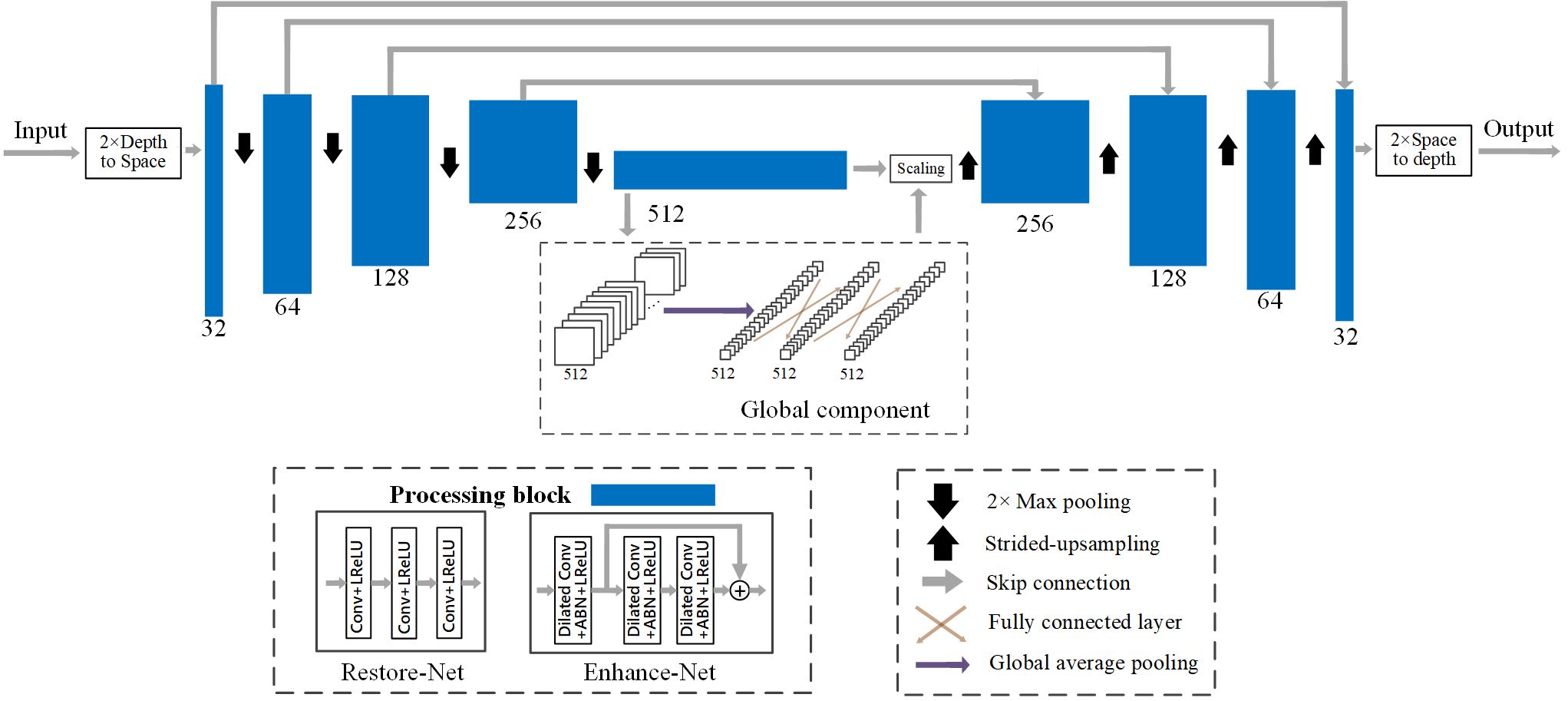}\hfill
	\end{center}
	\vspace{-0.1in}
	\caption{The U-Net model for Restore-Net and Enhance-Net modules in the proposed CameraNet system.}
	\label{fig4}
	\vspace{-0.15in}
\end{figure*}

\vspace{-0.05in}

\subsection{Two-stage Grouping}

As discussed in the previous subsection, we want the CNN model $M_{isp}$ to have a good representation of the various ISP components. However, this is not trivial because each of the ISP subtasks could be a complicated stand-alone research topic in the literature. One possible way to enhance the representational capability is to deploy a CNN subnetwork to address each ISP subtask and chain them in sequence \cite{Lin2011,chakrabarti2009}. However, this method makes the whole network cumbersome and difficult to train. Actually, it has been demonstrated that some ISP subtasks, e.g., demosaicking and denoising, are correlated and can be jointly addressed \cite{HirakawaTIP2006,Gharbi2016}. On the other hand, if we deploy a single CNN module to model the whole ISP pipeline, the reconstruction performance may not be satisfactory, either. This is because some ISP subtasks are weakly correlated and are hard to use a single module to represent.

Our idea for solving this problem is to group the set of ISP pipeline subtasks into several weakly correlated clusters, while each cluster consists of correlated subtasks. Accordingly, a CNN module is deployed for each cluster of subtasks. In this way, we can allow joint learning within each cluster to gain model compactness while adopting independent learning across different clusters to increase the representational capability. Based on the existing works in low-level vision, we group the ISP subtasks into image restoration and enhancement clusters. Specifically, image restoration cluster is located at the front position of an ISP pipeline, mainly including demosaicking, denoising and white balance. In contrast, the enhancement cluster contains the following ISP operations, including exposure adjustment, tone mapping and color enhancement. According to the influences on image content, the two clusters of operations have very different algorithm behaviors. The restoration operations aim to faithfully reconstruct the linear scene irradiance. They usually maintain the image distribution without largely changing the brightness and contrast style of an image. In contrast, the cluster of enhancement operations attempt to nonlinearly exaggerate the visual appearance of an image in aspects of color style, brightness and local contrast, producing images fitting better human perception. 

To visualize the differences between restoration and enhancement operators, we perform a test where the influences of different operators on image distribution are evaluated. These operators include demosaicking, denoising ($\sigma=20$), super-resolution and contrast enhancement. We preclude white balance because it can be simply accounted for by per-channel global scaling. We denote an image before and after operation $f(.)$ as $L$ and $f(L)$, respectively. The histogram vectors of $L$ and $f(L)$ with 256 bins are calculated. Then, the $\ell_{1}$ norm of histogram error vectors between $L$ and $f(L)$ are recorded to indicate the amount of change on the data distribution by operator $f(.)$. We use the BSD100 images for evaluation \cite{MartinICCV2001bsd}. For the restoration operators, we use the original images as $f(L)$ and manually degrade them to obtain $L$. For enhancement operator, we use the enhancement algorithm in \cite{WangTIP2013}. Fig.\ \ref{fig2} shows the $\ell_{1}$ norm of histogram errors per image. We can see that the enhancement operator imposes a much stronger change on the image distribution than the other restoration operators. The fact that the enhancement and restoration operations have substantially different impacts on the image distribution motivates us to separate them in the ISP pipeline learning.

\begin{figure*}[htp]
	\begin{center}
		\subfigure{
			\centering
			\includegraphics[width=7.2in]{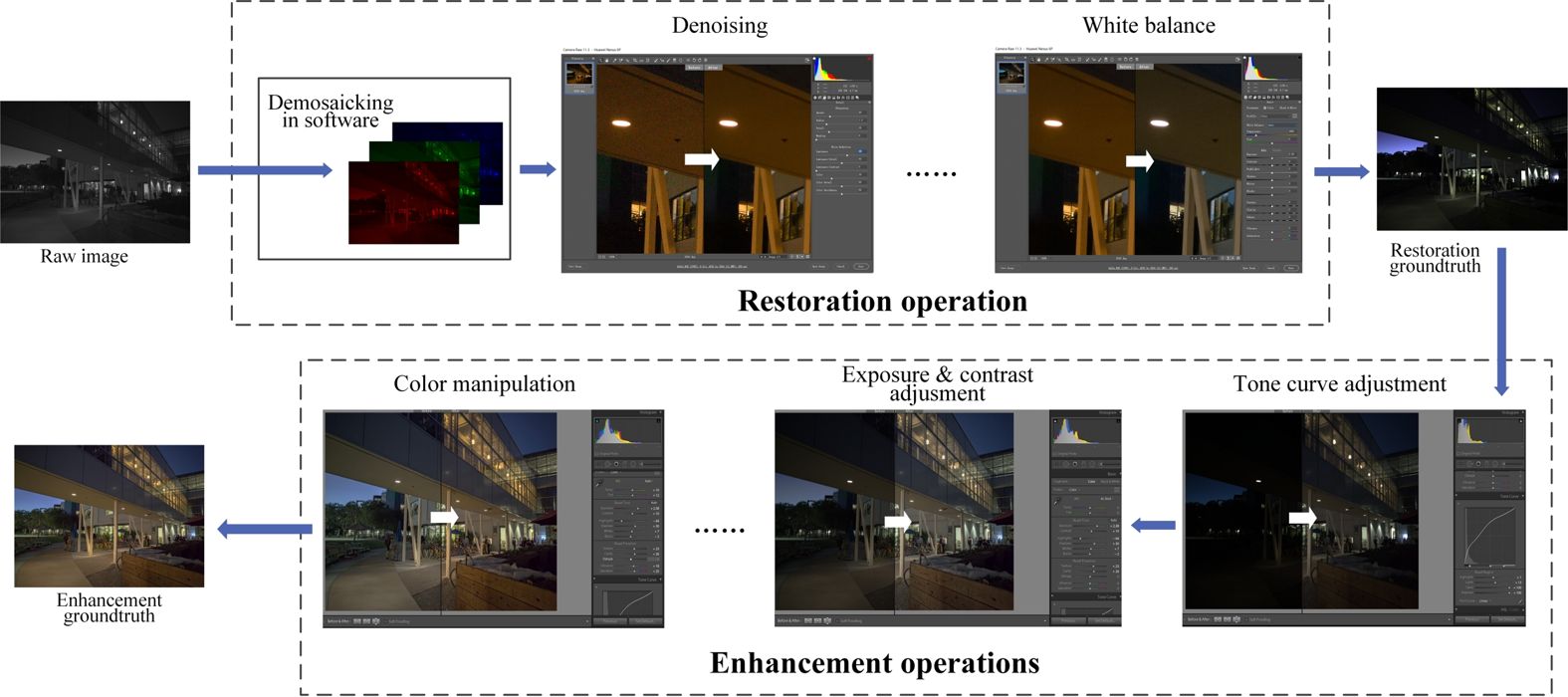}\hfill
		}
	\end{center}
	\caption{The workflow of creating two groundtruths with Adobe software for CameraNet training. The restoration groundtruth is created in Adobe Camera Raw, while the enhancement groundtruth is created in Lightroom.}
	\label{fig5}
\end{figure*}

\vspace{-0.02in}

\subsection{Network Structure Design}

Following the two-stage grouping strategy, the proposed CNN system, which is called CameraNet, is illustrated in Fig.\ \ref{fig3}. It has a data preparation module, a restoration module called Restore-Net and an enhancement module called Enhance-Net. 

The data preparation module $M_{prep}^{rgb}$ applies several pre-processing operations on the input raw image $I_{raw}^{cfa}$, including bad pixel removal, dark and white level normalization, vignetting compensation and initial demosaicking, which can be described as

\vspace{-0.15in}

\begin{equation}
\label{fn2}
\vspace{-0.01in}
\begin{split}
I_{raw}^{rgb} = M_{prep}^{rgb}(I_{raw}^{cfa}).
\end{split}
\end{equation}

\noindent For Bayer-pattern-based CFA, we use the interpolation kernels proposed in \cite{Tan2017ColorID} for initial demosaicking. The role of the data preparation module is to rule out the fixed camera-specific operations from the training process to reduce the learning burden. Then, $I_{raw}^{rgb}$ is converted to CIE XYZ color space with color matrix $C_{xyz}$, denoted as $I_{raw}^{xyz}=C_{xyz}\cdot I_{raw}^{rgb}$, as the input of Restore-Net. For this color conversion process, we use the color matrix recorded in camera metadata\footnote{There are two RGB-to-XYZ color matrices in the metadata, recorded as ColorMatrix1 and ColorMatrix2 tags. Although more sophisticated illumination-specific interpolation of the matrices can be adopted, we average the two matrices and take its inverse as the conversion matrix.}.

Then Restore-Net $M_{rest}^{xyz}$ applies the restoration-related operations on $I_{raw}^{xyz}$:

\vspace{-0.1in}

\begin{equation}
\label{fn3}
\vspace{-0.01in}
\begin{split}
I_{rest}^{xyz}& = M_{rest}^{xyz}(I_{raw}^{xyz};\theta_1) \\
&=
M_{rest}^{xyz}\Big(C_{xyz}\cdot M_{prep}^{rgb}\big( I_{raw}^{cfa}\big);\theta_1 \Big),
\end{split}
\end{equation}

\noindent where $\theta_1$ denotes the parameters to be optimized. The Restore-Net produces a restored intermediate image $I_{rest}^{xyz}$ that is white balanced, demosaicking-refined and denoised in the XYZ space. We operate $M_{rest}^{xyz}$ in XYZ space because it is a general color space which relates to human perception (the Y channel is designed to match the luminance response of human vision) \cite{Nguyen2017cvpr}. Then $I_{rest}^{xyz}$ is transformed to sRGB space with matrix $C_{srgb}$, denoted as $I_{rest}^{srgb}=C_{srgb}\cdot I_{rest}^{xyz}$. We use the standard XYZ-to-sRGB conversion matrix for this transformation. Finally, the Enhance-Net $M_{enh}^{srgb}$ performs tone mapping, detail enhancement and color style manipulation on $I_{rest}^{srgb}$ to produce the final sRGB image $I_{enh}^{srgb}$. This process can be described as:

\vspace{-0.08in}

\begin{equation}
\label{fn4}
\vspace{-0.01in}
\begin{split}
I_{enh}^{srgb} &= M_{enh}^{srgb}(I_{rest}^{srgb};\theta_2) \\
&= M_{enh}^{srgb}\Big(C_{srgb}\cdot M_{rest}^{xyz}\Big(C_{xyz}\cdot M_{prep}^{rgb}\big(I_{raw}^{cfa}\big),\theta_1\Big);\theta_2\Big),
\end{split}
\end{equation}

\noindent where $\theta_2$ denotes the parameters of Enhance-Net. We choose sRGB space for enhancement operations because it directly relates to display color space.

{\bf CNN modules}. We then describe the CNN architectures of Restore-Net and Enhance-Net. Although there could be many design choices for these two modules, we consider a simple yet effective one. We deploy two U-Nets with 5 scales for Restore-Net and Enhance-Net, respectively. As shown in Fig.\ \ref{fig4}, a U-Net has a contracting path containing progressive downsamplings to enlarge the spatial scale, followed by an expanding path with progressive upsampling to the original resolution \cite{Ronneberger2015unet}. Image structures are preserved by the skip connections of feature maps from the contracting path to the expanding path at the same scale. The advantages of a U-Net lie in the multiscale manipulation of image structures.

To account for the global processing in both modules (the white balance in Restore-Net and global enhancement in Enhance-Net), we deploy an extra global component in the U-Net modules, as shown in Fig.\ \ref{fig4}. The global component firstly applies global averaging pooling at the input feature maps on the 5th (largest) scale, followed by 2 fully connected layers to obtain the global scaling features as a 1-D vector. Finally, the global features are multiplied to the output feature maps on the 5th scale in a per-channel manner. This process can be described as:

\vspace{-0.1in}
\begin{equation}
\label{fn5}
\vspace{-0.01in}
\begin{split}
H_{5,out} = U_5(H_{5,in})\otimes F_c(F_c(P_{av}(H_{5,in}))),
\end{split}
\end{equation}

\noindent where $H_{5,out}$ and $H_{5,in}$ denote the output and input feature maps on the 5th scale, respectively. $U_5(.)$, $F_c(.)$ and $P_{av}(.)$ denote the scale-5 operation blocks of U-Net, the fully connected layer and the global pooling layer, respectively. $``\otimes"$ denotes per-channel multiplication.

To account for the differences between restoration and enhancement operations, there are several differences between the processing blocks of Restore-Net and Enhance-Net, as shown in the specification of Fig.\ \ref{fig4}. First, Restore-Net contains plain convolutional blocks, while Enhance-Net deploys a residual connection within a convolutional block to facilitate detail enhancement. Second, the convolution dilation rates of Enhance-Net are set to $\{$1,2,2,4,8$\}$ from the 1st scale to the 5th scale to enlarge the receptive field, whereas the dilation rates of Restore-Net remain 1. Finally, the Enhance-Net employs adaptive batch normalization (ABN) \cite{Chen2017ICCV} to facilitate enhancement learning, whereas it is not adopted in Restore-Net because we find that it will amplify noise.

\begin{figure*}[htp]
	\begin{center}
		\subfigure[Raw image]{
			\centering
			\begin{minipage}[b]{1.3in}
				\includegraphics[width=1.3in]{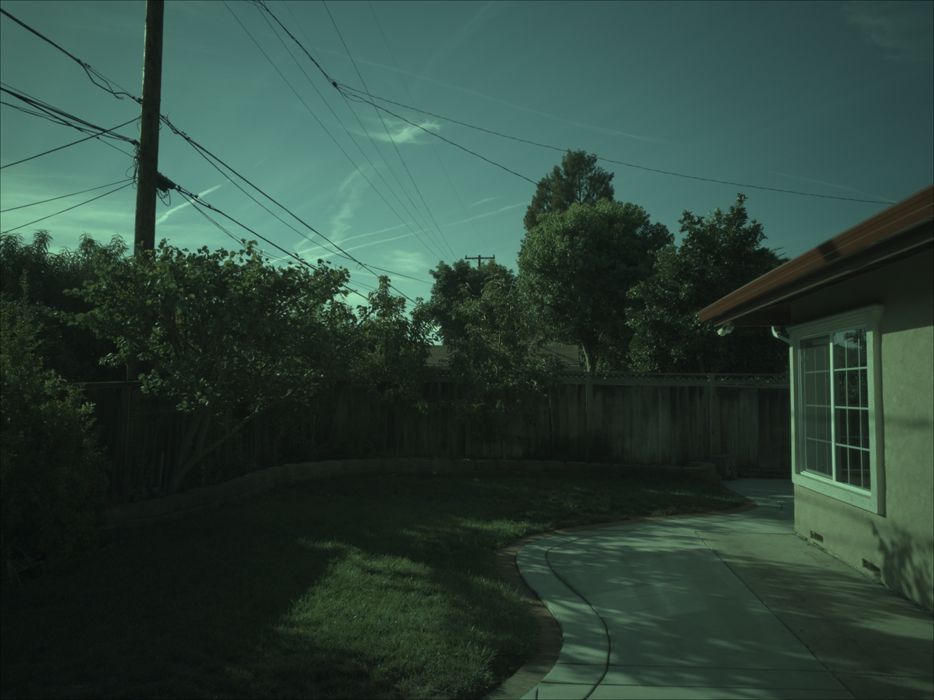}\hfill 
				\vspace{4pt}
				\includegraphics[width=1.3in]{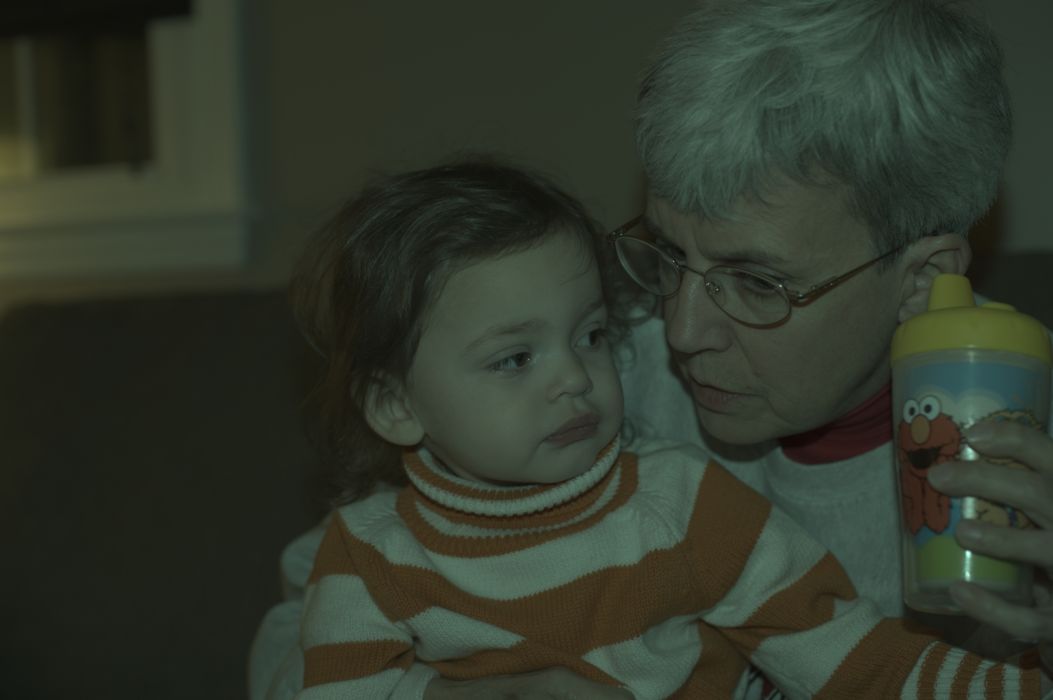}\hfill
			\end{minipage}
		}
		\hspace{-0.1in}
		\subfigure[Restored by Restore-Net]{
			\centering
			\begin{minipage}[b]{1.3in}
				\includegraphics[width=1.3in]{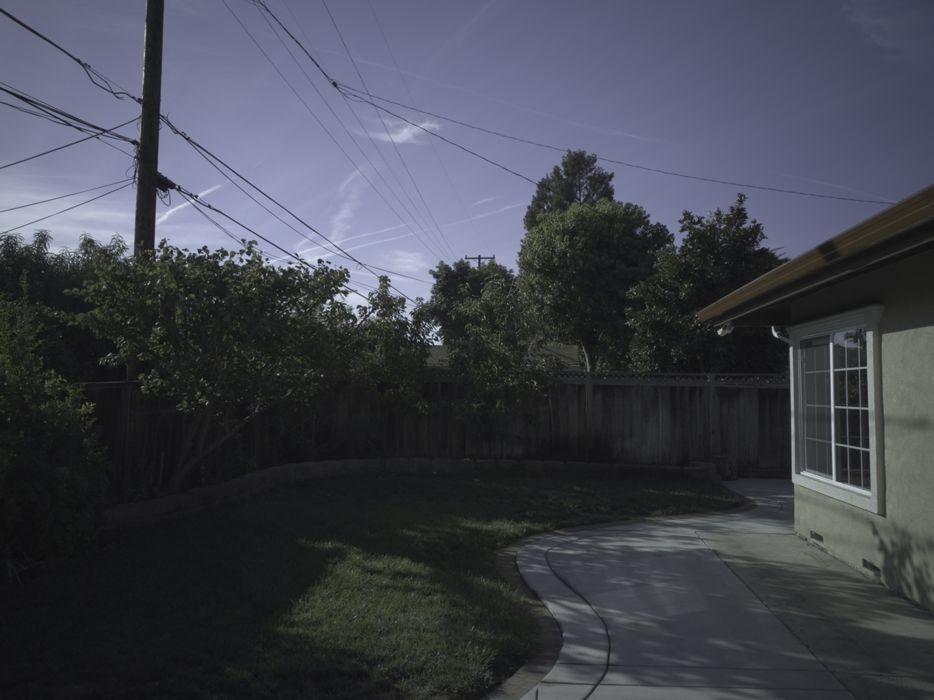}\hfill 
			    \vspace{4pt}
				\includegraphics[width=1.3in]{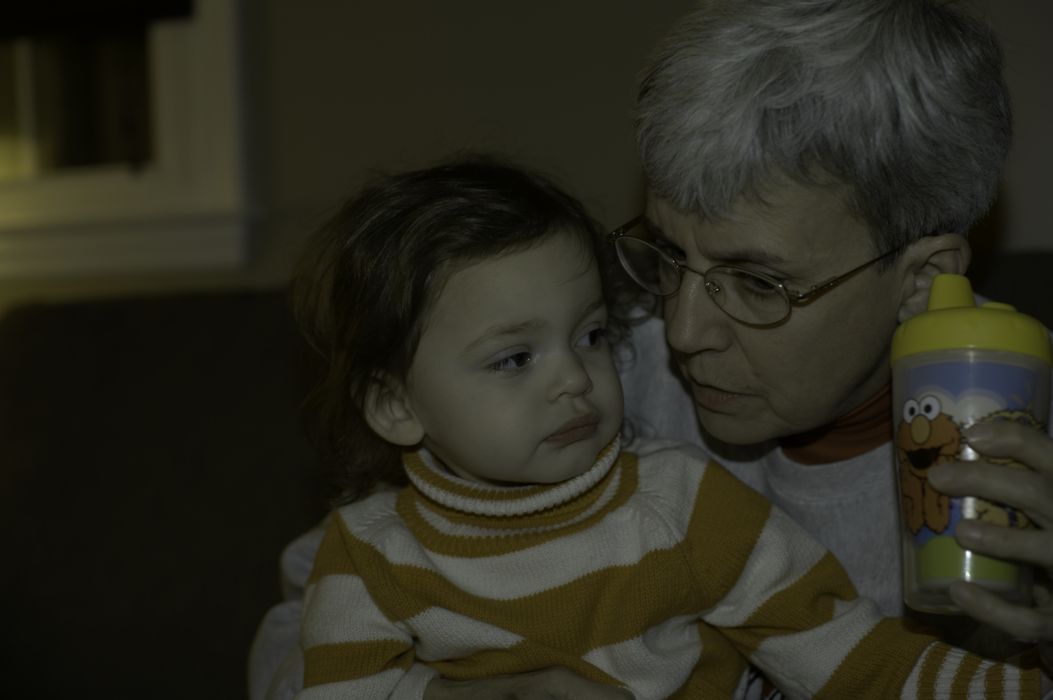}\hfill
		\end{minipage}
		}
	\hspace{-0.1in}
		\subfigure[Restoration groundtruth]{
			\centering
			\begin{minipage}[b]{1.3in}
				\includegraphics[width=1.3in]{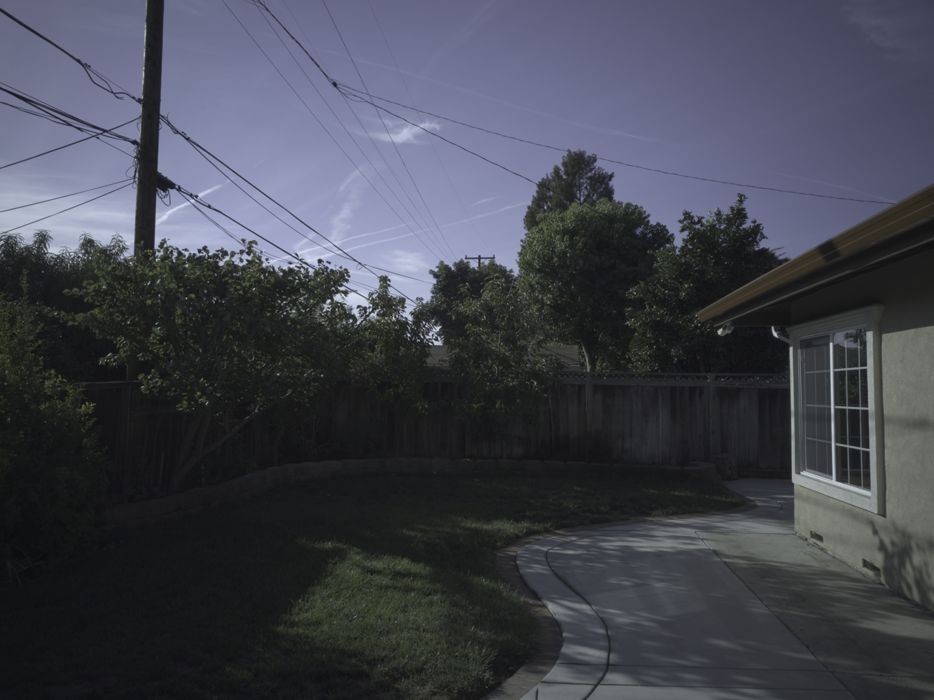}\hfill 
				\vspace{4pt}
				\includegraphics[width=1.3in]{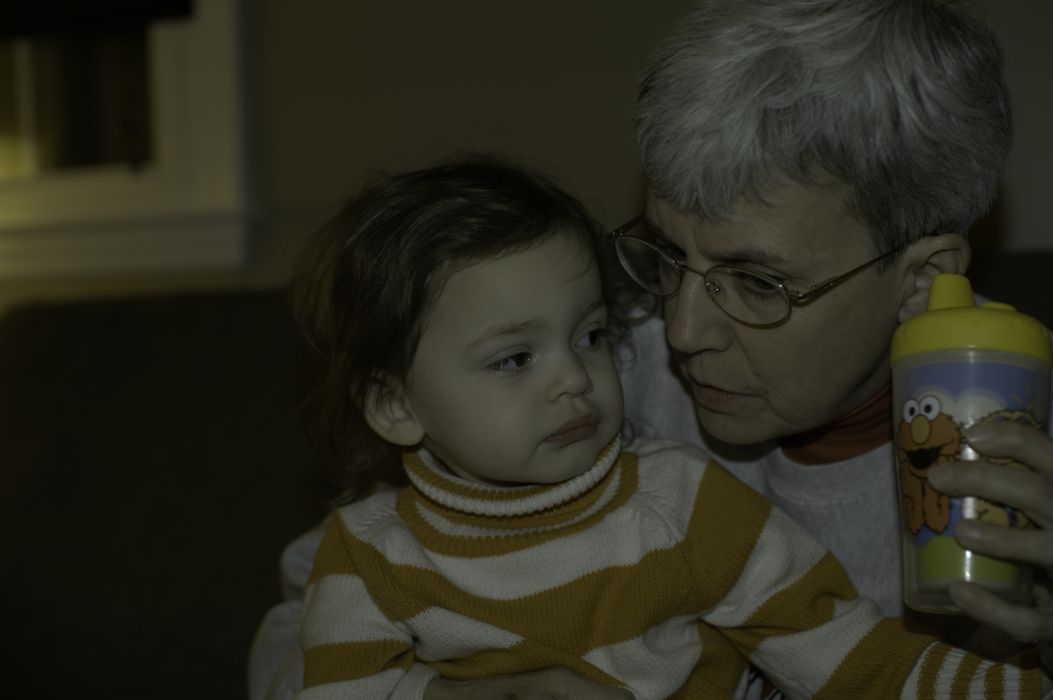}\hfill
			\end{minipage}
		}
	\hspace{-0.1in}
		\subfigure[Enhanced by Enhance-Net]{
			\centering
			\begin{minipage}[b]{1.3in}
				\includegraphics[width=1.3in]{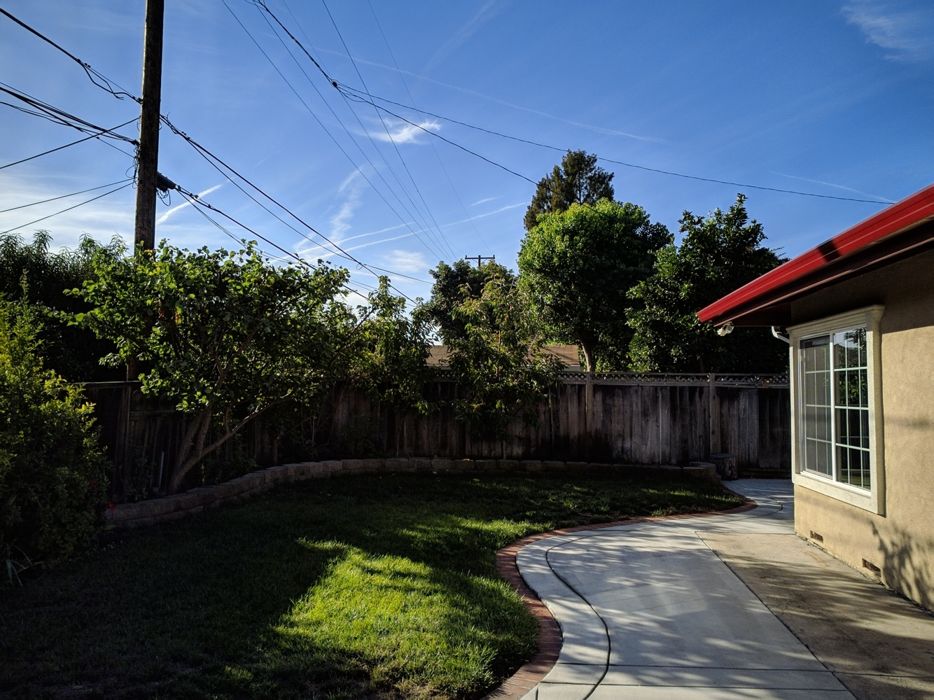}\hfill 
				\vspace{4pt}
				\includegraphics[width=1.3in]{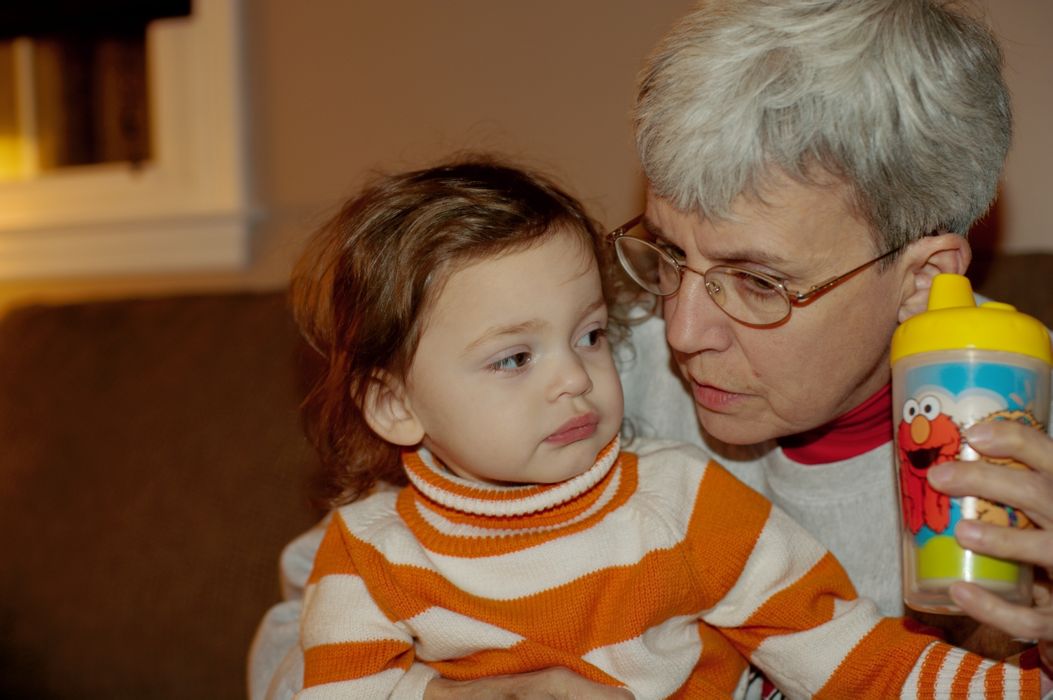}\hfill
			\end{minipage}
		}
	\hspace{-0.1in}
		\subfigure[Enhancement groundtruth]{
			\centering
			\begin{minipage}[b]{1.3in}
				\includegraphics[width=1.3in]{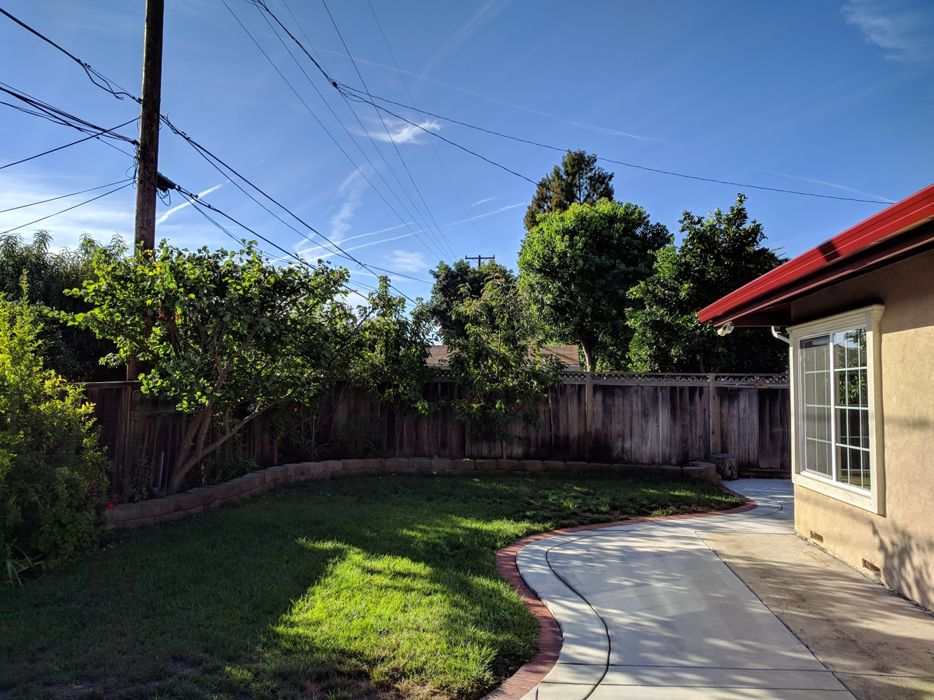}\hfill 
				\vspace{4pt}
				\includegraphics[width=1.3in]{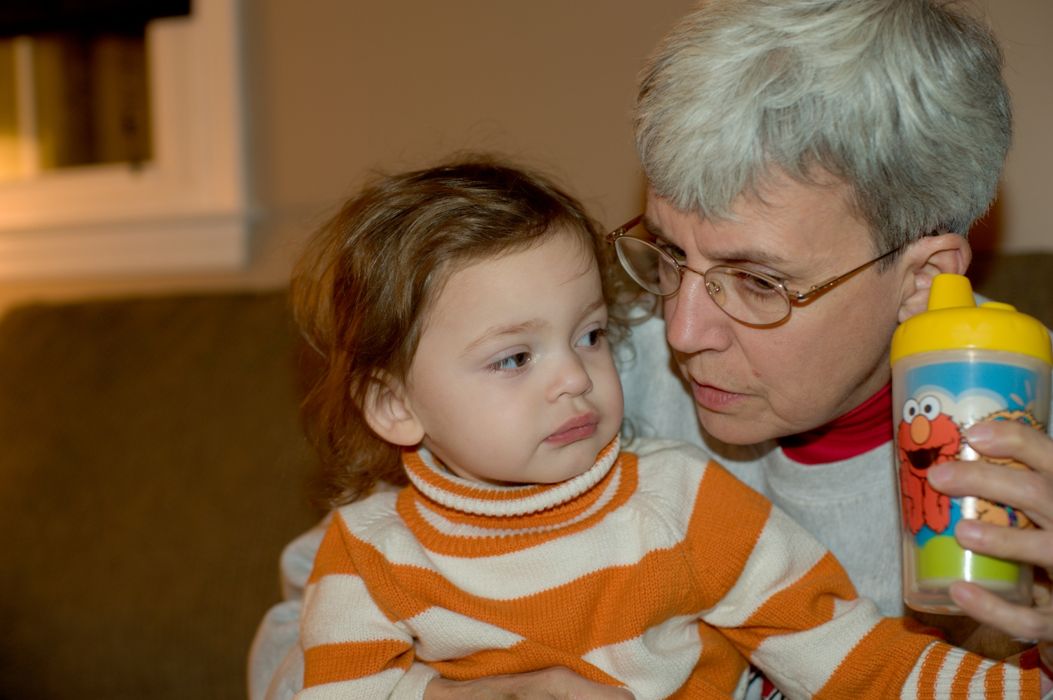}\hfill
			\end{minipage}
		}
	\end{center}
	\vspace{-0.08in}
	\caption{Illustration of the learned two-stage network outputs and groundtruths. The image in the first row is from the HDR+ dataset \cite{hasinoff2016}, while the image in the second row is from FiveK dataset \cite{Vladimir2011fivek}. A gamma transform with parameter 2.2 is applied to the raw images and restoration groundtruths for display.}
	\vspace{-0.08in}
	\label{fig6}
\end{figure*}

\vspace{-0.08in}
\subsection{Groundtruth Generation}

The proposed CNN system exports two images sequentially, the restored image $I_{rest}^{xyz}$ in XYZ space and the enhanced image $I_{enh}^{srgb}$ in sRGB space. The two groundtruth images $G_{rest}^{xyz}$ and $G_{enh}^{srgb}$ should be properly generated to train the CNN model.

The two groundtruth images can be easily generated, as shown in Fig.\ \ref{fig5}. The restoration groundtruth $G_{rest}^{xyz}$ is firstly created by applying restoration-related operations on the input raw images, including demosaicking, denoising, white balancing and color conversion into XYZ space. The creation of $G_{rest}^{xyz}$ involves only restoration operations to maintain the image distribution. Because of the objective nature of restoration tasks, the space of restoration groundtruth for a raw image is small. To obtain $G_{enh}^{srgb}$, enhancement-related retouching should be applied on top of $G_{rest}^{xyz}$, including contrast adjustment, tone mapping, color enhancement and color conversion into sRGB space. In contrast to $G_{rest}^{xyz}$, the creation of $G_{enh}^{srgb}$ mainly consists of subjective image manipulations. Thus, given a restored image, there could be various enhanced versions of the image by different human subjects. Fig.\ \ref{fig6} shows the image triplets from the HDR+ dataset and FiveK dataset, including an interpolated raw image, the restoration and the enhancement groundtruths. We also show the reconstructed images in the two stages by our CameraNet. One can see that the two restoration groundtruths possess certain similar visual attributes, whereas the two enhancement groundtruths are substantially different. The enhancement groundtruth in HDR+ dataset emphasizes on detail enhancements while that in FiveK+ dataset focuses on color style manipulation.

One can make use of various capturing and retouching methods to create the two groundtruths depending on the imaging environment. The simplest creation method is to use Adobe software or DCRaw to process the input raw image and sequentially obtain a restored image and an enhanced image as groundtruths. Besides this simple method, some sophisticated creation approach can be adopted. For example, in nighttime imaging, a burst of noisy raw images can be captured and fused into one clean raw image, which then goes through the restoration-related processing to obtain the restoration groundtruth \cite{hasinoff2016}. 

\begin{figure*}[ht]
	\begin{center}
		\subfigure[Raw image]{
			\begin{minipage}[b]{1.6in}
				\centering
				\includegraphics[width=1.6in]{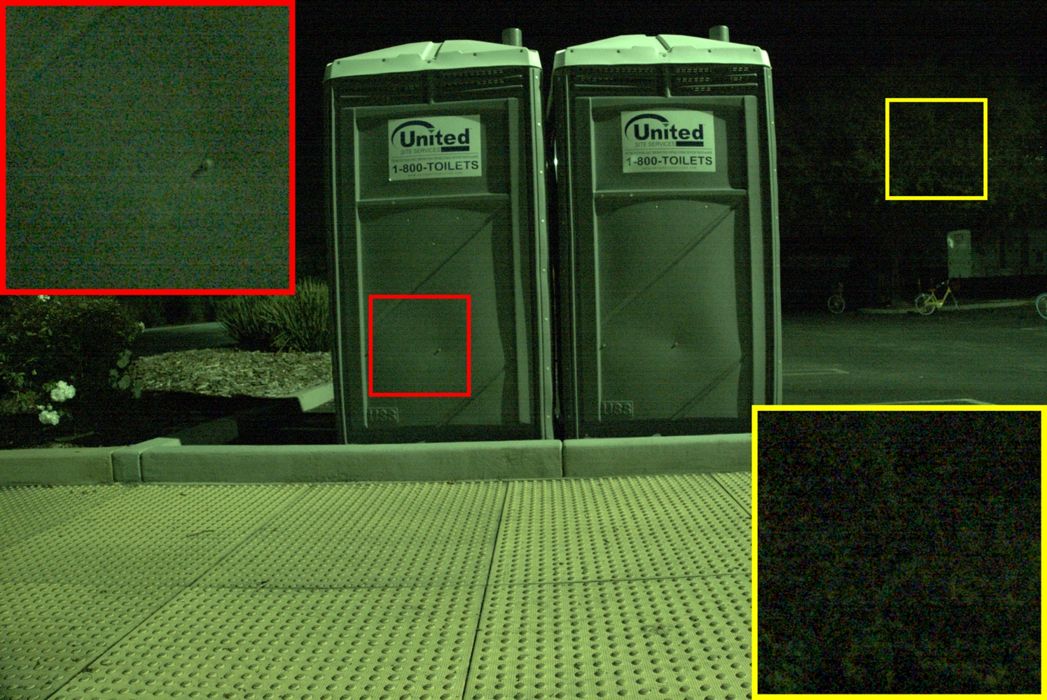}\hfill 
				\vspace{4pt}
				\includegraphics[width=1.6in]{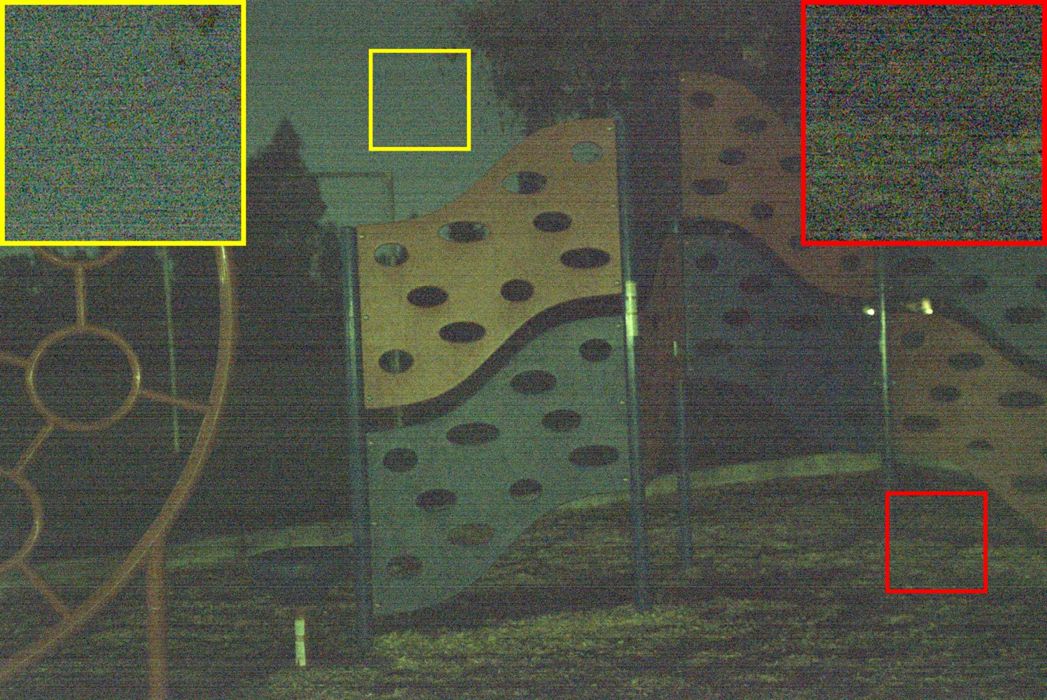}\hfill
			\end{minipage}	
		}
		\hspace{-0.1in}
		\subfigure[Result by one-stage setting]{
			\centering
			\begin{minipage}[b]{1.6in}
				\includegraphics[width=1.6in]{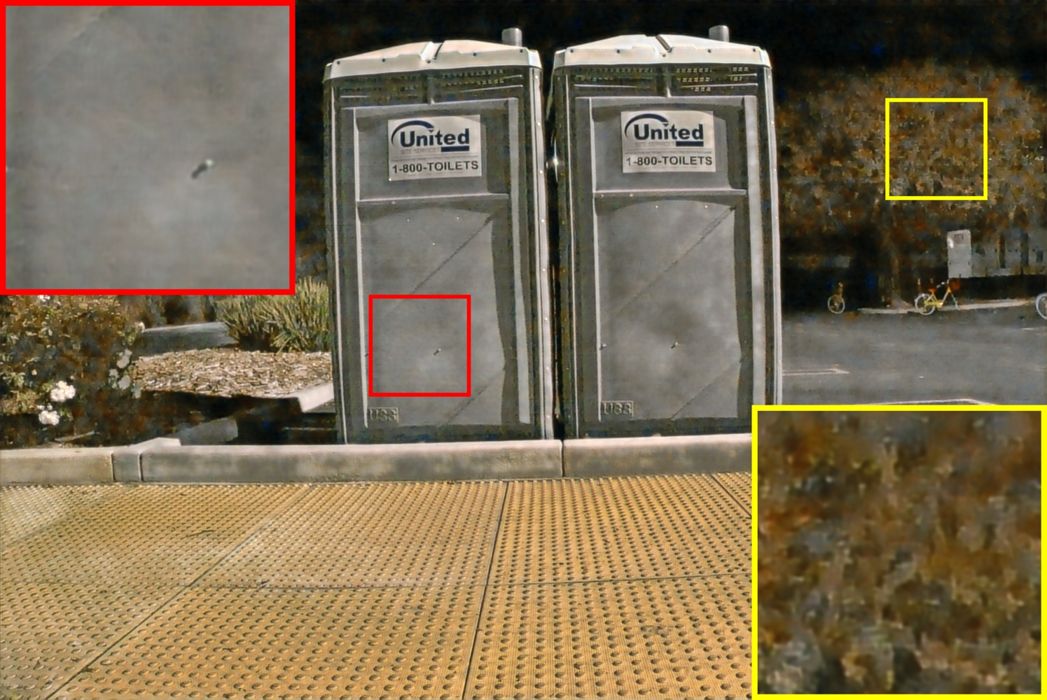}\hfill 
				\vspace{4pt}
				\includegraphics[width=1.6in]{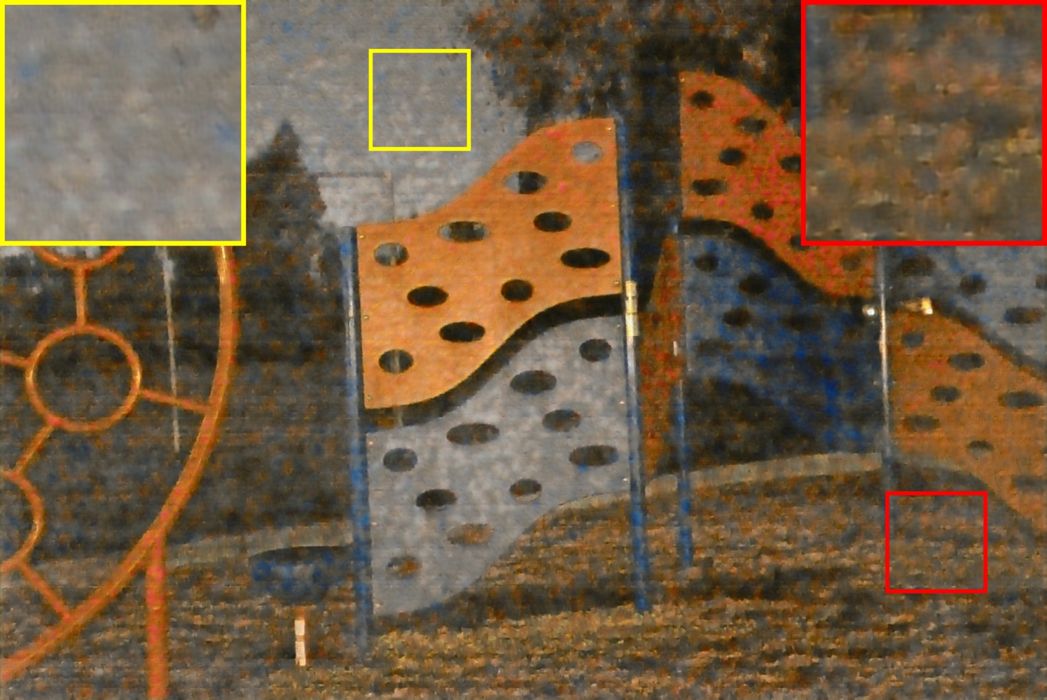}\hfill
			\end{minipage}
		}
		\hspace{-0.1in}
		\subfigure[Result by two-stage setting]{
			\centering
			\begin{minipage}[b]{1.6in}
				\includegraphics[width=1.6in]{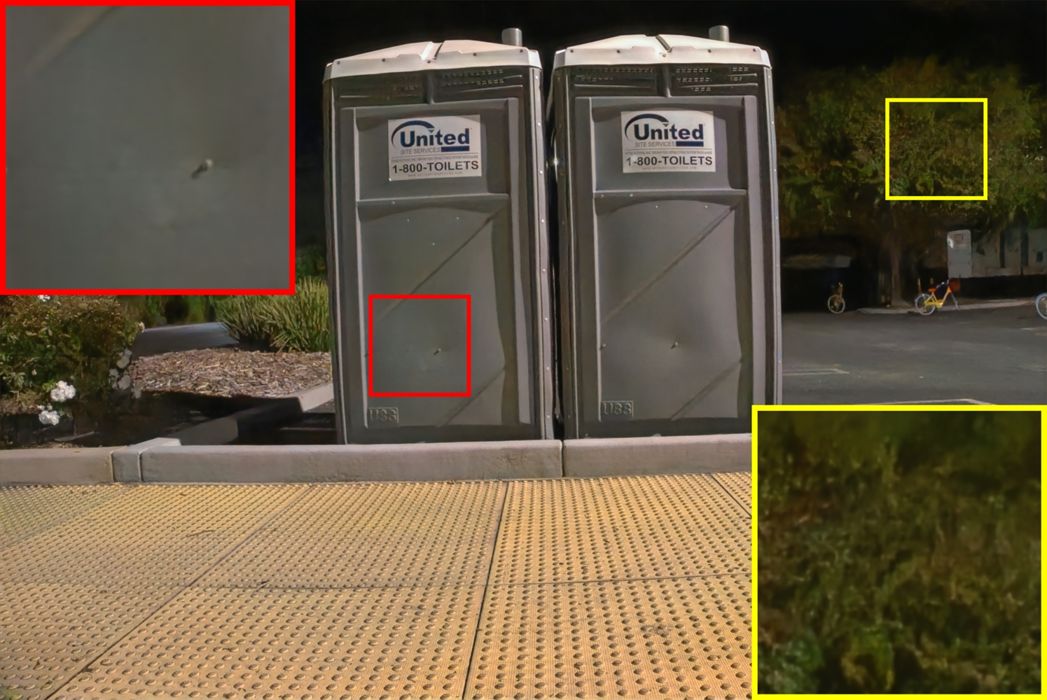}\hfill 
				\vspace{4pt}
				\includegraphics[width=1.6in]{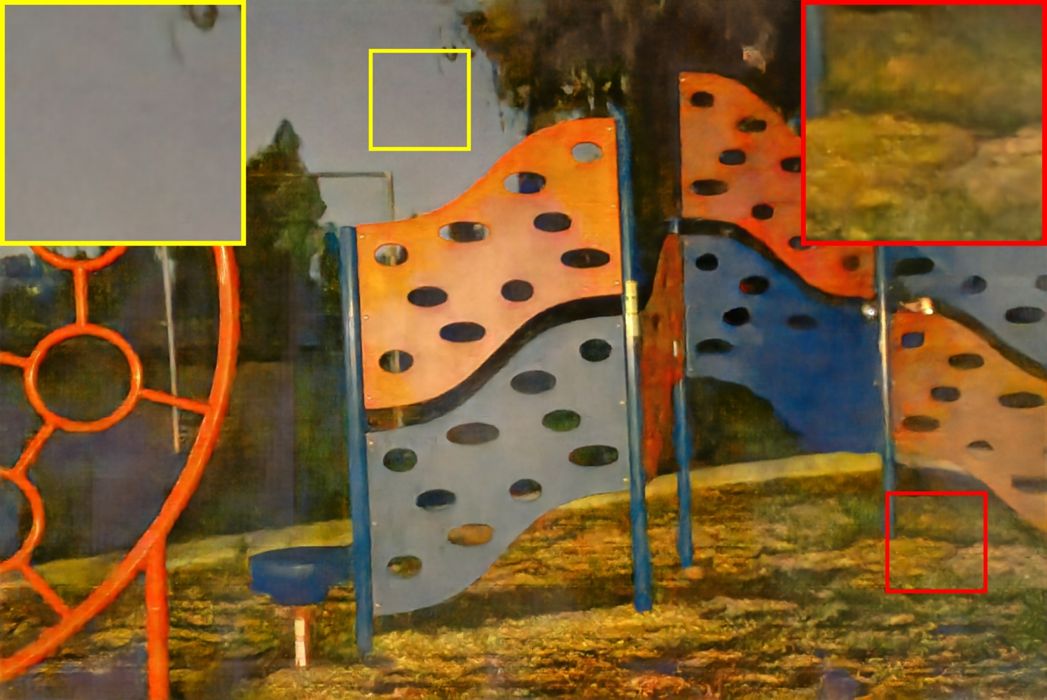}\hfill
			\end{minipage}
		}
		\hspace{-0.1in}
		\subfigure[Groundtruth]{
			\centering
			\begin{minipage}[b]{1.6in}
				\includegraphics[width=1.6in]{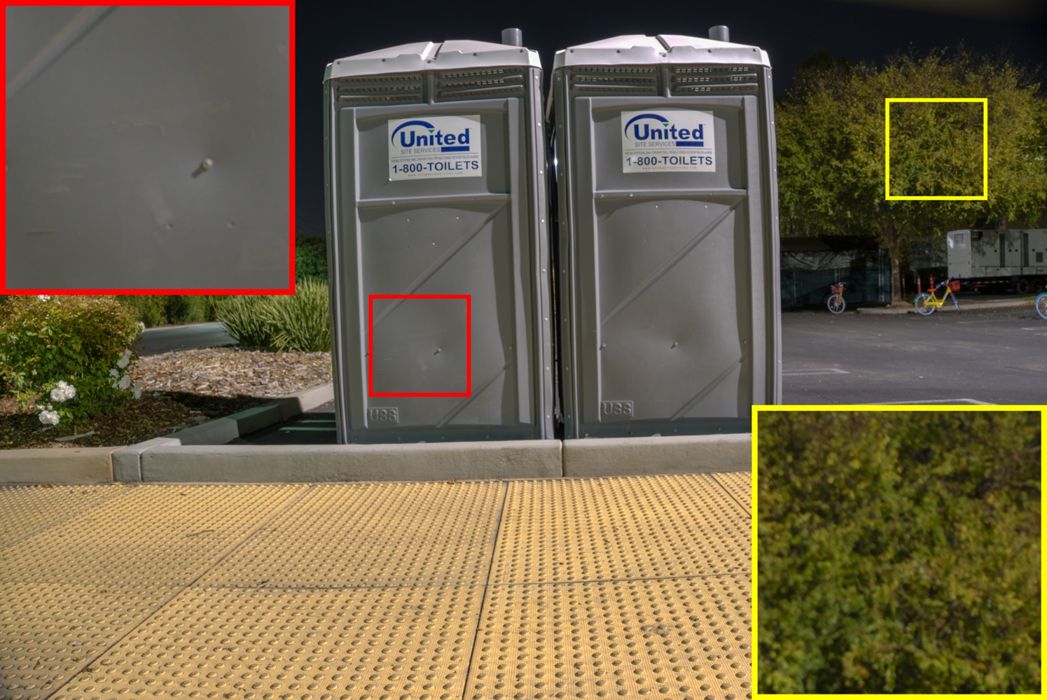}\hfill 
				\vspace{4pt}
				\includegraphics[width=1.6in]{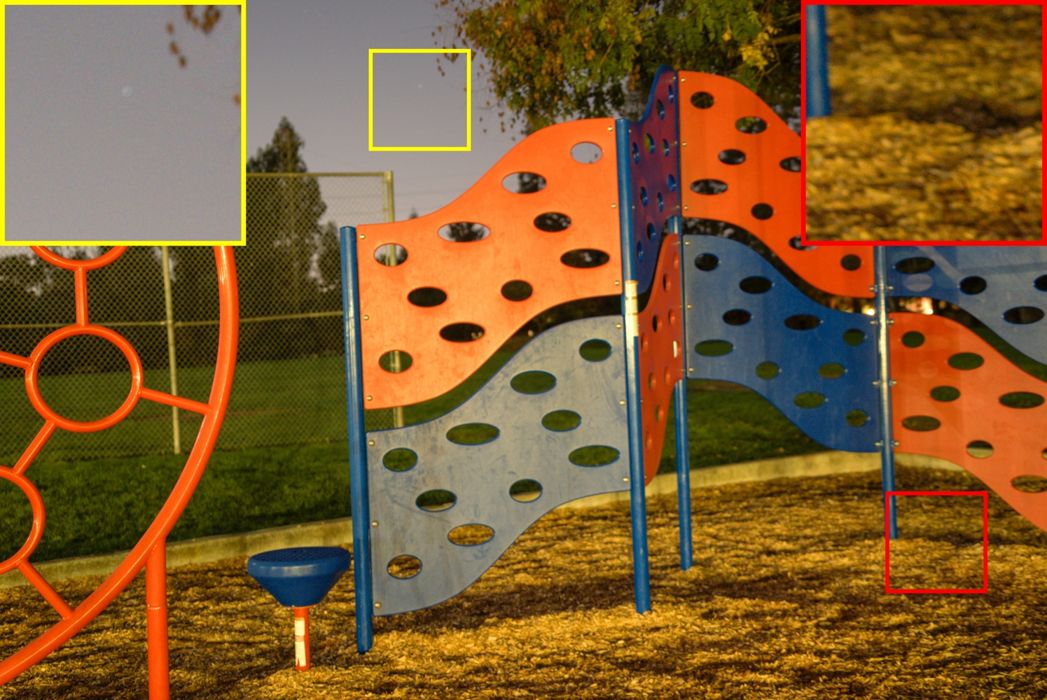}\hfill
			\end{minipage}
		}
	\end{center}
	\vspace{-0.05in}
	\caption{Results by one-stage and two-stage CNN models. The two sets of images are from the SID dataset \cite{Chen2018CVPR}. A gamma transform with parameter 2.2 is added on the raw images and restoration groundtruths for display.}
	\label{fig7}
\end{figure*}

\begin{table*}[th]
	\centering	
	\caption{Ablation study on HDR+ and SID datasets. The best and second best scores are highlighted in red and blue at each column.}
	\label{tab1}
	\begin{tabular}{|m{3.8cm}| m{0.6cm}<{\centering}| m{0.6cm}<{\centering}| m{1.4cm}<{\centering}| m{0.6cm}<{\centering} |m{0.6cm}<{\centering}| m{1.4cm}<{\centering}|}
		\hline \multirow{2}{*}{}& 
		\multicolumn{3}{c|}{HDR+ dataset}  & \multicolumn{3}{c|}{SID dataset}  \\
		\cline{2-7} 
		& PSNR & SSIM & Color error & PSNR & SSIM & Color error  \\
		\hline
		{\bf Default setting} & {\color{red}24.98}  & {\color{red}0.858} & {\color{red}$4.95^\circ$}   & {\color{red}22.47} & {\color{red}0.744} & {\color{red}$6.97^\circ$}   \\
		One-stage setting & 21.60 & 0.819 & $5.82^\circ$ & 19.00& 0.691&  $7.96^\circ$\\
		Training without step 1, 2 & 22.06 & 0.826 & {$6.12^\circ$}  &21.82 &0.721 & $7.20^\circ$ \\ 
		Training without step 3 & {\color{blue}23.95} & {\color{blue}0.841} & {\color{blue}$4.98^\circ$} & {\color{blue}22.16} & {\color{blue}0.737} & {\color{blue}$7.07^\circ$} \\ 
		\hline
		One-stage SRGAN+CAN24 & 21.72&0.801 &$5.77^\circ$ & 19.85& 0.682& $8.10^\circ$  \\ 
		Two-stage SRGAN+CAN24 & 22.31& 0.815&$5.45^\circ$ &20.96 &0.714 &$7.66^\circ$  \\
		\hline
	\end{tabular}
\end{table*}

\subsection{Training Scheme}

Although there are many possible losses to train our CameraNet system, e.g., perceptual loss \cite{JohnsonAF16} and adversarial loss \cite{LedigTHCCAATTWS17}, we consider the simplest case with a set of $\ell_{1}$ losses because it is easy to calculate and converges to a good local minima that correlates to human perception \cite{Zhao2017loss}. Our training scheme is divided into three steps. The first two steps conduct independent trainings of Restore-Net and Enhance-Net to obtain their initial estimates, followed by a joint fine-tuning of the two modules in the last step.

In the first step, the Restore-Net is trained with a restoration loss that measures the $\ell_{1}$ errors in linear and log domains. This loss can be described as:

\begin{equation}
\label{fn6}
\begin{split}
\mathcal{L}_{rest}(I_{rest}^{xyz},&G_{rest}^{xyz})
=
\|I_{rest}^{xyz}-G_{rest}^{xyz}\|_1
\\
&
+\|log(max(I_{rest}^{xyz}),\epsilon))-log(max(G_{rest}^{xyz},\epsilon))\|_1,
\end{split}
\end{equation}

\noindent where $\epsilon$ is a small number to avoid log infinity. The settlement of loss in log domain is to penalize the image differences in terms of human perception \cite{Mildenhall2018CVPR,EilertsenKDMU17}. Otherwise, the CNN model may receive larger gradients to restore highlight areas than lowlight areas. 

In the second step, the Enhance-Net is trained with an enhancement loss, which can be denoted as:

\vspace{-0.03in}
\begin{equation}
\label{fn7}
\begin{split}
\mathcal{L}_{enh}(F_{enh}^{srgb},G_{enh}^{srgb})=
\|F_{enh}^{srgb}-G_{enh}^{srgb}\|_1,
\end{split}
\end{equation}

\noindent where $F_{enh}^{srgb}$ denotes the output of Enhance-Net that takes the restoration groundtruth as input:

\begin{equation}
\label{fn8}
\begin{split}
F_{enh}^{srgb} = M_{enh}^{srgb}(G_{rest}^{srgb})=
 M_{enh}^{srgb}(C_{srgb}\cdot G_{rest}^{xyz}).
\end{split}
\end{equation}

\noindent This setting has the merit that the training of Enhance-Net does not rely on the output of Restore-Net. Thus, the first two training steps for the two modules can be conducted in parallel. 

In the final step, the Restore-Net and Enhance-Net are jointly fine-tuned with two losses, described as:

\begin{equation}
\label{fn9}
\begin{split}
\mathcal{L}_{joint}=
\lambda\cdot\mathcal{L}_{rest}(I_{rest}^{xyz},G_{rest}^{xyz}) + 
(1-\lambda)\cdot\mathcal{L}_{enh}(I_{enh}^{srgb},G_{enh}^{srgb})
\end{split}
\end{equation}

\noindent Note that the enhancement sub-loss in (\ref{fn9}) takes $I_{enh}^{srgb}$ for loss calculation rather than $F_{enh}^{srgb}$ in Eq. (\ref{fn7}). This joint fine-tuning has two merits. First, the Enhance-Net receives the gradients from the enhancement sub-loss, while the Restore-Net receives the gradients from both restoration and enhancement sub-losses, weighted by $\lambda$ and $1-\lambda$, respectively. Thus, this setting allows the Restore-Net to contribute to the final sRGB image reconstruction by a factor of $1-\lambda$, which leads to higher reconstruction quality. Additionally, since the two modules are trained independently in the first two steps, reconstruction errors may occur due to the gap in the intermediate results. Joint fine-tuning can reduce such errors by interacting the two modules during the training. The setting of parameter $\lambda$ is scenario-specific. If the restoration subtasks dominate the ISP pipeline, e.g., in an low-light scenario, $\lambda$ should be set larger to maintain the restoration functionality of Restore-Net, and vice versa.

\section{Experiments}

In this section we present extensive experimental results to verify the learning capability and image reconstruction performance of our CameraNet system by using indices such as PSNR, SSIM and Color Error. The Color Error measures the mean difference of color angle (measured in degree, the smaller the better) between two images\footnote{We only measure the color difference in image regions within the luminance intensity range (0.05, 0.95) because the underexposed and overexposed regions have very large color errors that bias the comparison.}.

\begin{figure*}[ht]
	\begin{center}
		\subfigure[Raw image]{
			\centering
			\includegraphics[width=2in]{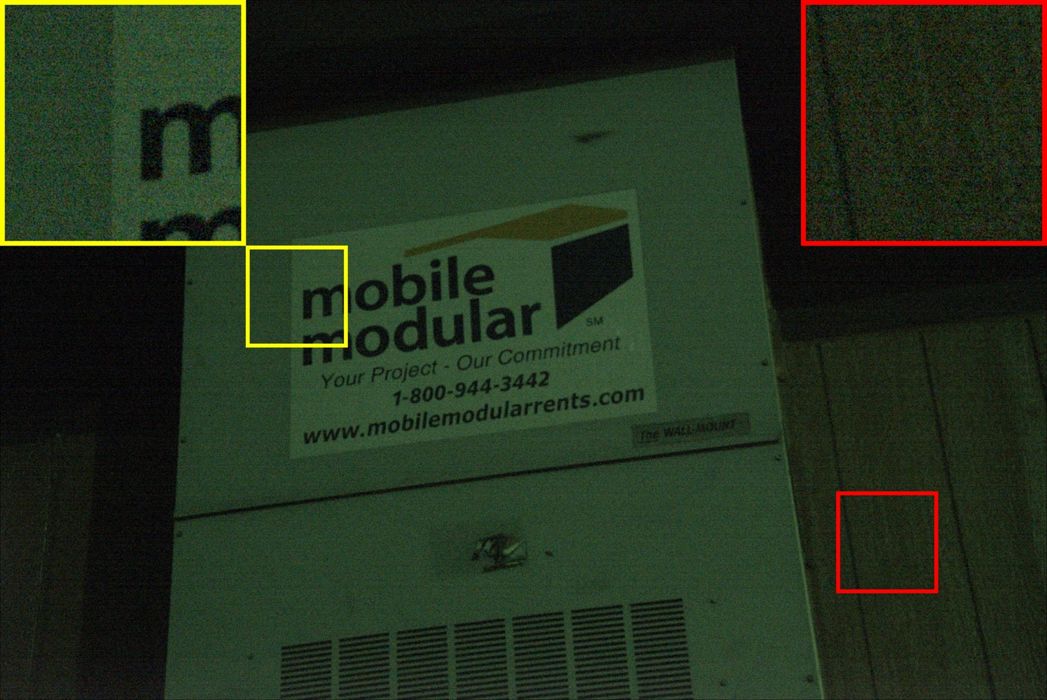}\hfill
		}
		\subfigure[One-stage SRGAN+CAN24]{
			\centering
			\includegraphics[width=2in]{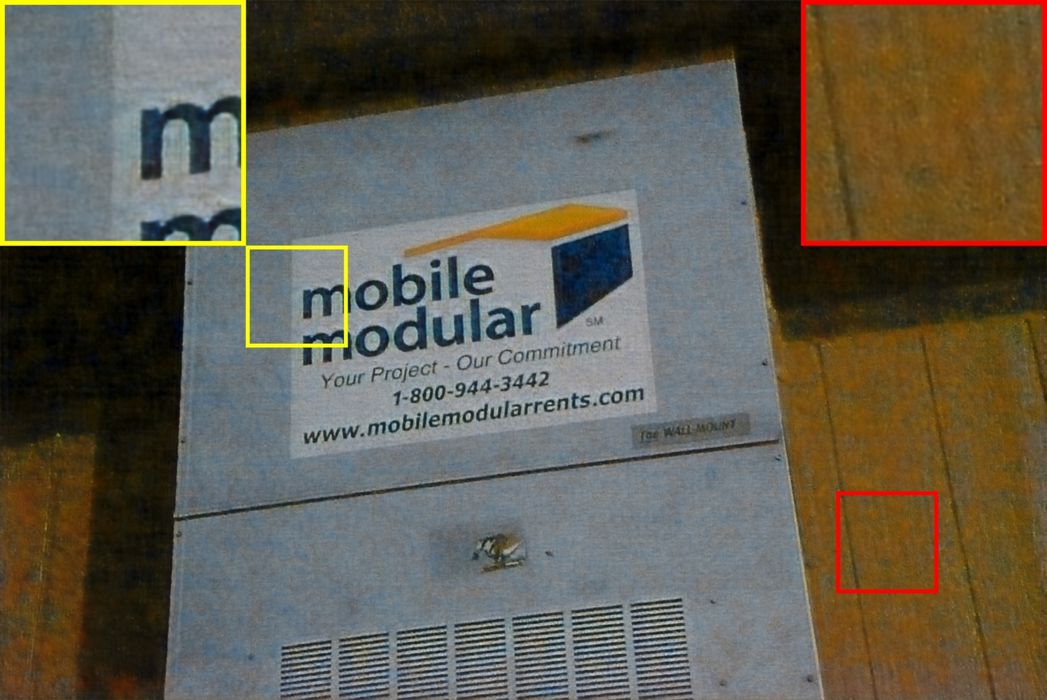}\hfill
		}
		\subfigure[Two-stage SRGAN+CAN24]{
			\centering
			\includegraphics[width=2in]{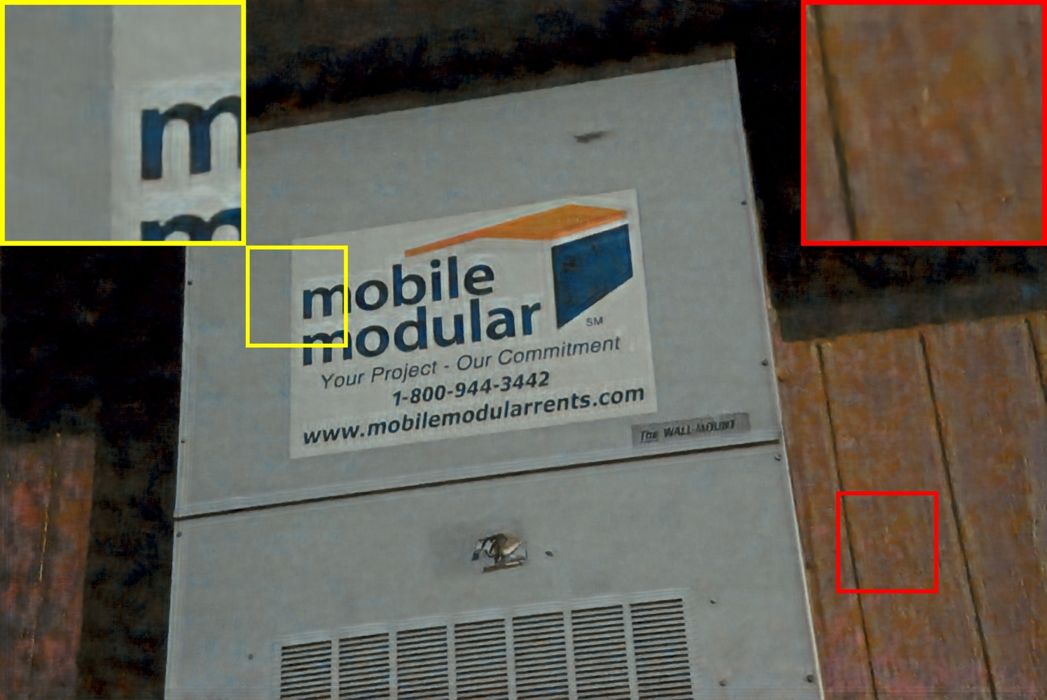}\hfill
		}
		\\
		\subfigure[CameraNet]{
			\centering
			\includegraphics[width=2in]{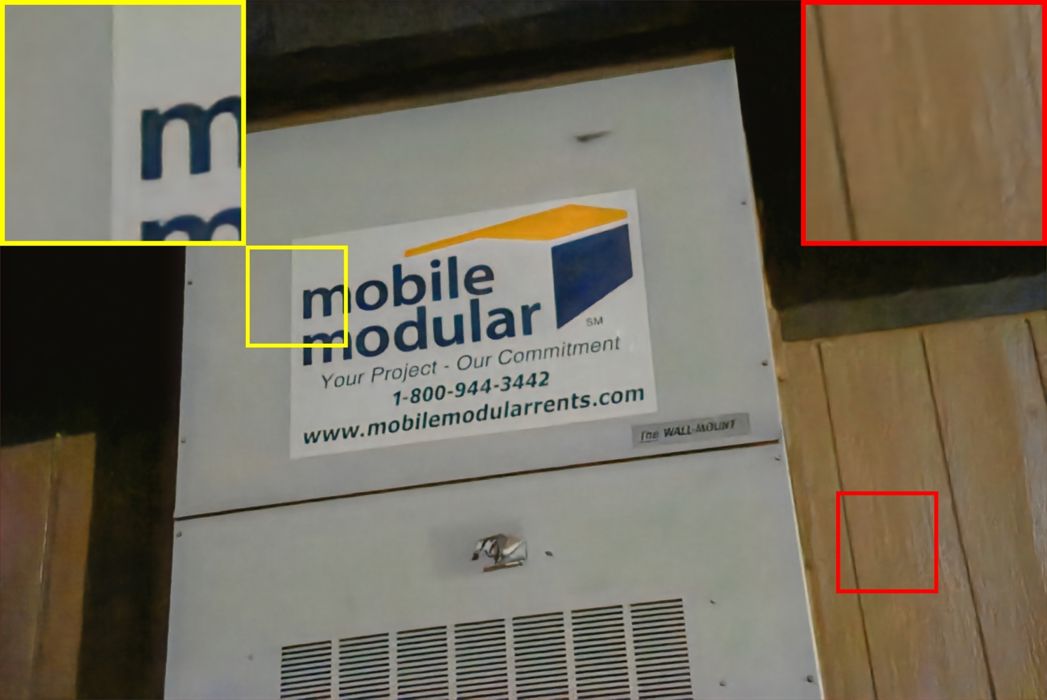}\hfill
		}
		\subfigure[Groundtruth]{
			\centering
			\includegraphics[width=2in]{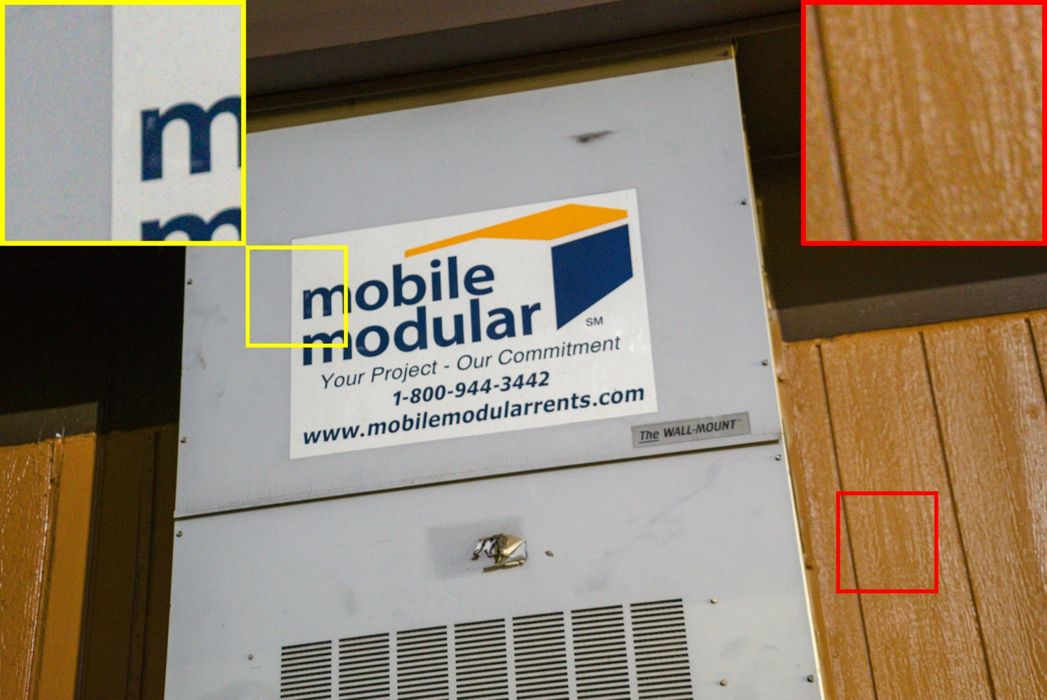}\hfill
		}
	\end{center}
	\vspace{-0.1in}
	\caption{Comparison of results between SRGAN+CAN24 model and CameraNet. A gamma transform with parameter 2.2 is applied to the raw images and restoration groundtruths for display.}
	\label{fig8}
\end{figure*}

\begin{figure}[thp]
	\begin{center}
		\subfigure[Without steps 1,2]{
			\centering
			\includegraphics[width=1in]{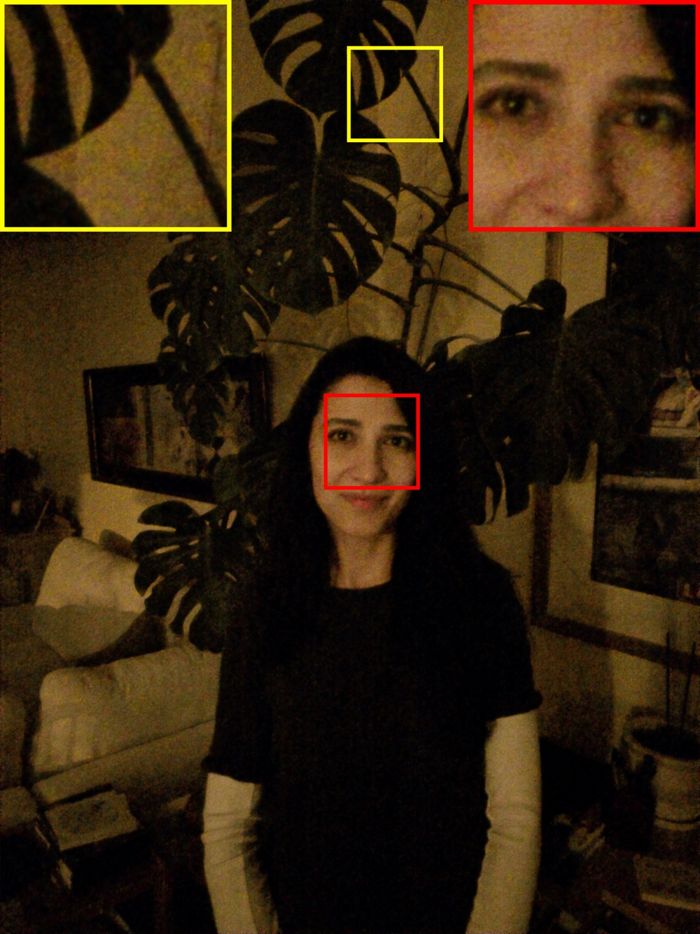}\hfill
		}
		\subfigure[Defualt setting]{
			\centering
			\includegraphics[width=1in]{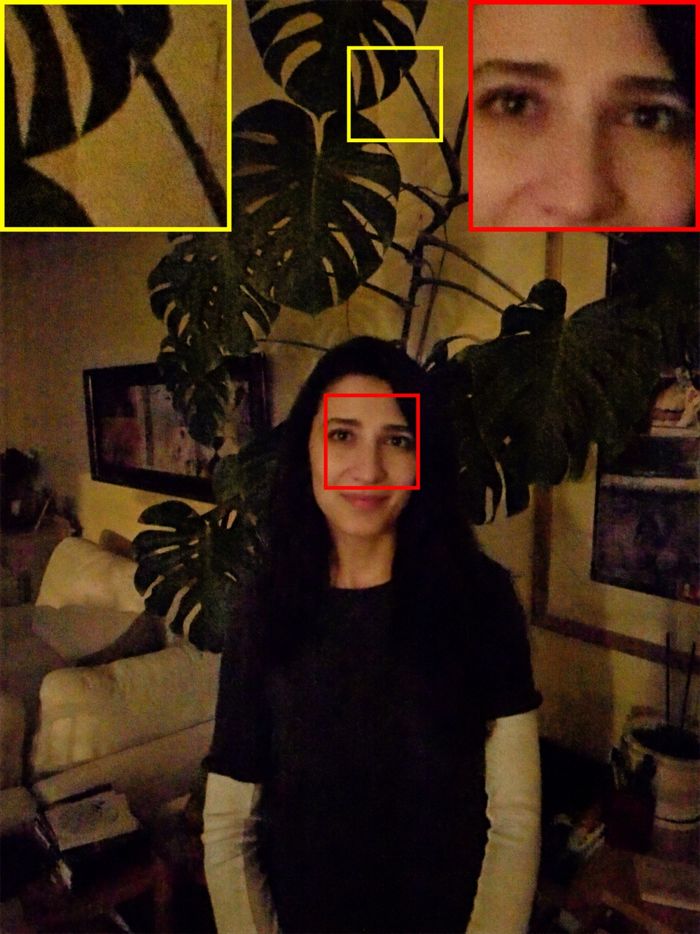}\hfill
		}
		\subfigure[Groundtruth]{
			\centering
			\includegraphics[width=1in]{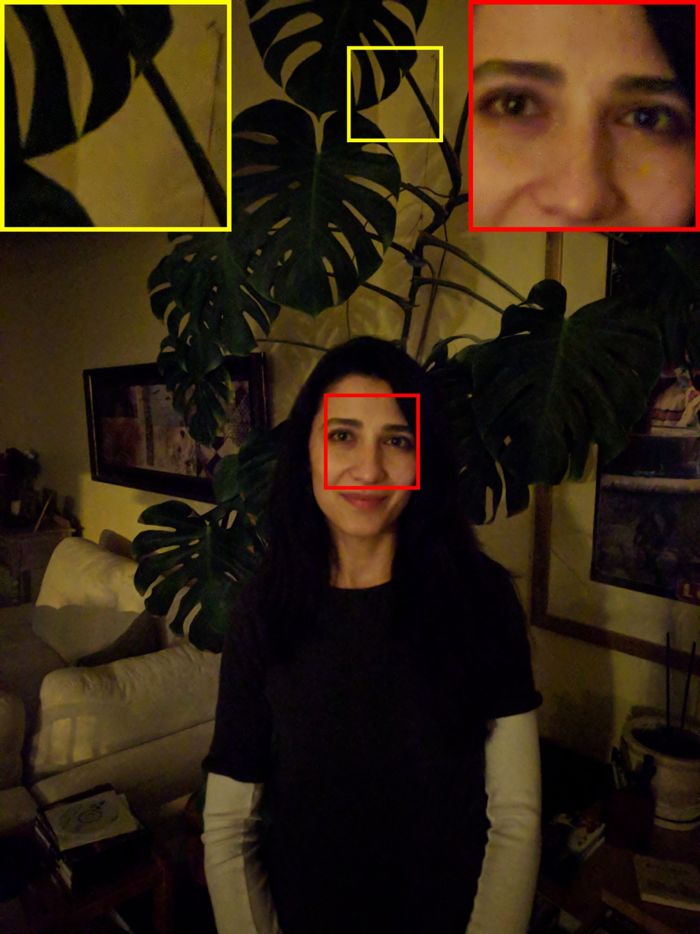}\hfill
		}
	\end{center}
	\vspace{-0.1in}
	\caption{Comparison between the default training setting and the setting without steps 1 and 2. The image is from the HDR+ dataset \cite{hasinoff2016}.}
	\label{fig9}
	\vspace{-0.1in}
\end{figure}

\begin{figure}[thp]
	\begin{center}
		\subfigure[Without step 3]{
			\centering
			\includegraphics[width=1.05in]{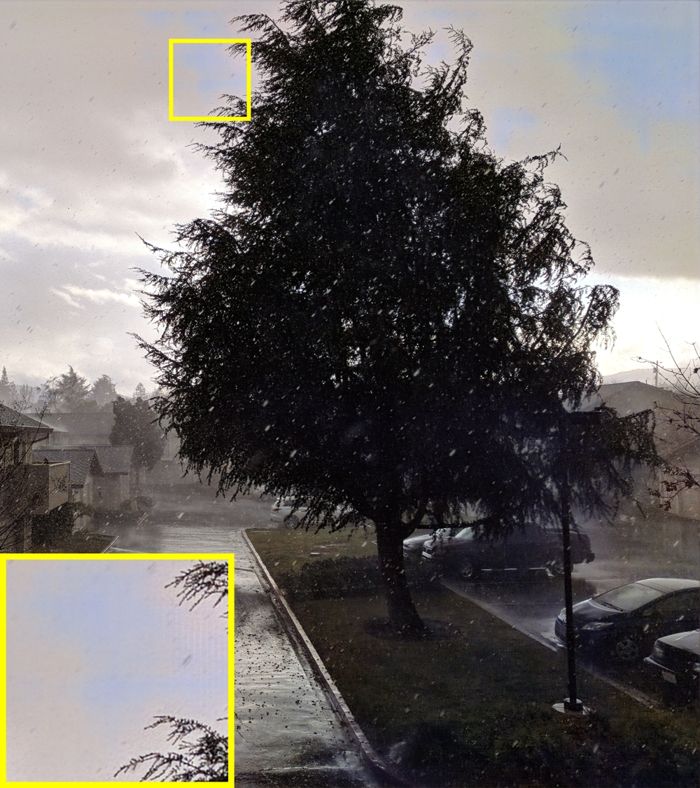}\hfill
		}
		\subfigure[Defualt setting]{
			\centering
			\includegraphics[width=1.05in]{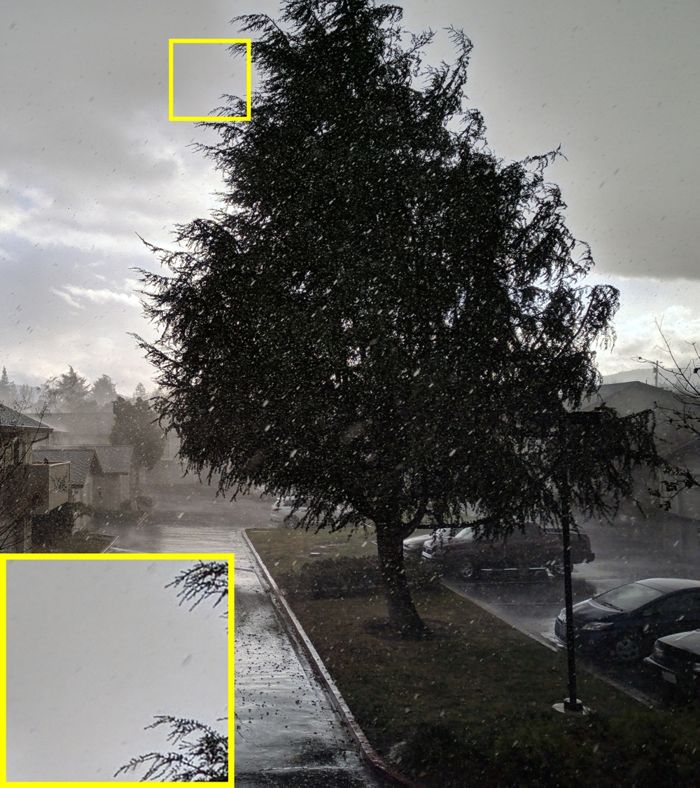}\hfill
		}
		\subfigure[Groundtruth]{
			\centering
			\includegraphics[width=1.05in]{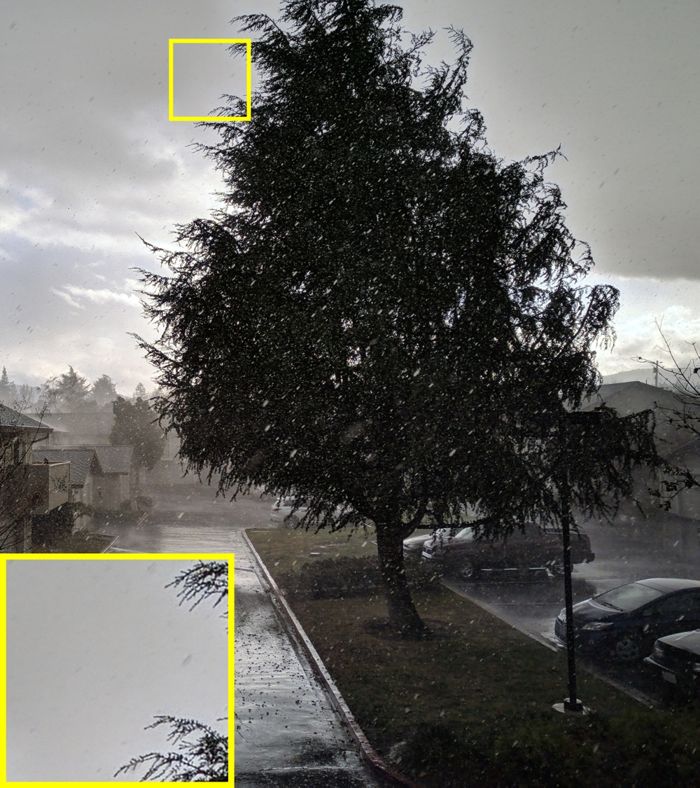}\hfill
		}
	\end{center}
	\vspace{-0.1in}
	\caption{Comparison between the default training setting and the setting without step 3. The image is from the HDR+ dataset \cite{hasinoff2016}.}
	\label{fig10}
	\vspace{-0.1in}
\end{figure}

\subsection{Datasets}

We use three ISP pipeline datasets in the experiments, including HDR+ dataset \cite{hasinoff2016}, FiveK dataset \cite{Vladimir2011fivek} and SID dataset \cite{Chen2018CVPR}, which are briefly described as follows.

The main features of HDR+ dataset \cite{hasinoff2016} include burst denoising and detail enhancement. For each scene, the HDR+ algorithm captures a burst of underexposed raw images and fuses them into a burst-denoised raw image. The DCRaw is used to process the denoised raw image to obatin the restoration groundtruth. Then, sophisticated tone mapping algorithm is applied on the denoised raw image to produce the final sRGB image, which is treated as the enhancement groundtruth. We take the reference input raw images, which are used for image fusion in the HDR+ algorithm, as the input of CameraNet. The HDR+ dataset contains 3600 scenes captured by different smartphone cameras. In our experiments, we use the Nexus 6P subset, which includes 675 scenes as training data and 250 scenes as testing data. Other camera data are not used because there are some misalignments between the input images and the groundtruths. 

The SID dataset \cite{Chen2018CVPR} is featured with denoising in low-light environment. For each scene, it captures a noisy raw image with short exposure and a clean raw image with long exposure. We use the DCRaw to process the long-exposed raw images to obtain restoration groundtruths. Since the SID dataset does not involve any enhancement operation, we further process the restoration groundtruth by using the auto-enhancement tool in Photoshop to obtain the enhancement groundtruth. We use the Sony subset for experiments, which includes 181 and 50 scenes for training and testing, respectively.

The FiveK dataset \cite{Vladimir2011fivek} is featured with strong manual retouching on image tone and color style. For each raw image of the 5000 scenes, five photographers are employed to adjust various visual attributes of the image via Lightroom software and produce 5 photographic styles. We take the expert-C retouched set of images as the enhancement groundtruth. Since FiveK dataset does not contain restoration groundtruth, we process the input raw image using DCRaw to obtain the restoration groundtruth. We use Nikon D700 subset for the experiments with 500 training images and 150 testing images. 

\begin{figure*}[htp]
	\begin{center}
		\subfigure[Raw image]{
			\centering
			\includegraphics[width=1.8in]{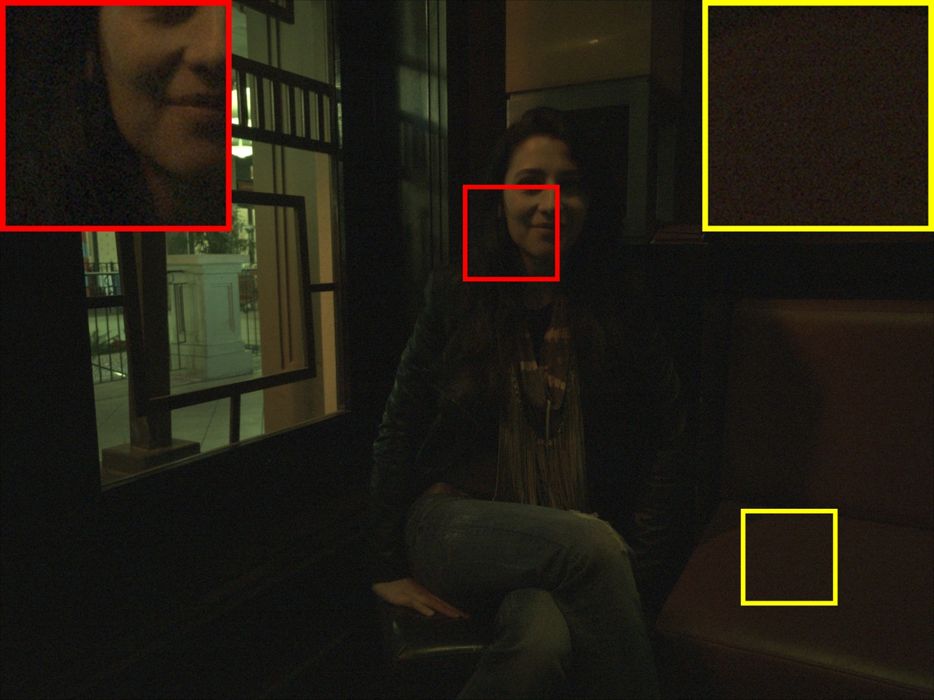}\hfill
		}
		\subfigure[Result by DCRaw]{
			\centering
			\includegraphics[width=1.8in]{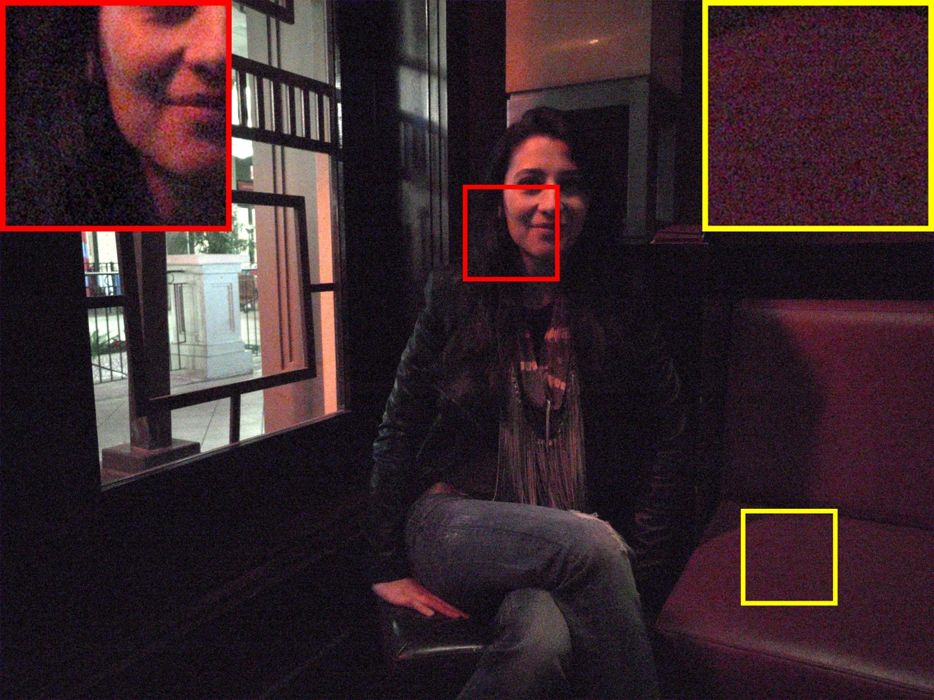}\hfill
		}
		\subfigure[Result by Camera Raw]{
			\centering
			\includegraphics[width=1.8in]{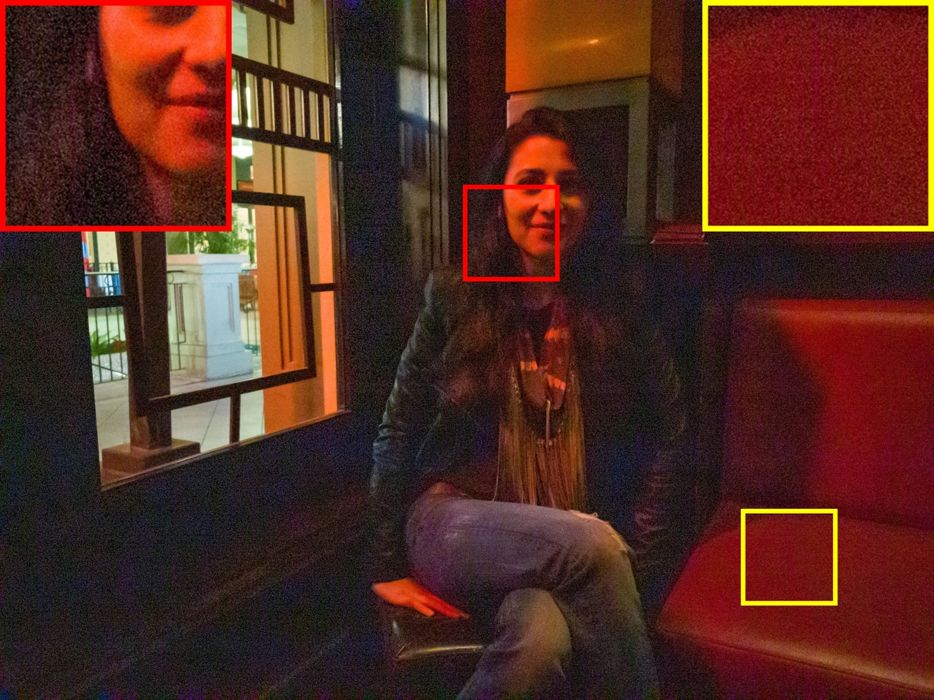}\hfill
		}
		\\
		\subfigure[Result by DeepISP-Net]{
			\centering
			\includegraphics[width=1.8in]{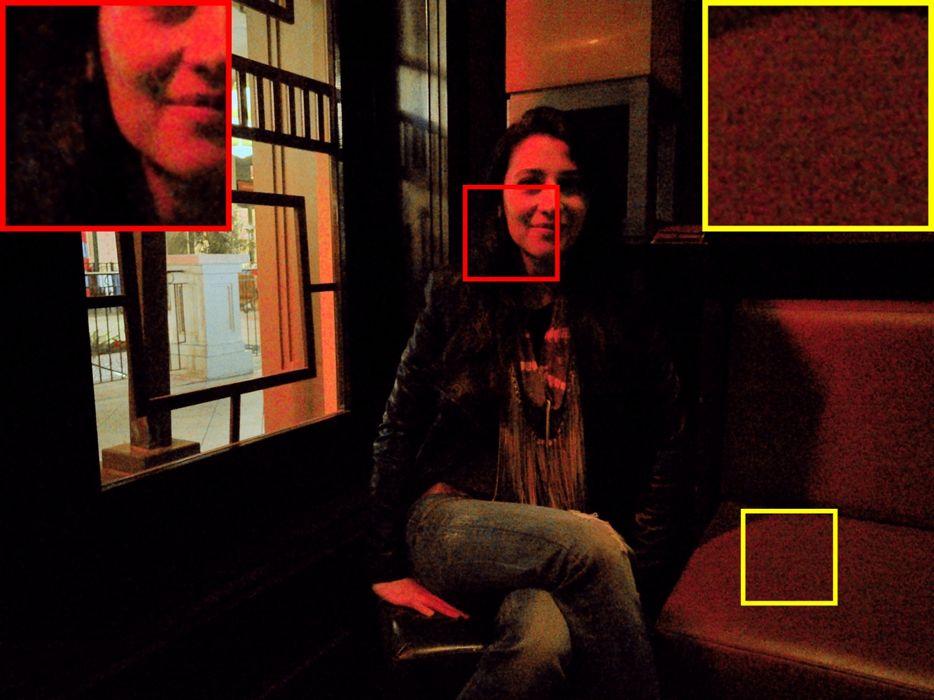}\hfill
		}
		\subfigure[Result by CameraNet]{
			\centering
			\includegraphics[width=1.8in]{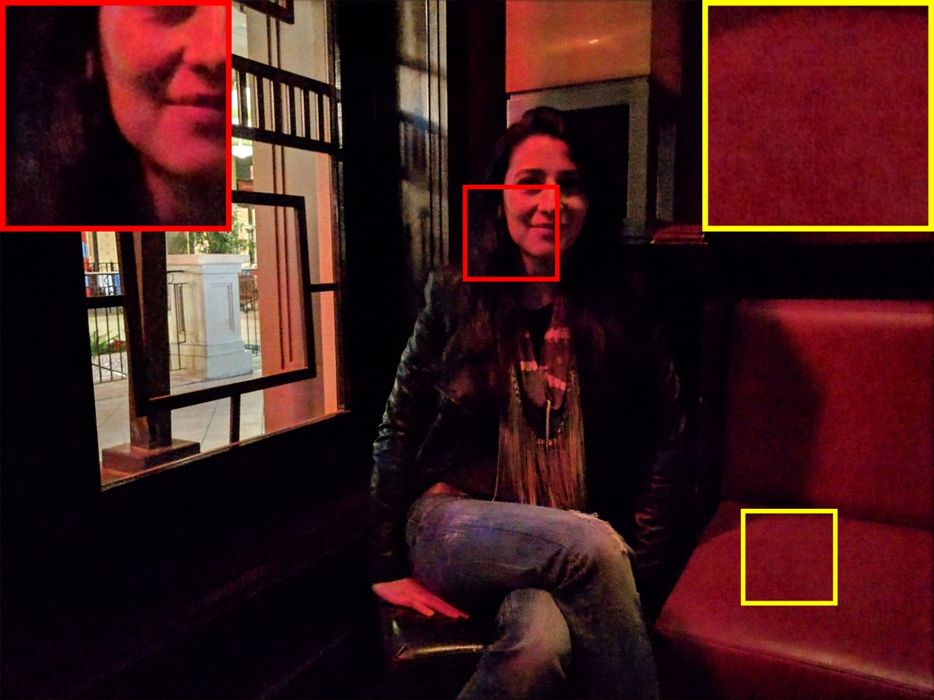}\hfill
		}
		\subfigure[Groundtruth]{
			\centering
			\includegraphics[width=1.8in]{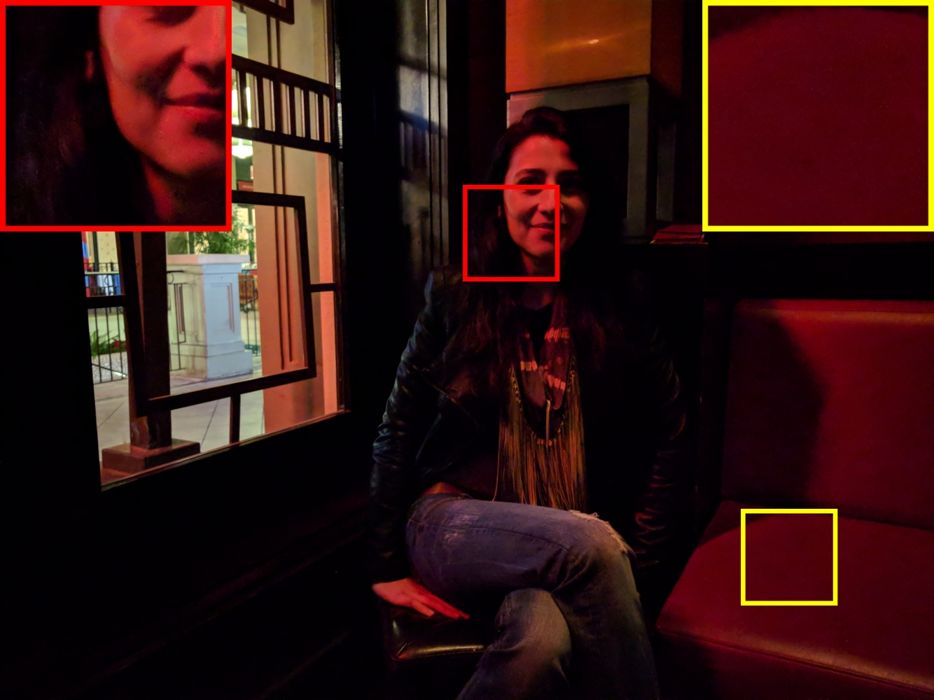}\hfill
		}
	\end{center}
	\caption{Results on a dark indoor image from the HDR+ dataset \cite{hasinoff2016} by the competing methods. A gamma transform with parameter 2.2 is applied to the raw image for better visualization. }
	\label{fig11}
\end{figure*}

\begin{table*}[th]
	\centering	
	\caption{Objective comparison of ISP pipelines.}
	\label{tab2}
	\begin{tabular}{|m{3cm}| m{0.8cm}<{\centering}| m{0.8cm}<{\centering}| m{1.4cm}<{\centering}| m{0.8cm}<{\centering} |m{0.8cm}<{\centering}| m{1.4cm}<{\centering}|m{0.8cm}<{\centering} |m{0.8cm}<{\centering}| m{1.4cm}<{\centering}|}
		\hline \multirow{2}{*}{}& 
		\multicolumn{3}{c|}{HDR+ dataset}  & \multicolumn{3}{c|}{SID dataset} & \multicolumn{3}{c|}{FiveK dataset}  \\
		\cline{2-10} 
		& PSNR & SSIM & Color error & PSNR & SSIM & Color error&PSNR & SSIM & Color error  \\
		\hline
		CameraNet & {\bf 24.98}  & {\bf0.858} & ${\bf4.95^\circ}$  & {\bf22.47} & {\bf0.744} & ${\bf6.97^\circ}$  &{\bf23.37}&{\bf0.848}& ${\bf6.04^\circ}$\\
		DeepISP-Net \cite{Schwartz2019tip} & 22.88 & 0.818 & $5.23^\circ$ & 18.26& 0.649&  $8.67^\circ$ &22.59&0.845&$6.31^\circ$\\
		DCRaw & 16.63 & 0.575 &$9.84^\circ$ & 12.49&0.153 & $22.48^\circ$ &19.46&0.801&$7.62^\circ$\\
		Camera Raw software & 19.84 & 0.698 & $8.45^\circ$& 13.36& 0.245& $20.8^\circ$ &21.55&0.813&$7.80^\circ$\\ 
		\hline
	\end{tabular}
\end{table*}

\subsection{Experimental Setting}

We use the Adam optimizer ($\beta_1=0.9,\beta_2=0.99$) to train CameraNet and all the compared CNN models. Depending on the task complexity for each of the three datasets, we train Restore-Net in the first step for 1000, 4000 and 1000 epochs on HDR+, SID and FiveK datasets, respectively. In the second step, the Enhance-Net is trained with the same 600 epochs on all of the datasets due to the comparable complexities of enhancement subtasks in these datasets. The last fine-tuning step lasts for 200 epochs. The initial learning rates for the first two training steps are $10^{-4}$, which exponentially decays by 0.1 at 3/4 epochs. The learning rate for the last fine-tuning step is reduced to a fixed value $10^{-5}$. Considering the importance of the restoration subtask, the parameter $\lambda$ in (\ref{fn9}) is set to 0.5, 0.9 and 0.1 in the last step for HDR+, SID and FiveK datasets, respectively. In all training steps the batch size is set to 1 and the patch size is set to 1536$\times$1536. Random rotations, vertical and horizontal flippings are applied for data augmentation. 

\subsection{Ablation Study}

We use the HDR+ and SID datasets for ablation study. All the compared models in this subsection are trained until convergence.

We first compare the default two-stage CameraNet with its one-stage counterpart, where we deploy one single U-Net with comparable number of parameters to the default two-stage setting. Specifically, the number of processing blocks is doubled on each scale of U-Net. The one-stage network is trained with the final enhancement loss. The results are shown in Table \ref{tab1}. One can see that the default two-stage setting achieves significantly higher PSNR, SSIM and lower Color Error than the one-stage setting. Some results are visualized in Fig.\ \ref{fig7}, from which we can see that the results by the default two-stage setting have better visual quality, whereas the results by the one-stage setting exhibit various visual artifacts. This phenomenon indicates that the weakly correlated restoration and enhancement subtasks can be better addressed explicitly by different CNN modules, as manifested by our two-stage CameraNet. However, under the mixed treatment of these two sets of tasks by a deep single network, noise in the image may not be completely removed and is amplified for contrast enhancement, leading to serious visual artifacts

Secondly, we compare our default CameraNet with two of its variants that have different training schemes. The first one removes the first two training steps and directly goes to the third step. This means that we only train the CameraNet with the simultaneous loss in Eq. (\ref{fn9}) with a learning rate of $10^{-4}$. The second variant remains the first two steps but removes the third joint fine-tuning step. From Table \ref{tab1}, we can see that without the first two training steps, the objective scores are significantly worse than the default training setting. From the results visualized in Fig.\ \ref{fig9}(a), we cant see that some noise remains in the reconstructed image. This indicates the importance of independent trainings of the two modules to obtain a good initial estimates. In contrast, the variant without the third training step has comparable objective scores with the default setting, as can be seen in Table \ref{tab1}. However, due to the lack of fine-tuning, reconstruction errors occur in a few images. From the example shown in Fig.\ \ref{fig10}(a), we can see that the sky area has a sudden color change and has unnatural appearance.

To verify whether the merits of the two-stage framework can be generalized to other CNN architectures, we further compare the one-stage and two-stage settings by using a different CNN architecture. We use SRGAN \cite{LedigTHCCAATTWS17} with 10 layers as the restoration subnetwork and CAN24 \cite{Chen2017ICCV} as the enhancement subnetwork. The PSNR, SSIM and Color Error indices are shown in Table \ref{tab1} and one example is shown in Fig.\ \ref{fig8}, from which we can see that the two-stage setting of SRGAN+CAN24 also outperforms the one-stage counterpart. Furthermore, the two-stage SRGAN+CAN24 is not as effective as the U-Net-based CameraNet in noise removal. We believe this is mainly because SRGAN+CAN24 lacks multiscale processing that facilitates the denoising task.  

\begin{figure*}[htp]
	\begin{center}
		\subfigure[Raw image]{
			\centering
			\includegraphics[width=1.8in]{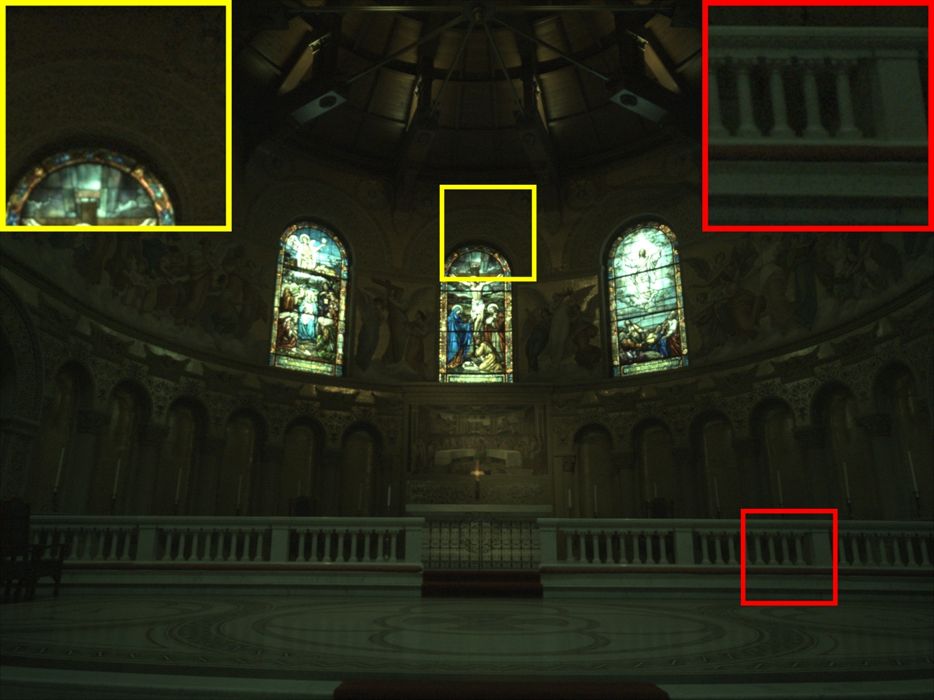}\hfill
		}
		\subfigure[Result by DCRaw]{
			\centering
			\includegraphics[width=1.8in]{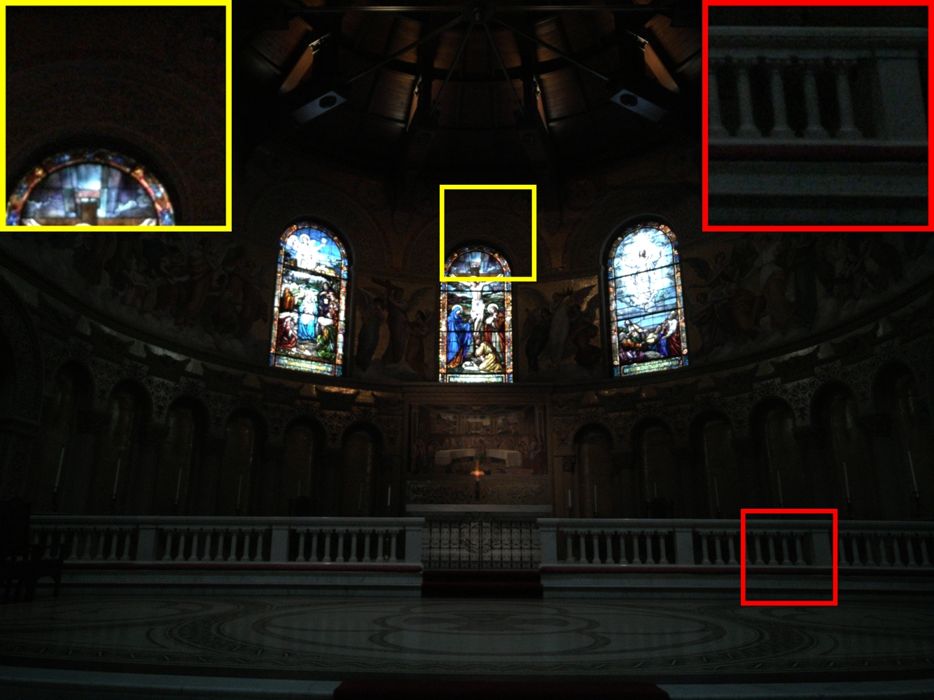}\hfill
		}
		\subfigure[Result by Camera Raw]{
			\centering
			\includegraphics[width=1.8in]{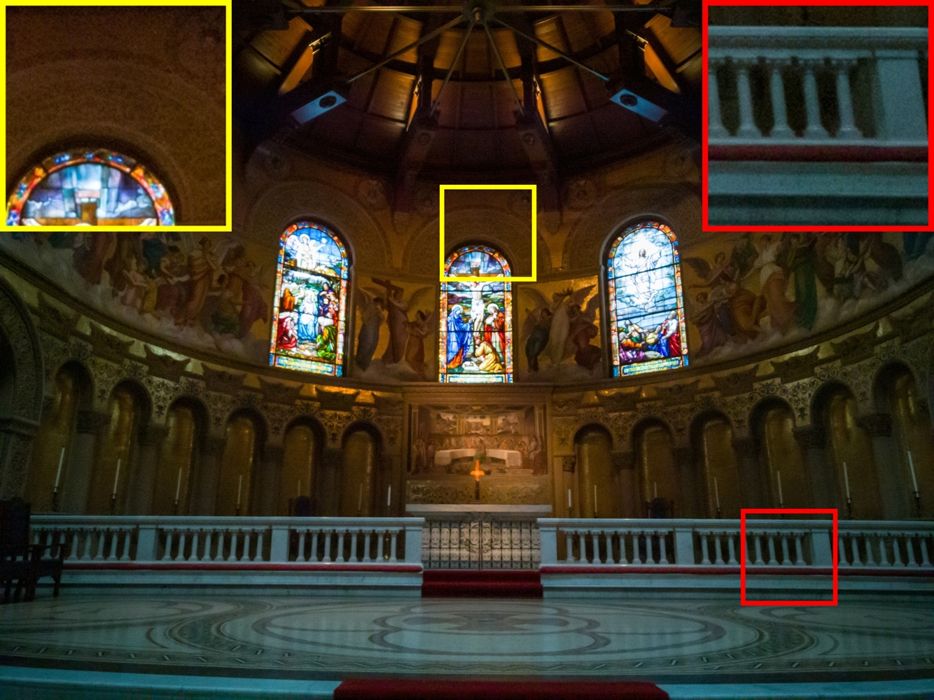}\hfill
		}
		\\
		\subfigure[Result by DeepISP-Net]{
			\centering
			\includegraphics[width=1.8in]{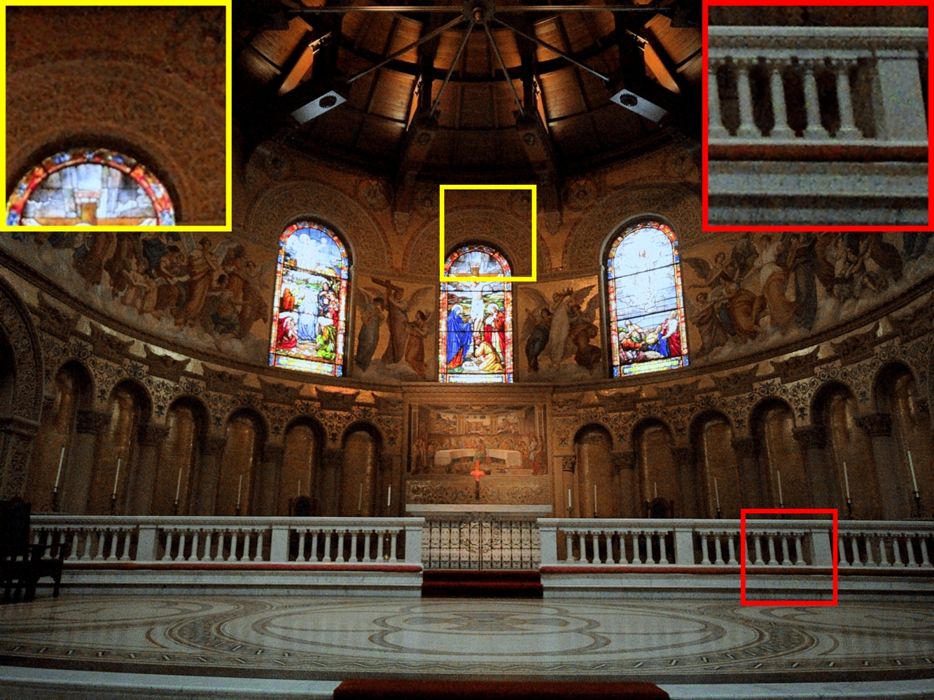}\hfill
		}
		\subfigure[Result by CameraNet]{
			\centering
			\includegraphics[width=1.8in]{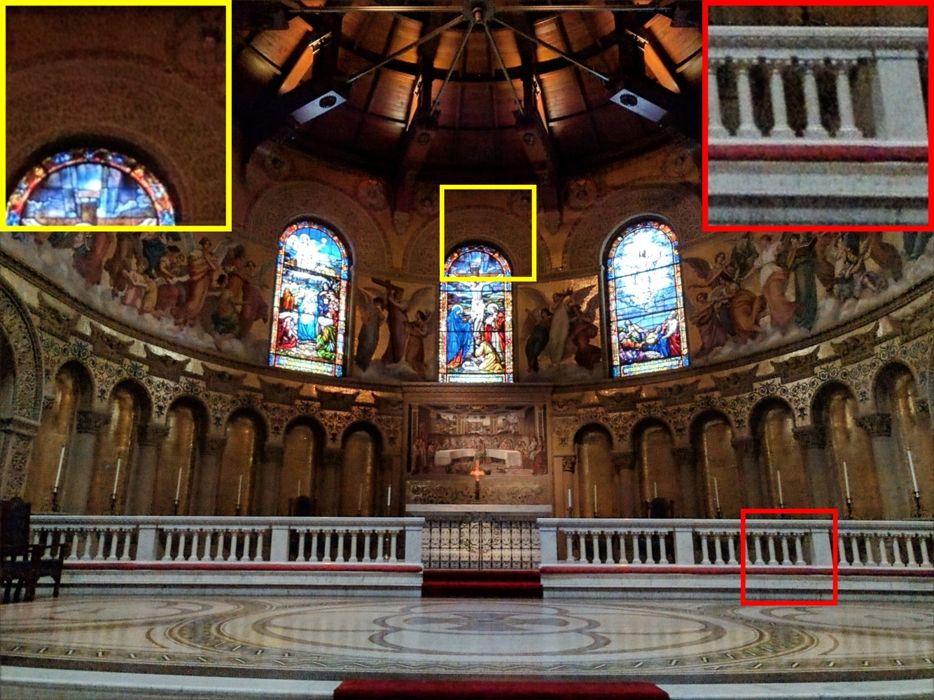}\hfill
		}
		\subfigure[Groundtruth]{
			\centering
			\includegraphics[width=1.8in]{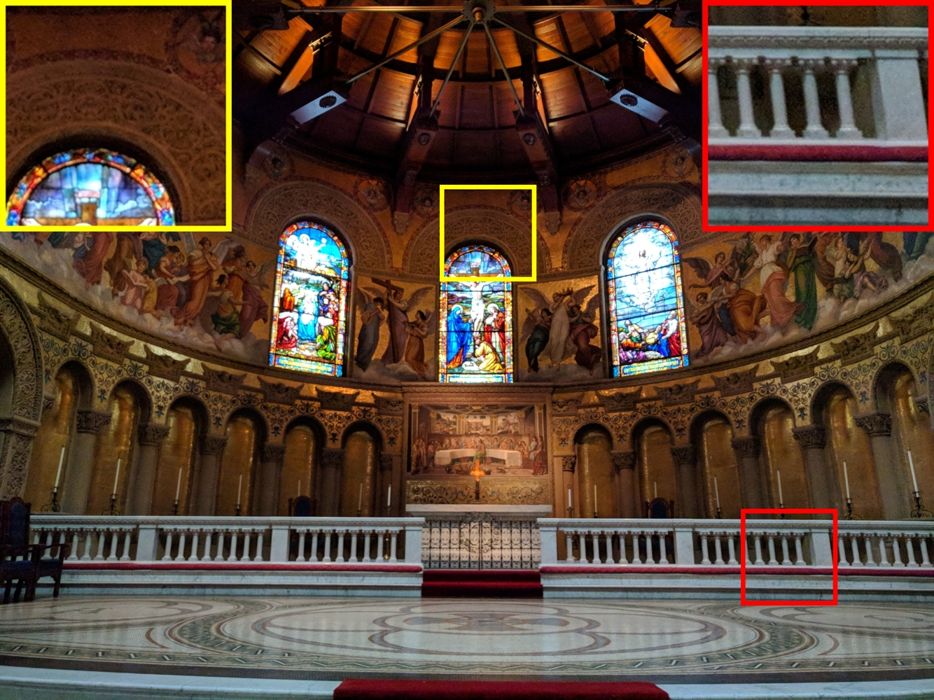}\hfill
		}
	\end{center}
	\vspace{-0.1in}
	\caption{Results on a church image from the HDR+ dataset \cite{hasinoff2016} by the competing methods. A gamma transform with parameter 2.2 is applied to the raw image for better visualization.}
	\label{fig12}
\end{figure*}

\begin{figure*}[htp]
	\begin{center}
		\subfigure[Raw image]{
			\centering
			\includegraphics[width=1.8in]{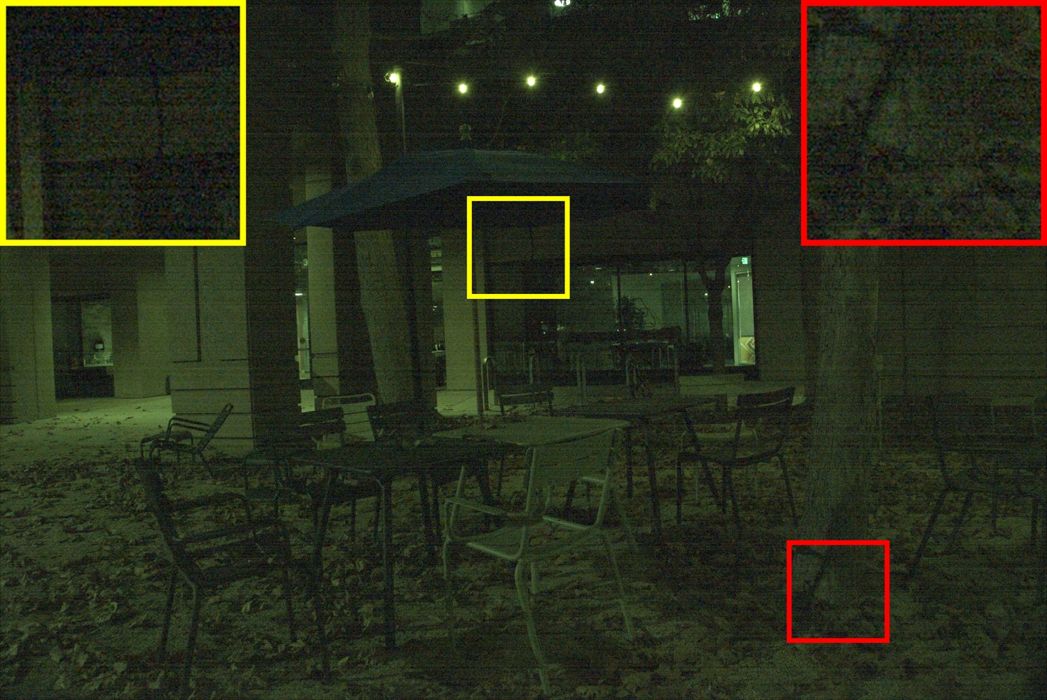}\hfill
		}
		\subfigure[Result by DCRaw]{
			\centering
			\includegraphics[width=1.8in]{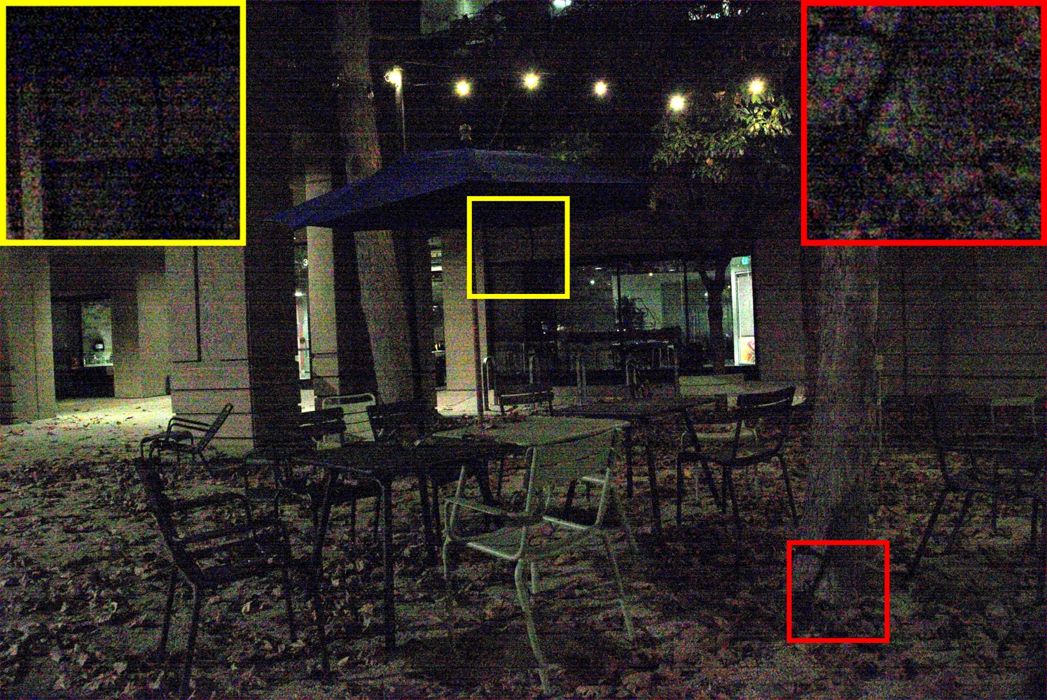}\hfill
		}
		\subfigure[Result by Camera Raw]{
			\centering
			\includegraphics[width=1.8in]{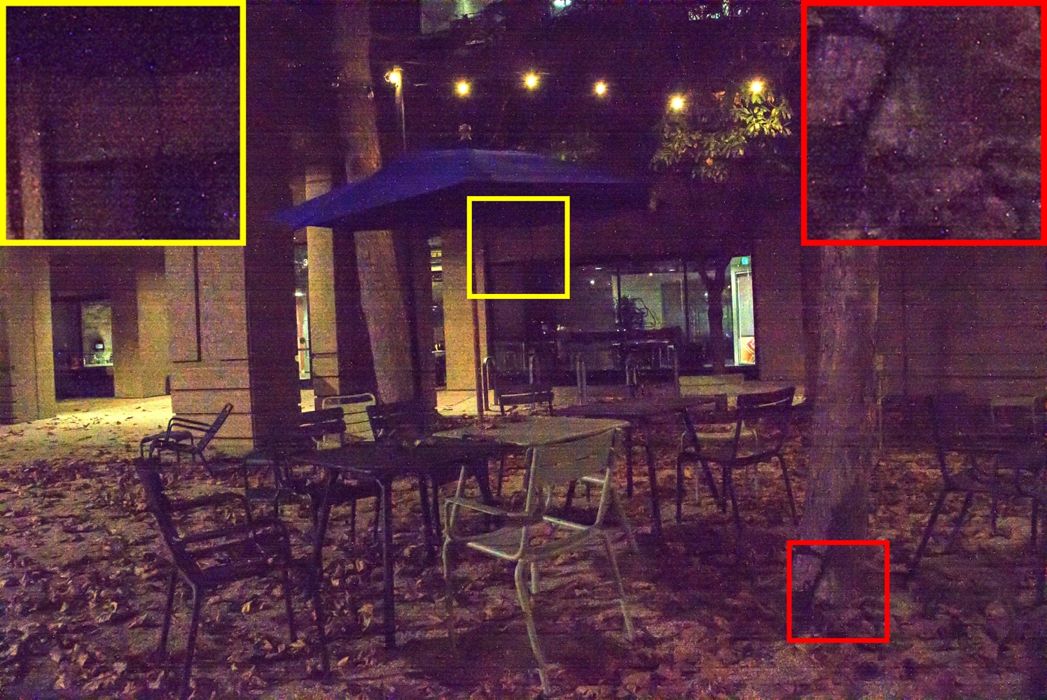}\hfill
		}
		\\
		\subfigure[Result by DeepISP-Net]{
			\centering
			\includegraphics[width=1.8in]{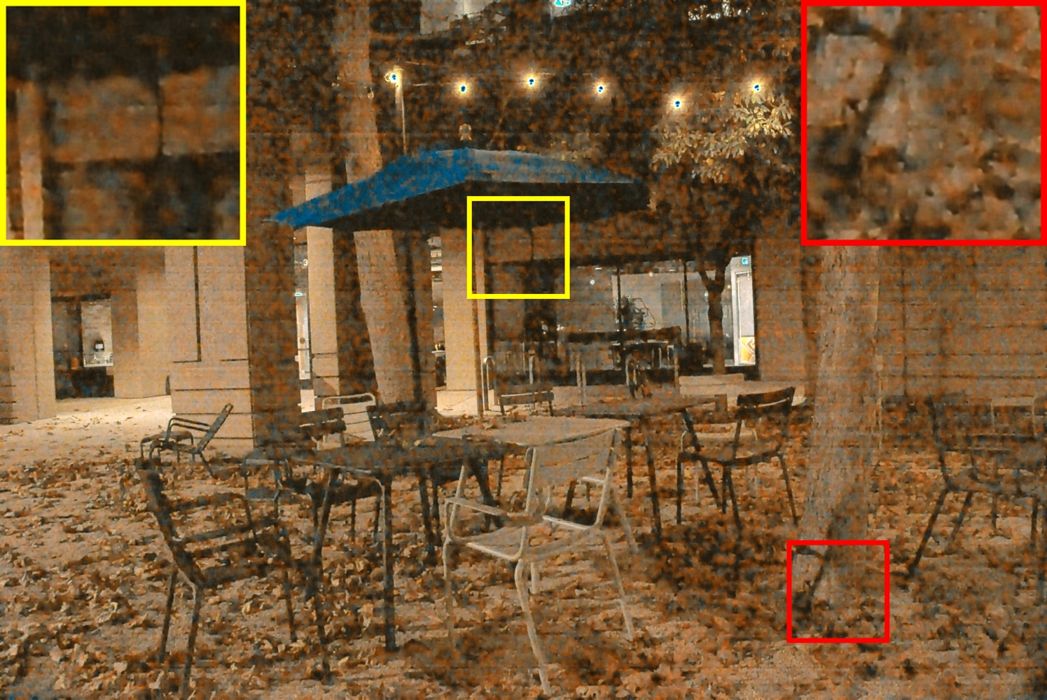}\hfill
		}
		\subfigure[Result by CameraNet]{
			\centering
			\includegraphics[width=1.8in]{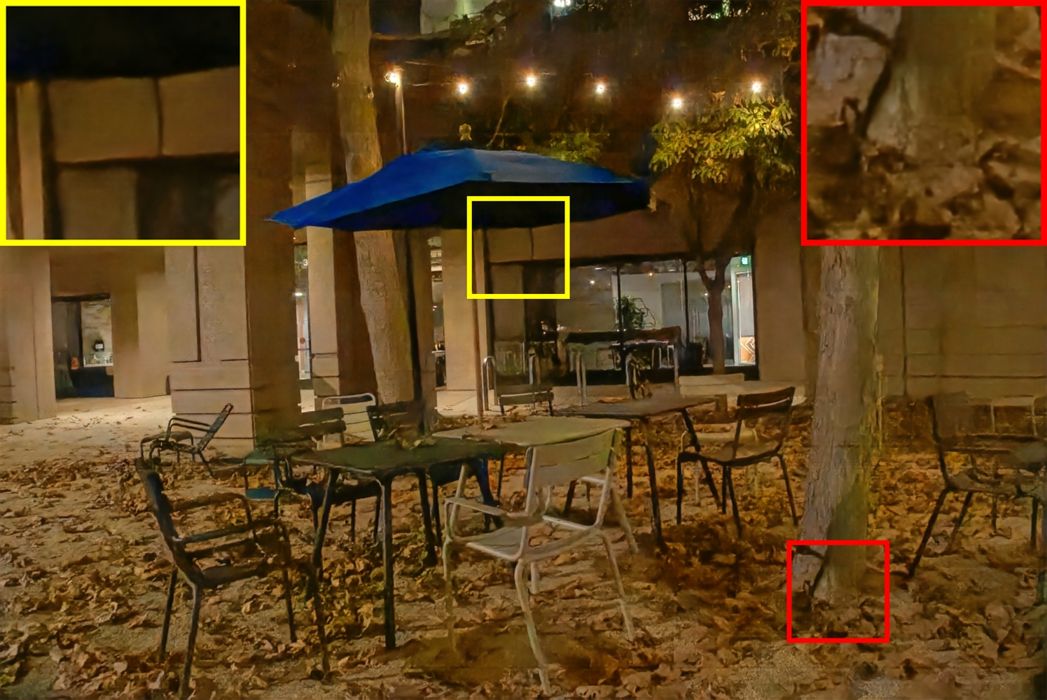}\hfill
		}
		\subfigure[Groundtruth]{
			\centering
			\includegraphics[width=1.8in]{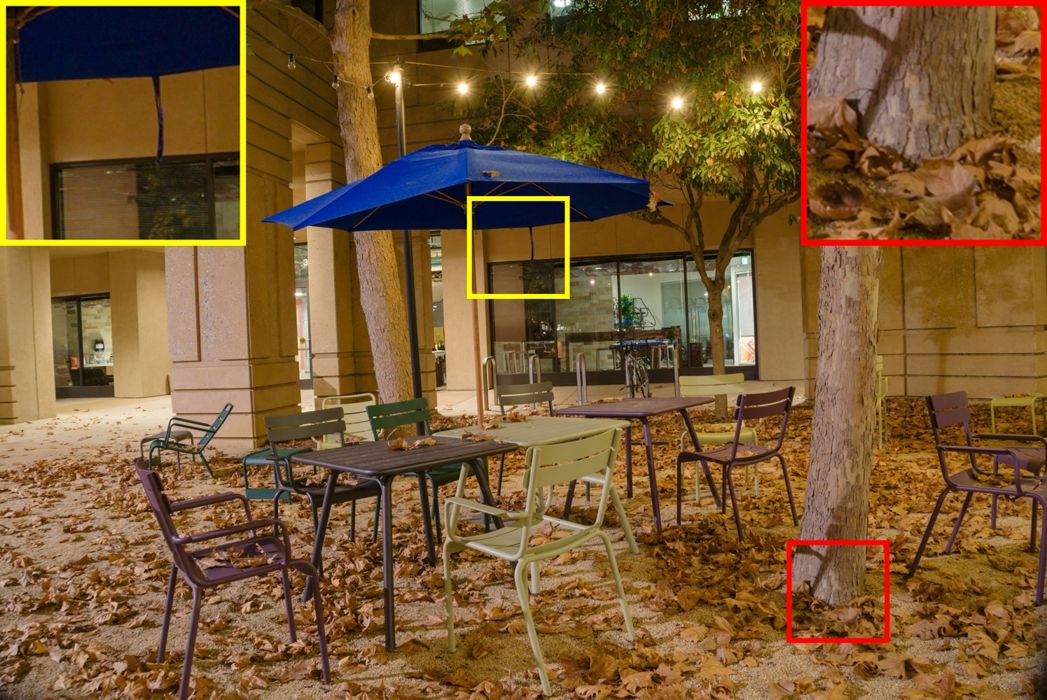}\hfill
		}
	\end{center}
	\vspace{-0.1in}
	\caption{Results on a pavilion image from the SID dataset \cite{Chen2018CVPR} by the competing methods. A gamma transform with parameter 2.2 is applied to the raw image for better visualization.}
	\label{fig13}
\end{figure*}

\begin{figure*}[htp]
	\begin{center}
		\subfigure[Raw image]{
			\centering
			\includegraphics[width=1.8in]{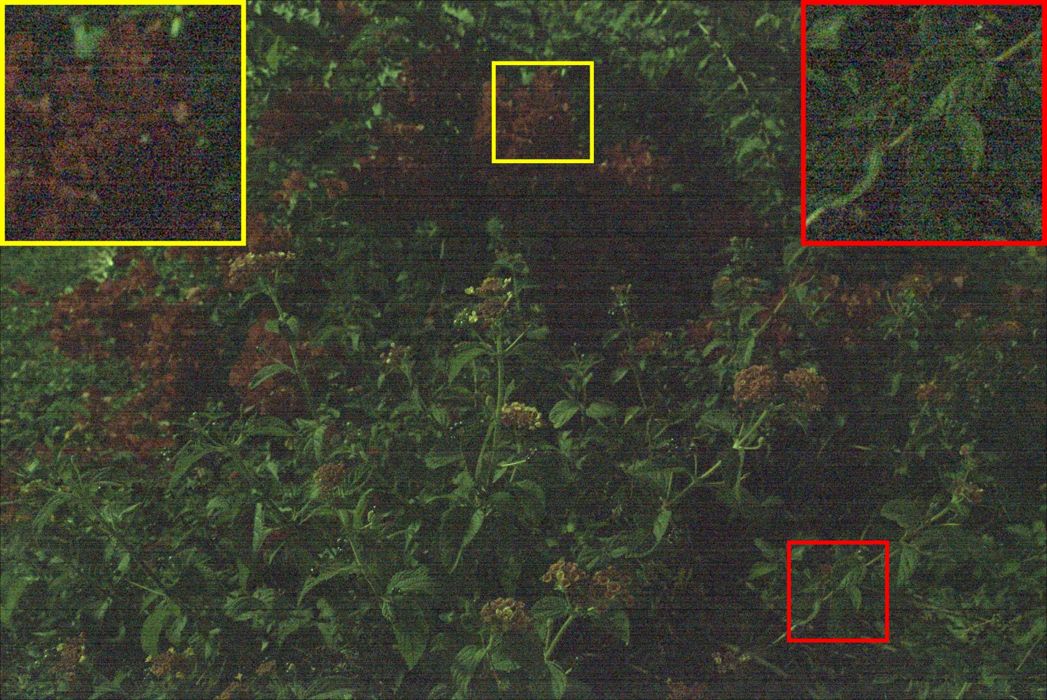}\hfill
		}
		\subfigure[Result by DCRaw]{
			\centering
			\includegraphics[width=1.8in]{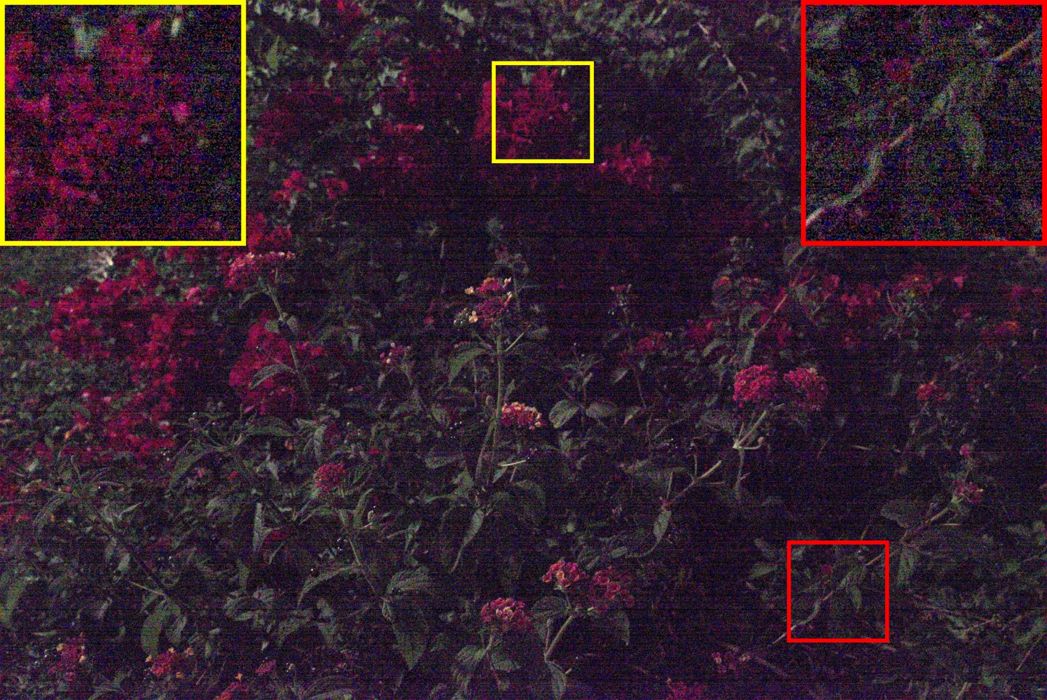}\hfill
		}
		\subfigure[Result by Camera Raw]{
			\centering
			\includegraphics[width=1.8in]{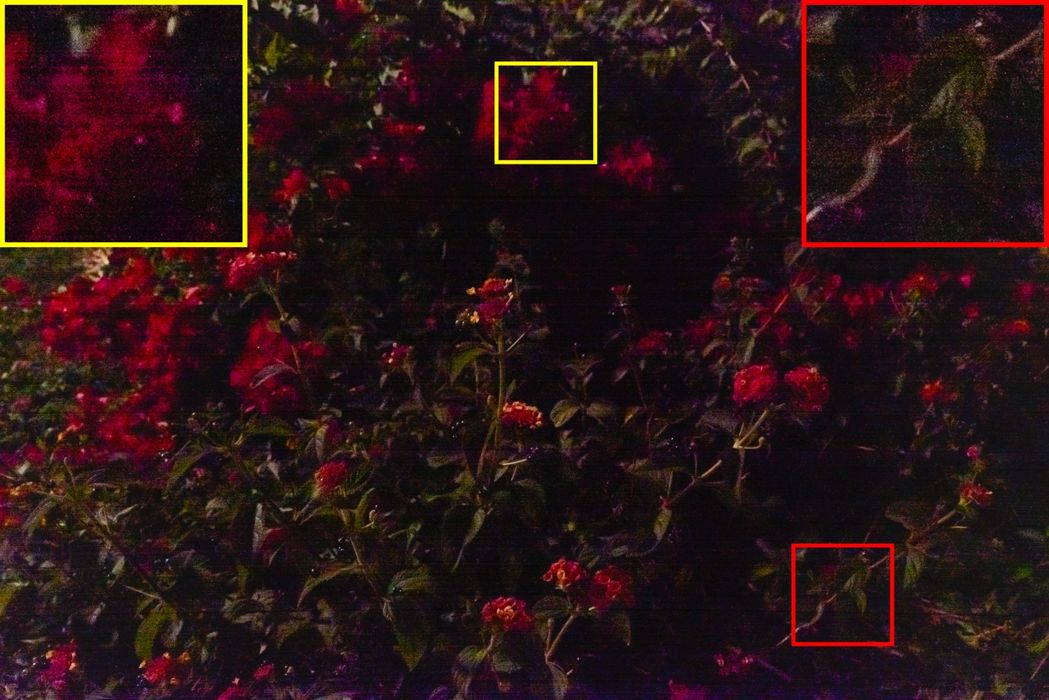}\hfill
		}
		\\
		\subfigure[Result by DeepISP-Net]{
			\centering
			\includegraphics[width=1.8in]{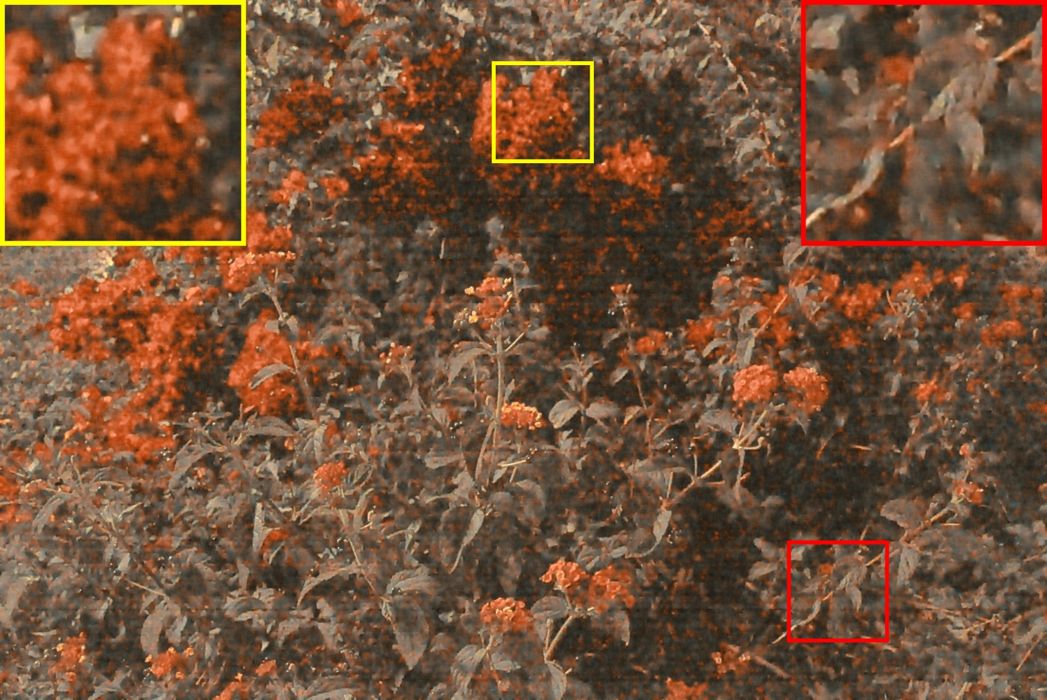}\hfill
		}
		\subfigure[Result by CameraNet]{
			\centering
			\includegraphics[width=1.8in]{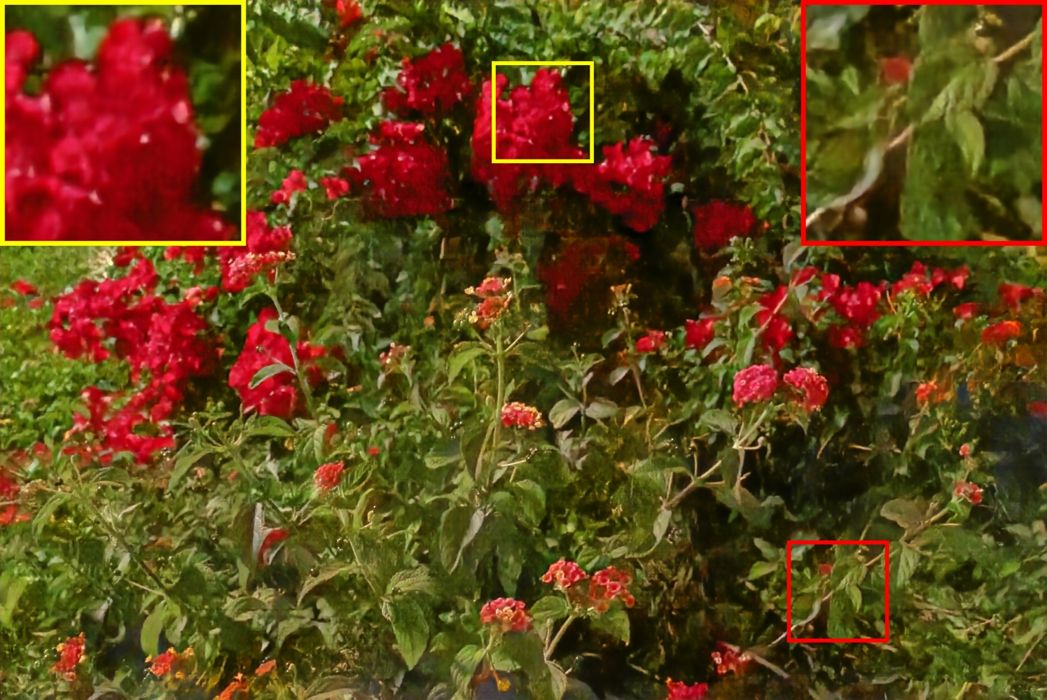}\hfill
		}
		\subfigure[Groundtruth]{
			\centering
			\includegraphics[width=1.8in]{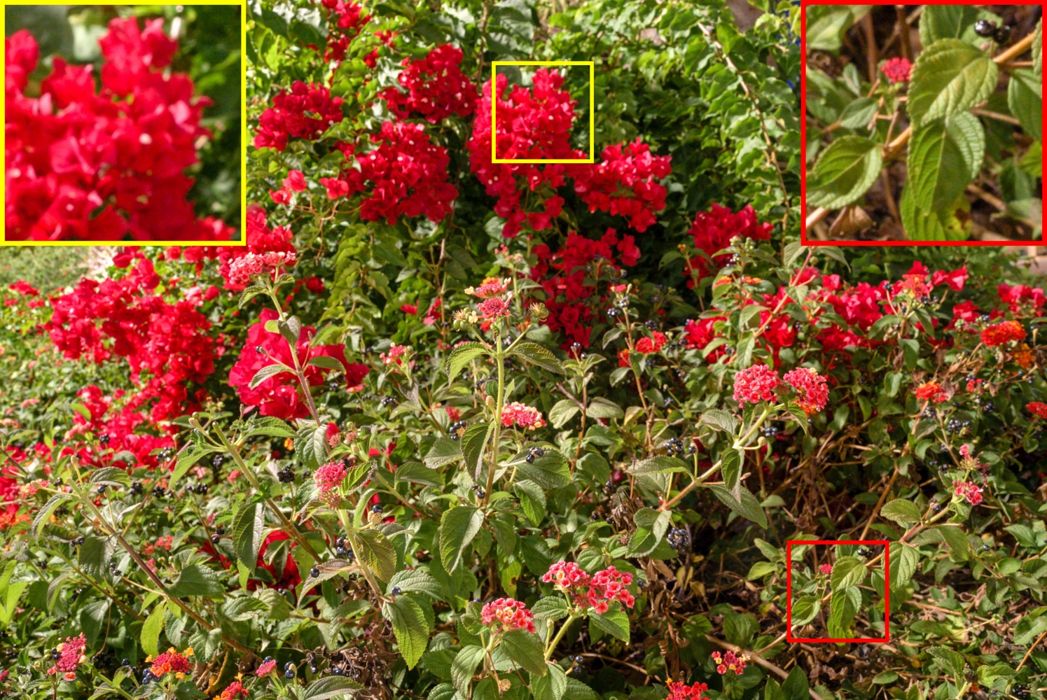}\hfill
		}
	\end{center}
	\vspace{-0.1in}
	\caption{Results on a flower image from the SID dataset \cite{Chen2018CVPR} by the competing methods. A gamma transform with parameter 2.2 is applied to the raw image for better visualization.}
	\label{fig14}
\end{figure*}

\begin{figure*}[htp]
	\begin{center}
		\subfigure[Raw image]{
			\centering
			\includegraphics[width=1.8in]{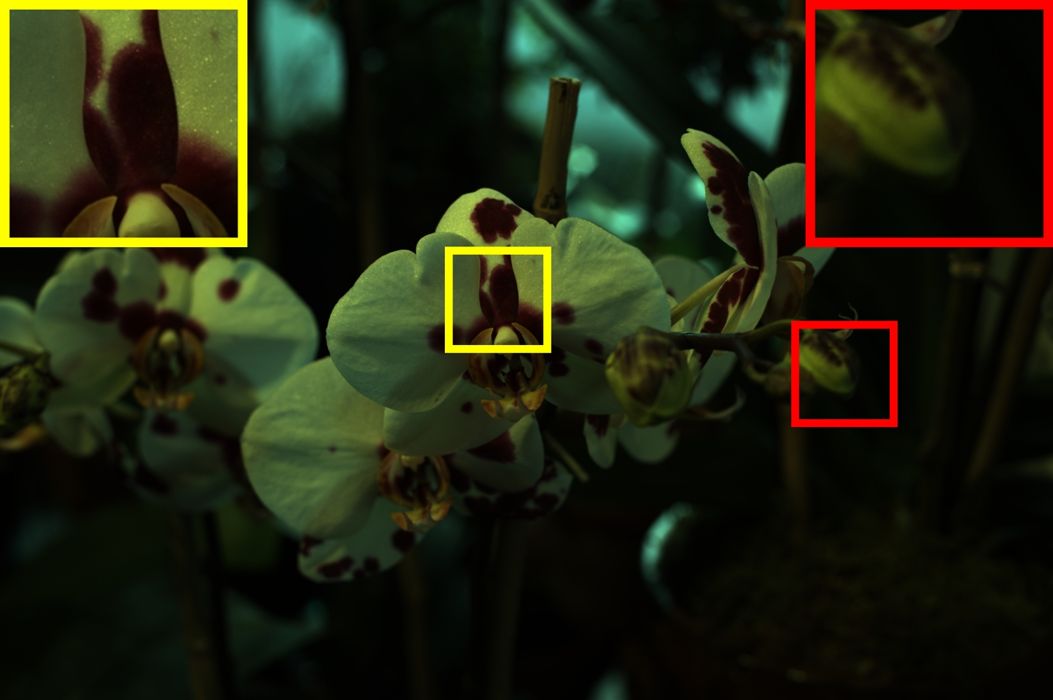}\hfill
		}
		\subfigure[Result by DCRaw]{
			\centering
			\includegraphics[width=1.8in]{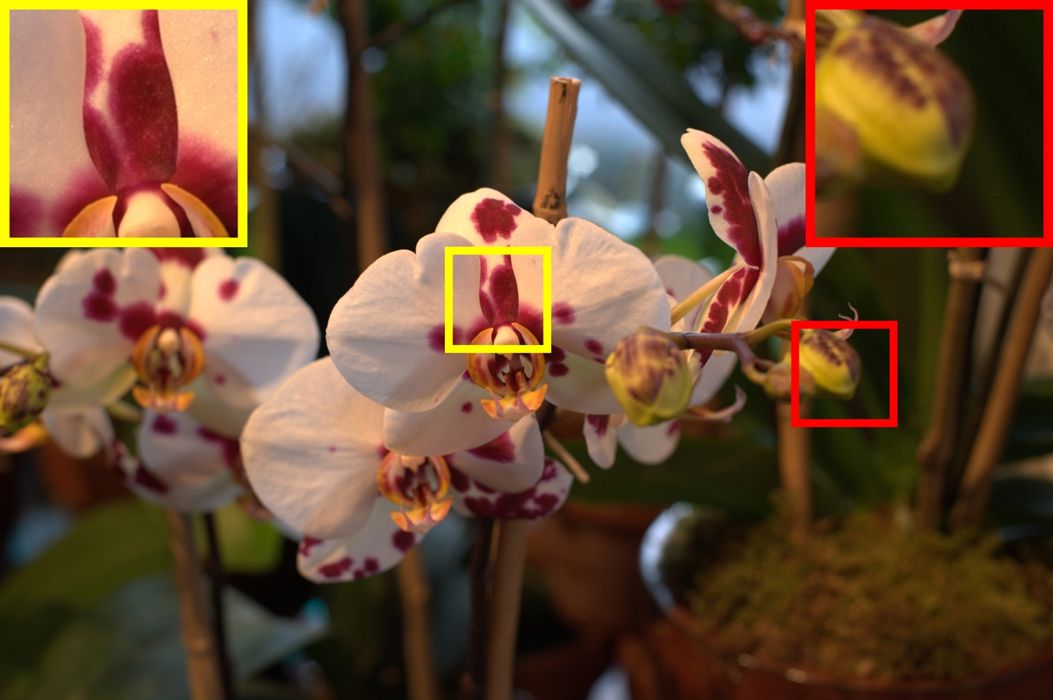}\hfill
		}
		\subfigure[Result by Camera Raw]{
			\centering
			\includegraphics[width=1.8in]{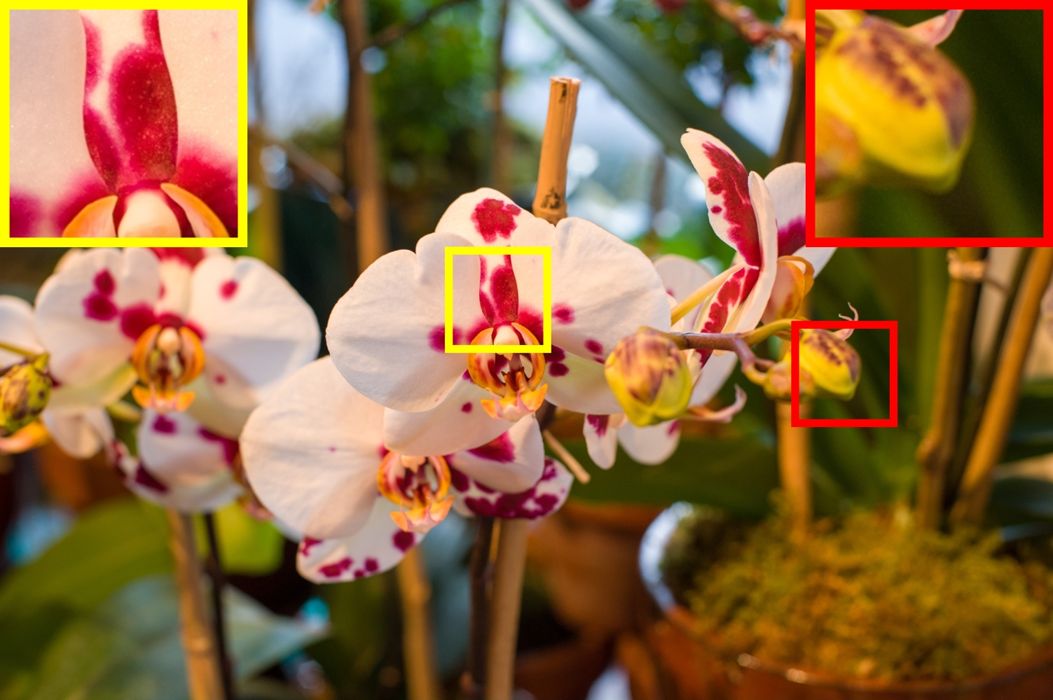}\hfill
		}
		\\
		\subfigure[Result by DeepISP-Net]{
			\centering
			\includegraphics[width=1.8in]{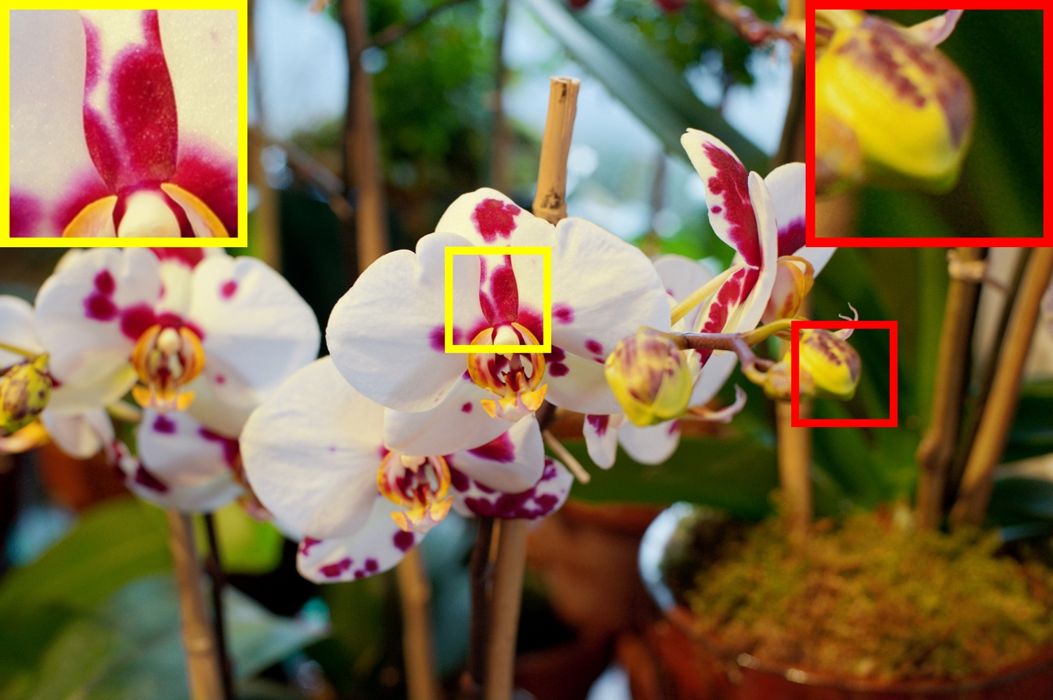}\hfill
		}
		\subfigure[Result by CameraNet]{
			\centering
			\includegraphics[width=1.8in]{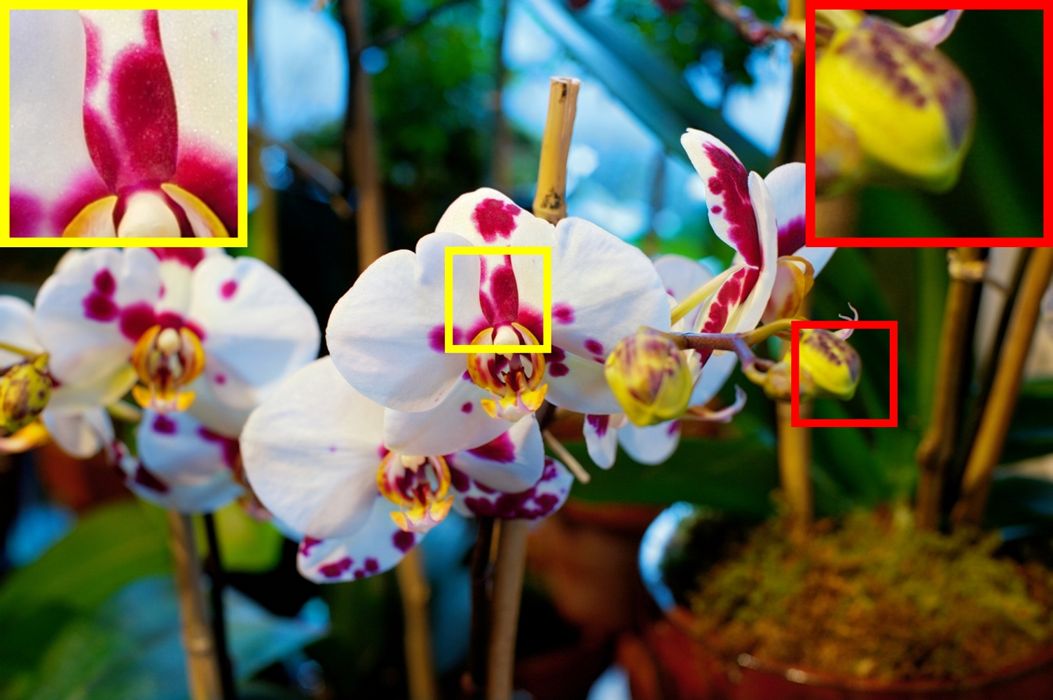}\hfill
		}
		\subfigure[Groundtruth]{
			\centering
			\includegraphics[width=1.8in]{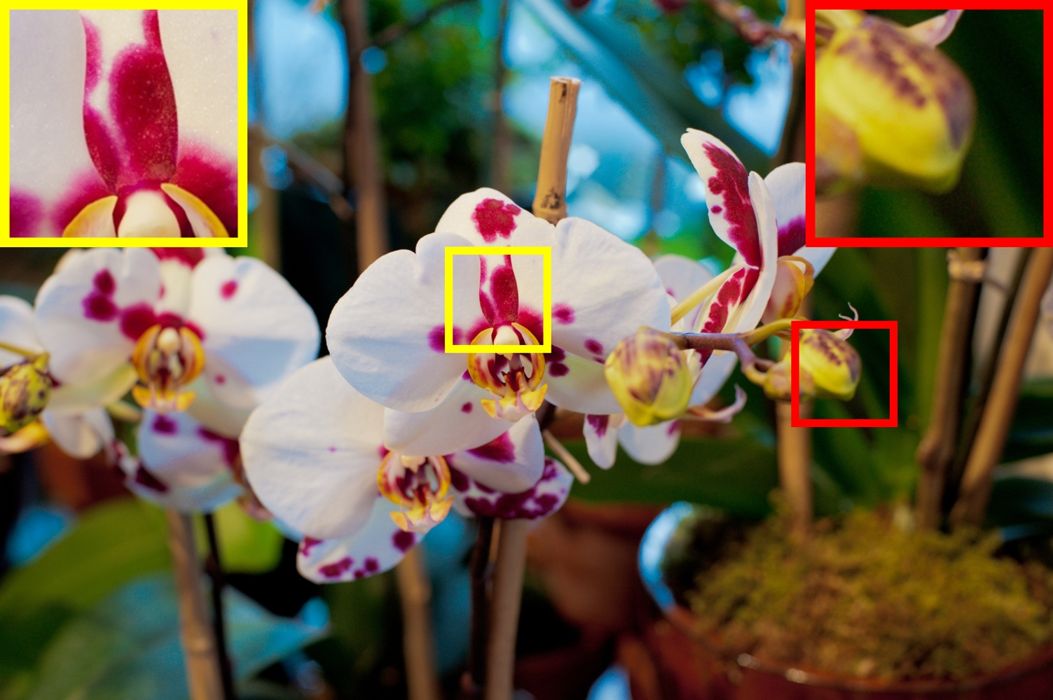}\hfill
		}
	\end{center}
	\vspace{-0.1in}
	\caption{Results on a flower image from the FiveK dataset \cite{Vladimir2011fivek} by the competing methods. A gamma transform with parameter 2.2 is applied to the raw image for better visualization.}
	\label{fig15}
\end{figure*}

\begin{figure*}[htp]
	\begin{center}
		\subfigure[Groundtruth]{
			\centering
			\begin{minipage}[b]{1.2in}
				\includegraphics[width=1.2in]{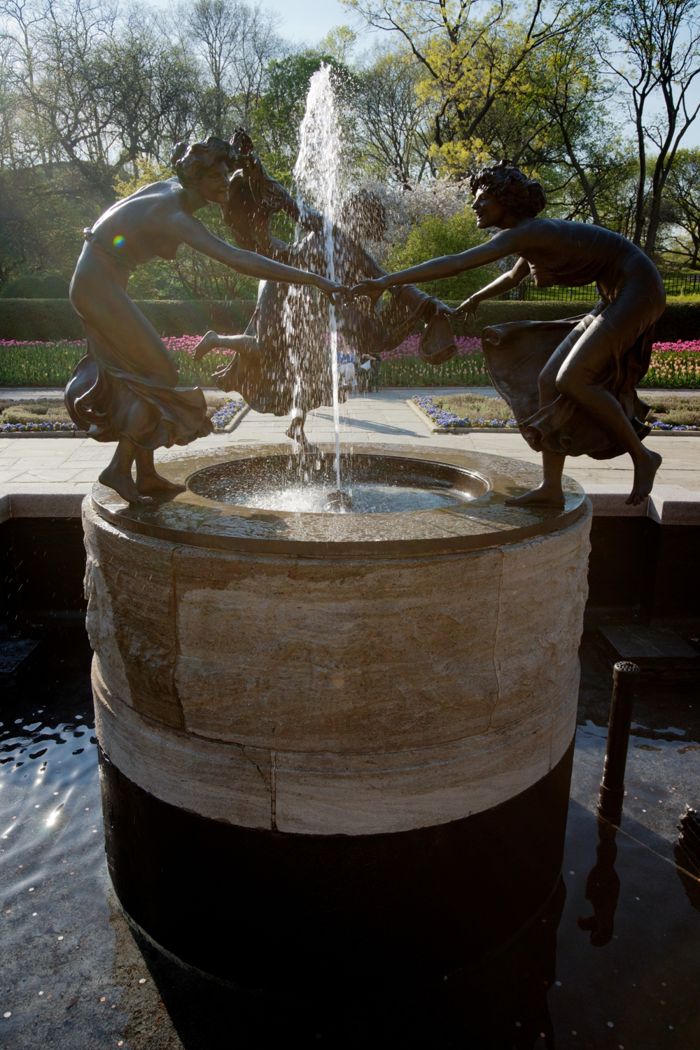}\hfill 
				\vspace{4pt}
				\includegraphics[width=1.2in]{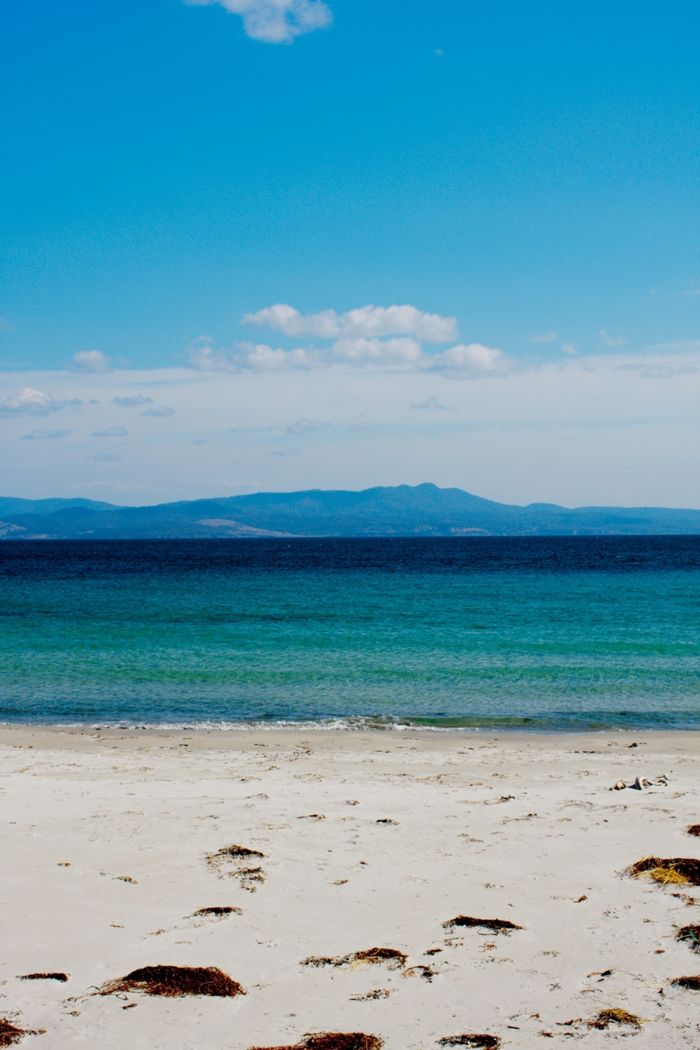}\hfill
			\end{minipage}
			
		}
		\subfigure[DeepISP-Net]{
			\centering
			\begin{minipage}[b]{1.2in}
				\includegraphics[width=1.2in]{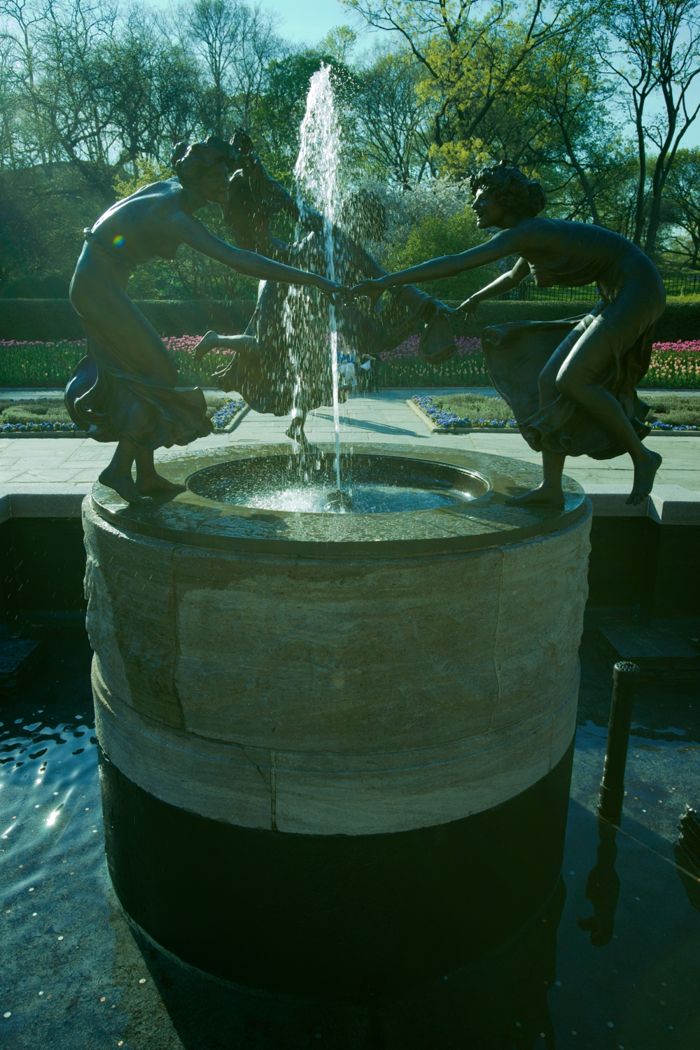}\hfill \vspace{4pt}
				\includegraphics[width=1.2in]{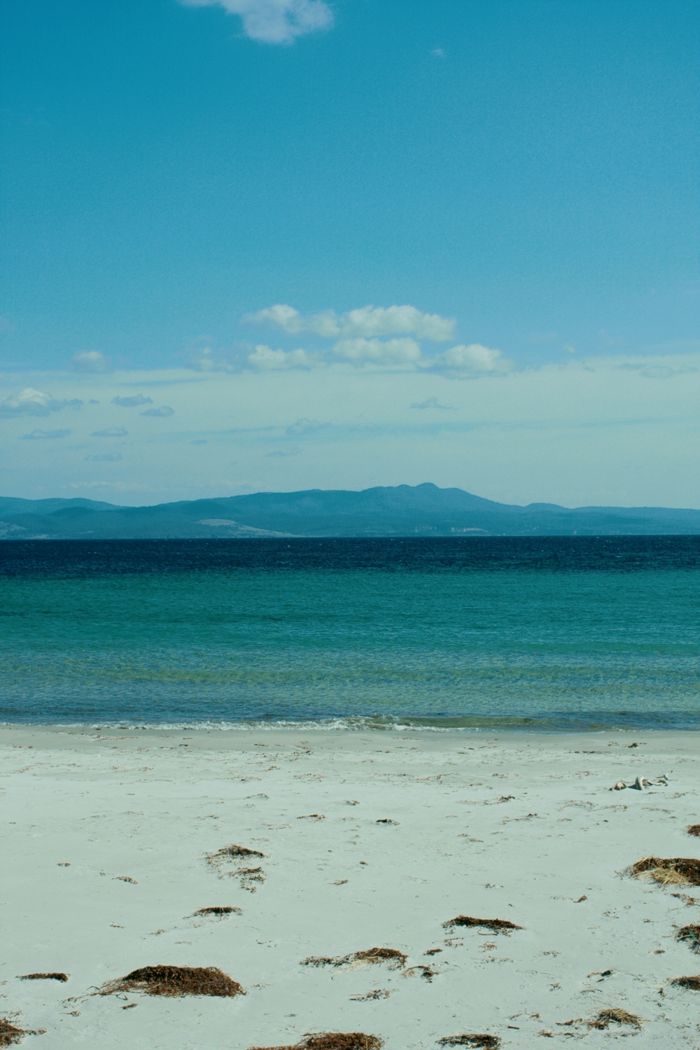}\hfill
			\end{minipage}
		}
		\subfigure[Color Error of (b)]{
			\centering
			\begin{minipage}[b]{1.2in}
				\includegraphics[width=1.2in]{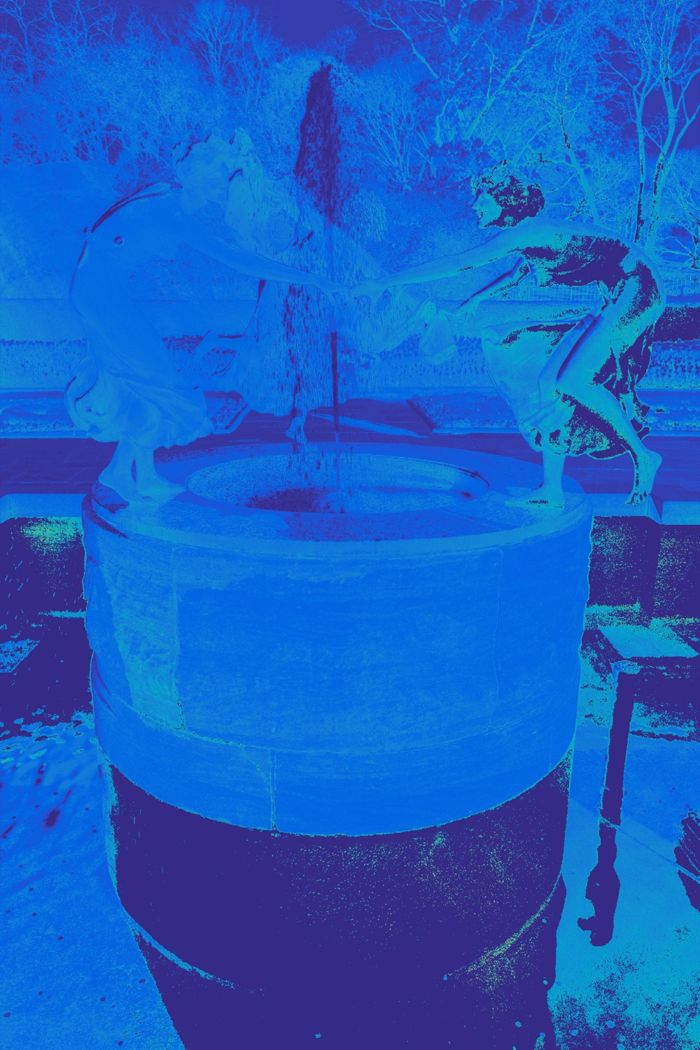}\hfill 
				\vspace{4pt}
				\includegraphics[width=1.2in]{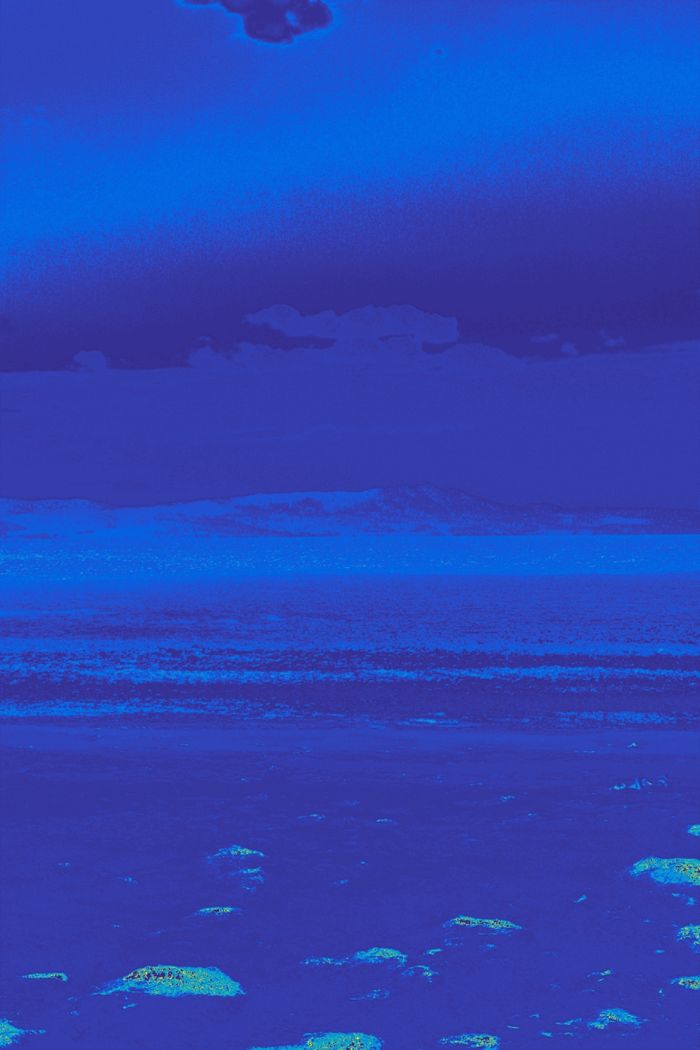}\hfill
			\end{minipage}
		}
		\subfigure[CameraNet]{
			\centering
			\begin{minipage}[b]{1.2in}
				\includegraphics[width=1.2in]{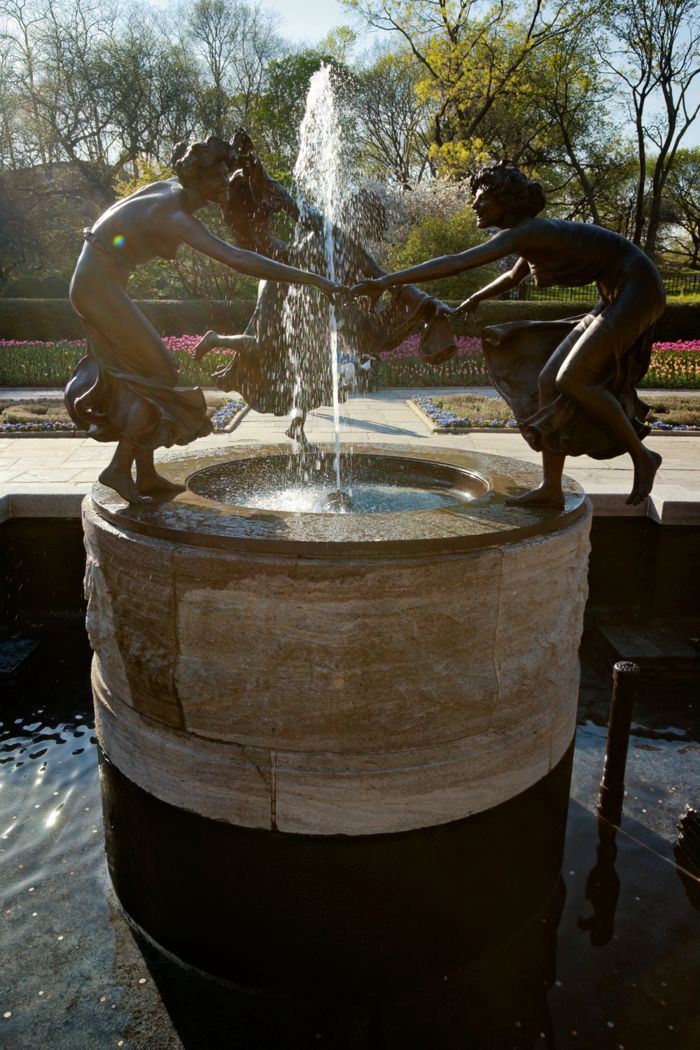}\hfill 
				\vspace{4pt}
				\includegraphics[width=1.2in]{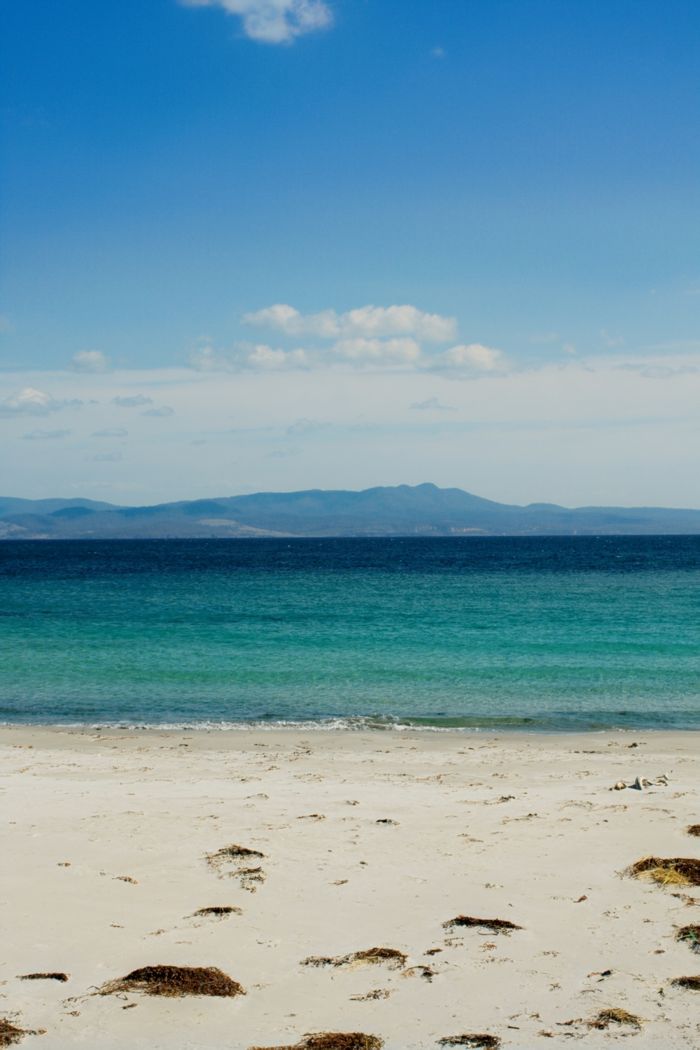}\hfill
			\end{minipage}
		}
		\subfigure[Color Error of (d)]{
			\centering
			\begin{minipage}[b]{1.2in}
				\includegraphics[width=1.2in]{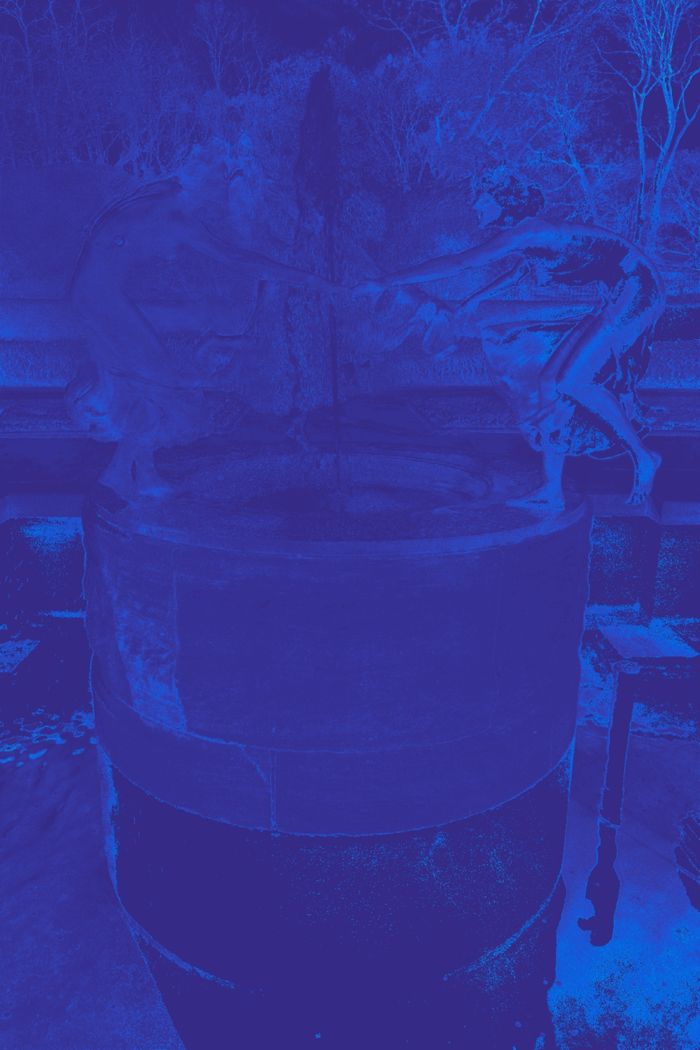}\hfill 
				\vspace{4pt}
				\includegraphics[width=1.2in]{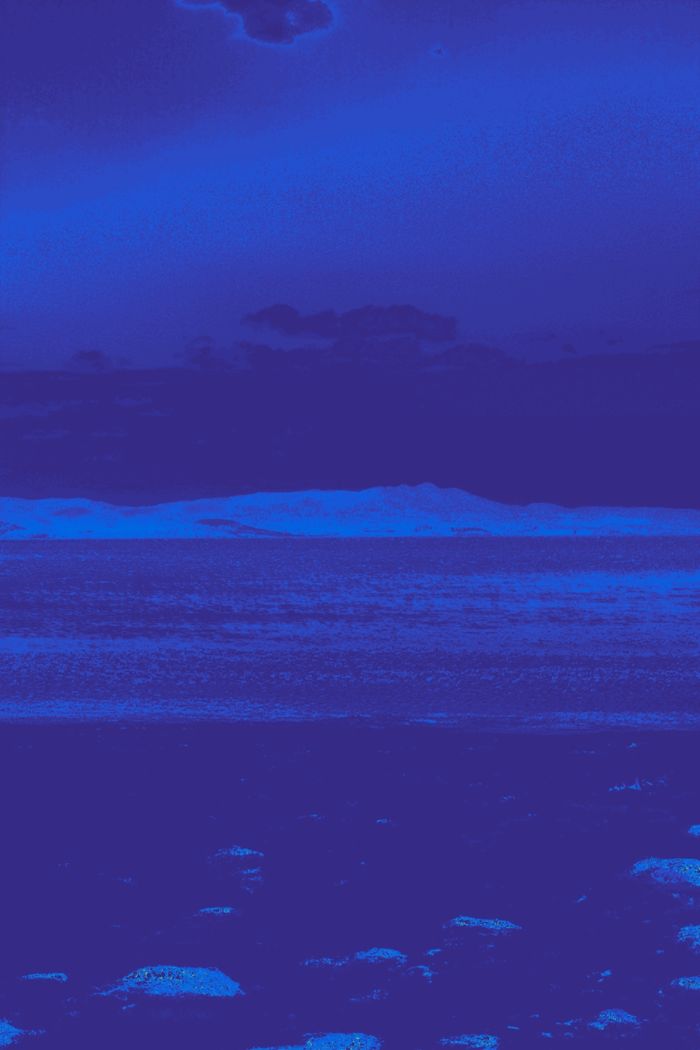}\hfill
			\end{minipage}
		}
		
	\end{center}
	\vspace{-0.1in}
	\caption{Results of cross camera testing. The first row shows the results of networks trained on Nikon D700 subset and tested on Sony A900 subset. The second row shows the results of networks trained on Nikon D700 subset and tested on Canon EOS 4D subset.}
	\label{fig16}
\end{figure*}

\begin{table*}[t]
	\centering	
	\caption{Comparison of cross-camera performance between DeepISP-Net and CameraNet. The indices are PSNR/SSIM/Color Error.}
	\label{tab3}
	\begin{tabular}{m{1.5cm}| m{2.2cm}<{\centering}| m{2.2cm}<{\centering}| m{2.5cm}<{\centering}|m{2.5cm}<{\centering}}
		\toprule[0.8pt] 
		& From Nikon D700 to its test set&From Nikon D700 to Sony A900 & From Nikon D700 to Canon EOS 5D& From Nikon D700 to Canon EOS 40D \\
		\hline
		DeepISP-Net & 22.59/0.845/$6.31^\circ$  & 17.39/0.758/$15.81^\circ$& 21.95/0.807/$7.93^\circ$&20.87/0.802/$8.24^\circ$\\
		\hline
		CameraNet & {\bf 23.37}/{\bf0.848}/${\bf6.04^\circ}$ & {\bf22.25}/{\bf0.825}/${\bf6.57^\circ}$& {\bf22.03}/{\bf0.811}/${\bf7.34^\circ}$&{\bf20.98}/{\bf0.805}/${\bf7.49^\circ}$ \\
		\bottomrule[0.8pt]
	\end{tabular}
	\vspace{-0.1in}
\end{table*}

\subsection{Comparison with Other ISP pipelines}

We then compare our CameraNet with the recently developed DeepISP-Net \cite{Schwartz2019tip} and two popular traditional ISP pipelines, including DCRaw and Adobe Camera Raw, on the HDR+, FiveK and SID datasets.
DeepISP-Net is proposed to address the ISP pipeline learning task in a one-stage manner. As the authors did not provide the source code, we implement DeepISP-Net and train it with enough epochs until convergence. DCRaw is an open-source library for ISP pipeline development. To obtain the sRGB image, we use the default settings in DCRaw. Adobe Camera Raw is a tunable ISP pipeline for flexible sRGB reconstruction. To obtain the sRGB image, we use the auto-mode in Camera Raw to adjust the image, and manually look for the best noise reduction setting for each compared image. The comparison results are shown in Table \ref{tab2}.

{\bf Results on HDR+ dataset.}  As can be seen from Table \ref{tab2}, the proposed CameraNet achieves substantially better objective scores than DeepISP-Net. This is because the two-stage nature of CameraNet effectively accounts for the restoration and enhancement tasks contained in HDR+ dataset. In contrast, the one-stage DeepISP-Net achieves lower scores due to the mixture of the two uncorrelated set of operations. Figs.\ \ref{fig11} and \ref{fig12} show the results of the compared methods, from which one can see that DeepISP-Net produces visual artifacts in the outputs. In contrast, the proposed CameraNet avoids this problem and produces visually pleasing results. Additionally, DCRaw and Camera Raw produce inferior results because of the inherent limitation of traditional algorithms.

{\bf Results on SID dataset.}  From Table \ref{tab2}, we can the that CameraNet outperforms DeepISP-Net by a large margin on the SID dataset. This is because the noise level in SID dataset is larger than that in HDR+ dataset. This requires the CNN model to have stronger denoising capability. Our CameraNet meets this requirement by explicitly expressing the denoising operation in the first stage, whereas the DeepISP-Net has weak capability of denoising by mixing all the ISP subtasks together, leading to inferior learning performance. Figs.\ \ref{fig13} and \ref{fig14} show the results of the compared methods. We can see that the visual quality of the reconstructed images by the proposed CameraNet is significantly higher than that of DeepISP-Net. DeepISP-Net not only amplifies noise in the raw image, but also produces inaccurate colors. In contrast, CameraNet effectively reduces the noise level and enhances the image structure. Moreover, we can see that the DCraw and Camera Raw have very low reconstruction quality. They are not able to effectively reduce the noise and restore the correct color in such low-light scenarios. 

{\bf Results on FiveK dataset.}  On this dataset, we can see that the SSIM scores of CameraNet and DeepISP-Net are comparable, while the PSNR and Color Error indices of CameraNet are better than those of DeepISP-Net. This is because the complexity of restoration-related tasks in FiveK dataset is not as high as the HDR+ and SID datasets. Indeed, FiveK dataset mainly contains high-end cameras with good sensors whose noise levels are not high. As a result, the dominant tasks in FiveK dataset are color manipulations, which can be learned well by the one-stage-based DeepISP-Net. Fig.\ \ref{fig15} shows the results of the compared methods, from which we can see that the results by CameraNet and DeepISP-Net are both satisfactory due to the simplicity of the tasks in FiveK.

{\bf Cross camera testing.} Finally, we compare the cross camera performance between CameraNet and DeepISP-Net. When trained on one camera device, we examine to what extent the network can be applied to a new camera device. We use the FiveK dataset for this experiment because it contains diverse camera models and the training image pairs are well-aligned. We apply the CameraNet and DeepISP-Net trained on Nikon D700 subset to Sony A900, Canon EOS 5D and Canon EOS 40D subsets. For every target Camera subset, 50 images are used to evaluate the performance.

Table \ref{tab3} illustrates the objective indices of the cross camera experiment. We have two observations. First, both DeepISP-Net and CameraNet encounter certain drops in objective indices when transfered to other devices. In spite of this, the objective scores of CameraNet are much better than DeepISP-Net. The performance of DeepISP-Net is unstable as it has a significant drop when transfered to Sony A900 subset. Second, in all cases CameraNet produces substantially lower color error than DeepISP-Net. This merit can be attributed to the fact that CameraNet adds a RGB-to-XYZ conversion preprocessing operation before CNN learning, which accounts for the color space differences in different camera sensors. As can be seen from Fig.\ \ref{fig16}, the results by CameraNet have similar color with the groundtruth, whereas the results by DeepISP-Net have strong global color bias because it does not account for the color space differences in different sensors. 

{\bf Computational complexity.} The proposed CameraNet is composed of two U-Nets, and it takes 3306.69 GFLOPS to process a 4032$\times$3024 sized image, while DeepISP-Net takes 12869.79 GFLOPS. The lower computational complexity of CameraNet can be attributed to its multi-scale operations. On the other hand, CameraNet has 26.53 million parameters while DeepISP-Net has 0.629 million parameters.

\section{Conclusion}

We proposed an effective and general two-stage CNN system, namely CameraNet, for data-driven ISP pipeline learning. We exploited the intrinsic correlations among the ISP components and categorized them into two sets of restoration and enhancement operations, which are weakly correlated. The proposed CameraNet system adopted a two-stage modules to account for the two independent operations, hence improving the learning capability while maintaining the model compactness. Two groundtruths were specified to train the two-stage model. Experiments showed that in terms of ISP pipeline learning, the proposed two-stage CNN framework significantly outperforms the traditional one-stage framework that is commonly used for deep learning. Additionally, the proposed CameraNet outperforms state-of-the-art ISP pipeline models on three benchmark ISP datasets in terms of both subjective and objective evaluations.

\ifCLASSOPTIONcaptionsoff
  \newpage
\fi

{
\footnotesize
\bibliographystyle{IEEEtran}
\bibliography{tipref}
}

\end{document}